\documentclass[pdflatex,sn-mathphys-num]{sn-jnl}

\usepackage{graphicx}
\usepackage{amsmath,amssymb,amsfonts,amsthm}
\usepackage{mathrsfs}
\usepackage[title]{appendix}
\usepackage{xcolor}
\usepackage[most]{tcolorbox} 
\usepackage{textcomp}
\usepackage{manyfoot}
\usepackage{booktabs}
\usepackage{paralist}
\usepackage{pifont}
\usepackage{svg}

\usepackage{algpseudocode}
\usepackage{listings}
\usepackage{tabularx,threeparttable,array}
\usepackage{pifont}

\usepackage{amsmath,amssymb}  \usepackage[ruled,vlined]{algorithm2e}
\usepackage{multicol,multirow}
\usepackage[table]{xcolor}
\usepackage{url}
\usepackage[strings]{underscore}

\usepackage[resetlabels]{multibib}
\newcites{SP}{Selected Papers}
\makeatletter
\let\OriginalCiteSP\citeSP
\renewcommand{\citeSP}[1]{%
  \begingroup\def\NAT@open{[SP}\OriginalCiteSP{#1}\endgroup
}
\makeatother

\newcolumntype{W}{>{\raggedright\arraybackslash}X} 
\newcolumntype{C}{>{\centering\arraybackslash}X}   

\newcommand{\cmark}{\ding{51}} 

\definecolor{codegreen}{rgb}{0,0.6,0}
\definecolor{codegray}{rgb}{0.5,0.5,0.5}
\definecolor{codepurple}{rgb}{0.58,0.3,0.82}
\definecolor{backcolour}{rgb}{0.95,0.95,0.95}
\definecolor{headbg}{gray}{0.93}

\newcommand*{\DiffAdded}{%
  \rlap{\color{green!20}\rule[-0.5ex]{\linewidth}{\baselineskip}}%
}

\newcommand*{\DiffRemoved}{%
  \rlap{\color{red!20}\rule[-0.5ex]{\linewidth}{\baselineskip}}%
}

\newcommand{\ccite}[1]{%
  \setlength{\fboxrule}{0.7pt}%
  \setlength{\fboxsep}{0.6pt}%
  \fcolorbox{green}{white}{\cite{#1}}%
}

\lstdefinestyle{mystyle}{
    language=python,
    backgroundcolor=\color{backcolour},
    commentstyle=\color{codegreen},
    keywordstyle=\color{magenta},
    numberstyle=\tiny\color{codegray},
    stringstyle=\color{codepurple},
    basicstyle=\ttfamily,
    breaklines=true,
    captionpos=b,
    keepspaces=true,
    numbers=left,
    numbersep=5pt,
    showspaces=false,
    showstringspaces=false,
    showtabs=false,
    tabsize=2,
    frame=tb,
    rulecolor=\color{gray},
    framerule=0.9pt,
    breaklines=true,
    breakatwhitespace=true,
    breakautoindent=true,
    postbreak=\mbox{\textcolor{gray}{\tiny$\hookrightarrow$}\space},
}

\theoremstyle{thmstyleone}

\theoremstyle{thmstyletwo}

\theoremstyle{thmstylethree}

\colorlet{blue}{black}
\begin{document}

\title[Article Title]{LLM Code Smells: A Taxonomy and Detection Approach}
\author*[1]{\fnm{Zacharie} \sur{CHENAIL-LARCHER}}\email{zacharie.chenail.larcher@gmail.com}
\equalcont{The first two authors contributed equally to this work and share first authorship.}

\author*[1]{\fnm{Brahim} \sur{MAHMOUDI}}\email{brahim.mahmoudi.1@ens.etsmtl.ca}
\equalcont{The first two authors contributed equally to this work and share first authorship.}

\author[1]{\fnm{Naouel} \sur{MOHA}}\email{naouel.moha@etsmtl.ca}

\author[2]{\fnm{Quentin} \sur{STIÉVENART}}\email{stievenart.quentin@uqam.ca}

\author[2]{\fnm{Florent} \sur{AVELLANEDA}}\email{avellaneda.florent@uqam.ca}

\affil*[1]{\orgname{École de technologie supérieure},
\orgaddress{\city{Montréal}, \state{Québec}, \country{Canada}}}

\affil[2]{\orgname{Université du Québec à Montréal},
\orgaddress{\city{Montréal}, \state{Québec}, \country{Canada}}}

\abstract{Large Language Models (LLMs) are increasingly integrated into software systems for diverse purposes, due to their versatility, flexibility, and ability to simulate human reasoning to some extent. However, poor integration of LLM inference in source code can undermine software system quality. Therefore, inadequate LLM integration coding practices must be documented to help developers mitigate such issues. Following our earlier work on LLM code smells, this paper consolidates and refines the concept by presenting a self-contained taxonomy and a catalog of nine LLM code smells. We also create SpecDetect4LLM, a static source code analysis tool for their detection, and conduct extensive empirical evaluations of its detection effectiveness (precision and recall) as well as the prevalence of LLM code smells across 692 open-source software projects (171,194 source files). Our results show that LLM code smells affect 73.5\% of the analyzed systems, with a detection precision of 91.3\% and a recall of 71.8\%.}


\maketitle

\section{Introduction}

In recent years, Large Language Models (LLMs) have revolutionized the way information is processed and have gained increasing importance in everyday life. Their impact is such that the number of LLM-related publications has grown rapidly year after year, increasing from 503 in 2019 to 7109 in 2024 across 77 related conferences~\ccite{xia2025analyzing16193llmpapers}.  They are also being integrated into a growing number of software systems~\ccite{shao2025llmscorrectlyintegratedsoftware}, due to their versatility, eloquence, flexibility, and their ability to simulate human reasoning to some extent.

However, LLMs are not always reliable~\ccite{khatun2024reliability}, and their integration into software systems can introduce new risks and challenges due to their fundamentally different nature compared to traditional software components, notably their non-determinism, distinct hyperparameters, and interaction through natural language. Moreover, their performance and behaviour can vary significantly depending on how they are used~\ccite{YANG2025113503}. To ensure the reliability, robustness, performance, and maintainability of LLM-integrating systems (systems that use LLMs as components, ranging from simple inference to extensive agentic logic), it is essential to properly integrate them, both at the architectural level and within the source code itself~\ccite{bucaioni2025functionalsoftwarereferencearchitecture}. Therefore, as previously done for general~\ccite{10.5555/311424} and machine learning-specific practices~\ccite{zhang2022codesmellsmachinelearning}, establishing coding guidelines, including the specification of code smells, is essential for LLM-integrating systems.

While prior studies have defined taxonomies of general defects in LLM-integrating systems~\ccite{shao2025llmscorrectlyintegratedsoftware} and prompt-related defects~\ccite{ tian2025taxonomypromptdefectsllm}, to our knowledge, there is no dedicated concept addressing code-specific poor practices for the integration of LLMs in software systems. \textcolor{blue}{To avoid ambiguity, we use the term LLM code smells to refer to poor coding practices in human-written source code that integrates or orchestrates LLM inference within software systems, rather than code smells in source code generated by LLMs, which belong to a distinct line of work on defects and smells in LLM-generated code~\ccite{paul2025smells}. This notion was first introduced in our previous ICSE 2026 New Ideas and Emerging Results (NIER) paper, \textit{Specification and Detection of LLM Code Smells}~\ccite{Mahmoudi2026LLMCodeSmells}, which provided an initial catalog of five LLM code smells, a first version of \textit{SpecDetect4LLM}, and preliminary empirical results on detection precision and prevalence across 200 open-source projects. Our previous paper provides preliminary results, while the present paper constitutes the full study.}

\textcolor{blue}{This paper substantially extends our earlier work in the following ways. First, we revisit, refine, and consolidate the concept of LLM code smells itself. Second, we expand the catalog from five to nine smells and organise it into a taxonomy. Third, we extend \textit{SpecDetect4LLM} to cover the newly identified smells and provide a more detailed description of its static analysis scope and design choices. Fourth, we substantially enlarge the empirical evaluation by evaluating recall in addition to precision. Fifth, we expand the evaluation corpus from 200 to 692 open-source LLM-integrating systems, which allows a broader and more robust prevalence assessment. Accordingly, the present paper builds on the earlier NIER paper but substantially extends it in scope, methodological detail, and empirical validation.}

The contributions of this journal paper are as follows:
\begin{enumerate}
    \item We provide a self-contained taxonomy and catalog of nine LLM code smells.
    \item We extend \textit{SpecDetect4LLM} to cover the enlarged catalog and clarify its current static analysis scope. \textit{SpecDetect4LLM} is a standalone static analysis tool for detecting LLM code smells. It is derived from the AI-specific code smells detection tool \textit{SpecDetect4AI}~\ccite{mahmoudi2025ai}.
    \item We empirically evaluate the detection effectiveness of \textit{SpecDetect4LLM} in terms of both precision and recall on a stratified random sample of 381 Python source code files drawn from a corpus of 171,196 files.
    \item We conduct a large-scale empirical assessment of the prevalence of LLM code smells across 692 open-source LLM-integrating systems.
\end{enumerate}



This paper is structured as follows:
\begin{itemize}
    \item \textbf{Background (~\ref{sec:background})} : Presents essential concepts and definitions.
    \item \textbf{General Methodology (~\ref{sec:study_design})}: Describes the overall structure and methodology followed in this work.
    \item \textbf{Catalog Construction (~\ref{sec:catalog_contruction})} : Presents the research methodology leveraged to construct the LLM code smell catalog, including a systematic review of scientific literature and structured mining of grey and empirical literature.
    \item \textbf{LLM Code Smells (~\ref{sec:catalog_taxo})}: Introduces the taxonomy and the catalog.
    \item \textbf{Detection Approach with \textit{SpecDetect4LLM} (~\ref{sec:detection_approach})}: Explains the design and detection workflow of \textit{SpecDetect4LLM}.
    \item \textbf{Validation Design and Results (~\ref{sec:resuls_and_eval})}: Empirically studies the detection correctness of \textit{SpecDetect4LLM}.
    \item \textbf{Prevalence Study (~\ref{sec:prevalence_study})}: Describes our empirical assessment of LLM code smells prevalence in open-source systems and the methodology followed.
    \item \textbf{Limitations and threats to validity (~\ref{sec:limit_threats})}: Documents the limitations and threats to validity that affect our work.
    \item \textbf{Related Work (~\ref{sec:related_works})}: Discusses the state of the art and related scientific work.
    \item \textbf{Conclusion (~\ref{sec:conclusion})}: Summarizes the paper and concludes it.
\end{itemize}

\label{sec:intro}

\section{Background}
\label{sec:background}

This section includes the definitions and concepts required to understand this paper, including the different types of LLMs, the concept of LLM code smells, and the software quality attributes affected by these smells.

\subsection{Classes of LLMs and Related Concepts}
\textcolor{blue}{
\textbf{Large Language Models (LLMs).} LLMs are large-scale text generation models that are generally based on the transformer architecture \ccite{10.5555/3295222.3295349} and generate one text token at a time during inference. Beyond these general-purpose language models, recent progress has led to specialised variants designed to extend the range of supported inputs and inference behaviours. In particular, two important classes are Vision-Language Models (VLMs) and Reasoning Language Models (RLMs).}

\textcolor{blue}{
\noindent\textbf{Vision-Language Models (VLMs).} VLMs, also referred to as Large Vision-Language Models (LVLMs), extend standard LLMs by incorporating visual tokens in addition to text tokens. This multimodal capability allows them to jointly process images and natural language, thereby enabling the association of visual and textual information for tasks such as visual understanding, captioning, and multimodal question answering \ccite{zhang2024vlmsurvey}. In this sense, VLMs can be seen as a natural extension of standard LLMs toward multimodal inputs.}

\textcolor{blue}{\noindent\textbf{Reasoning Language Models (RLMs).} While VLMs mainly extend LLMs along the input modality dimension, RLMs extend them along the inference dimension. RLMs, also referred to as Reasoning Models, are language models designed to solve tasks by generating intermediate reasoning steps before producing a final answer. Compared to general-purpose LLMs, RLMs can provide more accurate results on complex tasks \ccite{openai2024learning}, but they generally require longer processing time and higher token costs due to the additional intermediate steps.}

\textcolor{blue}{\noindent\textbf{Tokens.} Tokens are the basic units into which AI model inputs and outputs are divided for processing. Depending on the format, tokens may represent different types of information, such as text or visual content. In this work, unless explicitly stated otherwise, \textbf{the term token refers to text tokens}. \textbf{Text tokens} correspond to discrete \textbf{textual units} used as the basic elements for text generation and LLM inference. In VLMs, \textbf{visual tokens} represent encoded visual information that enables image processing.}

\subsection{LLM Code Smells}


\noindent\textbf{Code smells} refer to low-level coding practices that degrade software quality without necessarily representing explicit bugs or failures~\ccite{10.5555/311424}. They are typically granular source-code patterns that negatively affect software quality and may indicate deeper design issues within a system. Over time, this concept has been extended through the development of catalogs and taxonomies tailored to specific application domains. 

\noindent Therefore, \textbf{LLM code smells} are code-level poor practices specific to the integration and use of LLM and VLM inference within software systems. While they do not always directly cause bugs or failures, they undermine key software quality attributes such as maintainability, reliability, performance, and robustness.

\noindent Following this perspective, \textbf{LLM code smells} are code-level poor practices specific to the integration and use of LLM and VLM inference within software systems \textcolor{blue}{rather than code smells in source code generated by LLMs \ccite{paul2025smells}. The study of code smells in LLM generated code belongs to a separate scope of research, whereas our work focuses on the quality of software that integrates or coordinates LLM based functionality.} Although LLM code smells do not always directly lead to failures, they may still negatively affect key software quality attributes, including maintainability, reliability, performance, and robustness.

\subsection{Effects and Quality Attributes}

The LLM code smells presented in this paper may affect any of four software quality attributes: robustness, performance, maintainability, and reliability \ccite{iso24765_2017,iso25010_2023,ieee61012_1990}. A code smell associated with a given attribute undermines it.\\ 

\begin{itemize}
    \item \textbf{Robustness.} We use robustness to refer to the \textbf{operational resilience} and the ability of a software system to remain available in the presence errors, failures, unexpected conditions, or adverse situations, such as malicious user behaviors. It characterizes the system's capacity to remain functional in faulty scenarios that could otherwise lead to service outages. LLM code smells that undermine robustness may increase the likelihood of runtime failures, system crashes, or loss of service.\\
     
    \item \textbf{Performance.} We use performance to refer to the ability of a software system to \textbf{execute its tasks with minimal computational overhead and resource costs}, including execution time, memory usage, computational resources, and monetary cost. LLM code smells that affect performance may introduce unnecessary latency, memory consumption, or monetary costs during system execution.\\
     
    \item \textbf{Maintainability.} Maintainability is the \textbf{capacity to preserve the integrity} of the software artifact during both development and execution. Systems with low maintainability may exhibit unnecessary coupling between components, poorly documented settings, or insufficient logging and execution traces. We use maintainability to relate to \textbf{reproducibility, portability, traceability, and observability} of LLM-integrating systems. It concerns artifacts such as source code, configuration files and logs.\\
     
    \item \textbf{Reliability.} We use reliability to refer to the \textbf{ability of a software system and its LLM inference to consistently produce correct and effective results over time}. It captures both the quality of the system's outputs and the stability of its behavior. LLM code smells that undermine reliability may degrade the quality of inference results, introduce behavioral inconsistencies, or lead to silent and passive changes in system behavior over time. In contrast to maintainability, which concerns the ease of evolving and managing a system, reliability focuses on the correctness, consistency, and behavioral stability of the system during execution.
\end{itemize}









\section{General Methodology}
\label{sec:study_design}

\begin{figure}[t]
  \centering
  \includegraphics[width=0.8\linewidth]{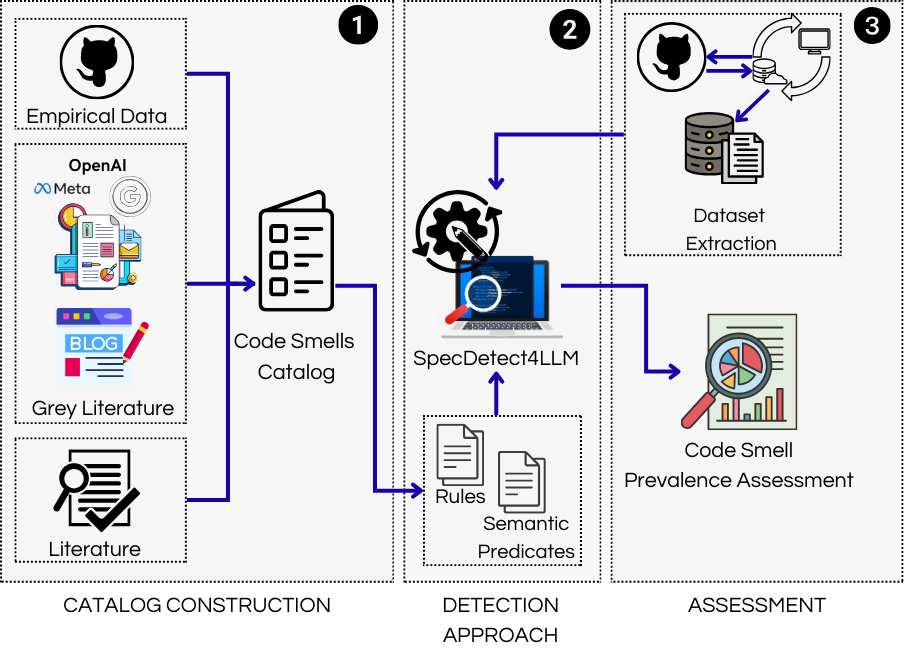}
  \caption{Research design and workflow of the study.}
  \label{fig:research-workflow}
\end{figure}

This study aims to propose a taxonomy and catalog of LLM code smells as theoretical guidelines to support the development of LLM-integrating systems, assess their prevalence in real systems to validate their legitimacy, and provide a detection tool to facilitate their practical use. As shown in Figure~\ref{fig:research-workflow}, this work follows an overall methodology composed of three main phases. This section provides an overview of each of these phases, which are described in much greater detail across their respective dedicated sections.

\subsection{Phase 1: Catalog Construction}

\noindent\textbf{Dedicated sections:} \textit{LLM Code Smells} (\ref{sec:catalog_taxo}) covers the taxonomy and presents the catalog, while \textit{Catalog Construction} (\ref{sec:catalog_contruction}) presents in greater details the methodology of this phase in detail.

\noindent\textbf{Goal:} Consolidate evidence to produce a taxonomy and catalog of LLM code smells.

\noindent\textbf{Inputs:} Academic literature search queries, and research prompts for empirical and grey literature.

\noindent\textbf{Outputs:} A taxonomy and a catalog of LLM code smells.

\noindent\textcolor{blue}{\textbf{Description:} We examined three complementary families of sources, namely academic, grey, and empirical literature, following established guidance for evidence based synthesis in software engineering~\ccite{kitchenham2007guidelines}. Candidate LLM code smells were retained when they were observable at the code level, supported by multiple sources, and associated with an accessible remediation strategy. They were then organised within the proposed taxonomy and documented using a fixed schema.}

\subsection{Phase 2: Detection Approach}

\noindent\textbf{Dedicated sections:} \textit{Detection Approach with SpecDetect4LLM} (\ref{sec:detection_approach}) describes the design of SpecDetect4LLM, while \textit{Validation Design and Result} (\ref{sec:resuls_and_eval}) evaluates it.

\noindent\textbf{Goal:} Develop a detection tool for LLM code smells based on static source code analysis, and validate it.

\noindent\textbf{Inputs:} The catalog of LLM code smells from the previous phase.

\noindent\textbf{Outputs:} \textit{SpecDetect4LLM}, a static detection tool, an empirical evaluation of its effectiveness, a dataset of open-source LLM-integrating systems and a a manually audited ground truth set comprising 381 files.

\noindent\textcolor{blue}{\textbf{Description:} We specified each catalogued smell as a declarative detection rule and implemented these rules using the domain specific language (DSL) of \textit{SpecDetect4AI}~\ccite{mahmoudi2025ai}. We then applied \textit{SpecDetect4LLM} to a corpus of 692 open source LLM-integrating systems and evaluated its effectiveness against manually audited annotations.}

\subsection{Phase 3: Prevalence Assessment}

\noindent\textbf{Dedicated sections:} \textit{Prevalence Study} (\ref{sec:prevalence_study})

\noindent\textbf{Goal:} Assess how prevalent LLM code smells and how they manifest in practice.

\noindent\textbf{Inputs:} The dataset of open-source LLM-integrating systems collected in the previous phase, and \textit{SpecDetect4LLM}.

\noindent\textbf{Outputs:} Prevalence and occurence metrics, statistics, and analysis of LLM code smells.

\noindent\textcolor{blue}{\textbf{Description:} Using \textit{SpecDetect4LLM}, we conducted a prevalence study on a dataset of 692 LLM integrating systems comprising 171,194 files. We measured how much LLM code smells affect files and projects, both overall and for each smell individually, and then examined the main patterns observed in practice.}

\section{Catalog Construction}
\label{sec:catalog_contruction}

This section explains how we constructed the catalog of LLM code smells by combining three complementary strands of evidence, following established guidelines for systematic mapping studies in software engineering and multi-source triangulation~\cite{kitchenham2007guidelines}.

First, we conducted a systematic review of research literature on defects, misuses, and bad practices that arise when integrating LLMs into software systems.

Second, we mined grey (blog posts, official documentation, etc.) and empirical (GitHub issues, open-source software projects, etc.) literature with GLiSE, a prompt-driven and machine-learning-based tool designed to automate the search, filtering, and extraction of relevant grey literature in software engineering~\ccite{cherief2025automatedgreyliteratureextraction} to capture practitioner experience and reports, as well as guidelines from sources such as provider documentation, engineering blogs, and Q\&A platforms. \textcolor{blue}{These types of literature were included because real-world practices in software engineering are not always reflected in peer-reviewed research, and because LLM integration is still recent enough that provider documentation and community reports often represent the most accurate sources on how to integrate and use LLMs in practice.}

For each candidate smell, we recorded which evidence types supported it and how many distinct sources contributed, which allows us to trace each smell back to the underlying the corpus.

\begin{table*}[t]
\centering
\footnotesize
\begin{threeparttable}
\caption{Systematic Literature Review and Grey Literature Sources}

\label{tab:sources-code-smells}
\renewcommand{\arraystretch}{1}

\begin{tabularx}{\textwidth}{@{} l *{2}{C}}
\toprule


\textbf{Code Smell} & \multicolumn{2}{c}{\textbf{Source}} \\

\cmidrule(lr){2-3}
& \textbf{Literature} & \textbf{Grey and Empirical} \\
\midrule
\textbf{No Structured Output (Section \ref{subsubsec:NSO}) }  & \citeSP{SP-liu2024structuredoutput,SP-papaioannou2024workload,SP-papaioannou2025llminference,SP-palla2025codegen,SP-sakai2024lowcost,SP-NeedStructuredOutput2024} & \citeSP{SP-Kharitonov2024EnforcingJSON,SP-DeveloperService2025PydanticLLM,SP-WymanBarber2024ValidateOutputs,SP-Modelmetry2024JSONSchema,SP-OpenAIDocs2025,SP-AzureOpenAIDocs2025,SP-OpenAIStructuredOutput,SP-LangChain2023Issue3709,SP-ArenaAI2025StructuredLogprobs,SP-nso_ghpr_dgy516_vllm_cibench_60,SP-nso_ghissue_vllm_23120,SP-nso_ghissue_ms_agent_framework_238,SP-nso_ghpr_berriai_litellm_14206,SP-karatasemr_structur_2025_b0762a,SP-dochertyan_structur_2025_c8f2b1,SP-kimdoil_guidedjs_2025_17f880,SP-thoughtson_structur_2024_fe4516,SP-blackburnp_explorin_2025_160a25,SP-gopenai_masterin_2025_8e3a5a,SP-vermakunal_structur_2024_e45f0e,SP-damodharan_llmbased_2024_b18c84,SP-desaimanoj_outputfo_2025_3bec38,SP-mychen76_practica_2023_cc45ce,SP-mete_controll_2024_028f0b,SP-devto_crafting_2024_60aaf6,SP-venkatrama_tamingll_2025_40c60e,SP-openai_usingthe_2024_dba98b,SP-pydantic_addllmst_2025_0cb1fb,SP-ksyeddata_usepydan_2025_c4edea,SP-pguso_newexamp_2025_38b8ab,SP-mhattingpe_structur_2025_e6808c,SP-hev_featurer_2025_a6abff,SP-aplassard_developc_2025_f85be3,SP-marcosomma_hardenll_2025_1356f2,SP-confidenta_bamlfore_2025_e9eedd,SP-bamresearc_integrat_2025_9379c0,SP-tandoorrec_addaisch_2025_2d47ab,SP-arakoodev_jsonllmo_2025_67484c,SP-gsindlinge_outputfo_2025_7bee28,SP-davidaguil_bugragas_2025_fc84f9,SP-lee0110_addtypef_2025_6979a1,SP-567labs_instruct_2025_bb8ae1,SP-jkrunal7_structur_2025_1ca4fc,SP-vchecha_vchechav_2025_25185c,SP-mt7180_mt7180ll_2025_8a0636,SP-vivekvjnk_heimdall_2025_cfc127,SP-jhd3197_promptur_2025_257736,SP-dani2112_dani2112_2025_ba4606,SP-petrukhaiv_petrukha_2025_8ec164,SP-emrekarata_bnfconst_2025_ed8ffa,SP-deepankarm_deepanka_2025_d32bc8,SP-oliverkwun_parsec_2025_a6d527,SP-yogeshkukr_yogeshku_2025_6fd671,SP-567labs_instruct_2025_77db03,SP-kapilreddy_instruct_2025_dc48e6,SP-kishoretvk_jsonai_2025_9c7288}\\
 \textbf{Unbounded Max Metrics (Section \ref{subsubsec:UMM})}  & ~\citeSP{SP-morishige2025reproducibility,SP-papaioannou2024workload,SP-xiao2024giot,SP-palla2025codegen,SP-buchicchio2024legalAI,SP-sakai2024lowcost,SP-themesLLM2025cain,SP-HanEtAl2025TokenBudgetAware,SP-chen2025adaptivelyrobustllminference} & ~\citeSP{SP-OpenAIDocs2025,SP-AnthropicDocs2025,SP-GoogleGeminiDocs2025,SP-AzureOpenAIDocs2025,SP-BigQuery2025Quotas,SP-aws2025timeouts,SP-openai_python2025,SP-umm_ghissue_datadog_ddtracepy_14688,SP-umm_ghissue_langchainjs_9088,SP-umm_ghpr_myyachtvalue_36,SP-umm_ghissue_bpmn_assistant_33,SP-umm_ghissue_ttmgsrv_19,SP-umm_ghpr_roocode_8483,SP-umm_so_77172214,SP-umm_so_77354317,SP-novitaai_whatarel_2024_914f73,SP-malhar_understa_2024_00e99a,SP-vectorshif_largelan_2024_9903eb,SP-hev_featurer_2025_c74807,SP-runllama_featurer_2025_07be8e,SP-intel_defaultv_2025_d4213a}\\
 \textbf{LLM Temperature Not Explicitly Set (Section \ref{subsubsec:TNES})} & ~\citeSP{SP-minh2024heat,SP-morishige2025reproducibility,SP-papaioannou2024workload,SP-papaioannou2025llminference,SP-palla2025codegen,SP-themesLLM2025cain,SP-MinhEtAl2025MinpSampling,SP-MontandonEtAl2025DABC} & ~\citeSP{SP-AnthropicDocs2025,SP-GoogleGeminiDocs2025,SP-OpenAIDocs2025,SP-VellumTemperature2025,SP-HFTransformersDocs2025,SP-OllamaModelfile2025,SP-tnes_ghissue_vllm_26806,SP-tnes_ghissue_langfuse_9566,SP-wangkelsey_acompreh_2025_7d4ce5,SP-novitaai_whatarel_2024_914f73,SP-malhar_understa_2024_00e99a,SP-pathakteju_setthete_2024_7a1bf6,SP-kaviyadhar_impactof_2024_5f226d,SP-tahirbalar_understa_2025_941da5,SP-uextremewa_whatisth_2025_d27670,SP-herrmannov_temperat_2024_1c137a,SP-ubonismart_zerotemp_2025_15a9b4,SP-vectorshif_largelan_2024_9903eb,SP-patelashis_llmtempe_2025_da6628,SP-hev_featurer_2025_c74807,SP-intel_defaultv_2025_d4213a,SP-huggingfac_thedocum_2025_b77968,SP-openhands_bugazure_2025_cfd769,SP-micz_exposemo_2025_d3e768,SP-penguoir_allowuse_2025_ee4b45,SP-pepelespoo_fixdarkm_2025_0ad199,SP-haktanceti_llmrando_2025_ec24a3,SP-openai_temp0_forum_2024,SP-promptengineering_temperature_top_p_2024} \\
\textbf{No Model Version Pinning (Section \ref{subsubsec:NMVP})} &\citeSP{SP-morishige2025reproducibility,SP-reyes2024bump,SP-venturini2023depended,SP-montandon2024defaultargs,SP-buchicchio2024legalAI,SP-simoes2024evaluating,SP-WilsonEtAl2014BestPractices,SP-VenturiniEtAl2023IDepended,SP-ReyesEtAl2024BUMP,SP-Morishige} & ~\citeSP{SP-AnthropicDocs2025,SP-Microsoft2025FoundationLifecycle,SP-HFTransformersDocs2025,SP-OpenRouterHome2025,SP-GoogleGeminiDocs2025} \\

 \textbf{No System Message (Section \ref{subsubsec:NSM})}  & ~\citeSP{SP-jeong2025systemmsg,SP-neumann2025position,SP-jeong2025messagegenerationuserpreferences,SP-Neumann_2025} & ~\citeSP{SP-PromptHub2025SystemMessages,SP-StackOverflow2023SystemRoleUseCase,SP-OpenAIDocs2025,SP-HFTransformersDocs2025,SP-openai_whatexac_2023_59a966,SP-openaicomm_question_2024_bdb2f3,SP-openai_understa_2023_f4b4d0,SP-openai_thesyste_2023_d84a9a,SP-openai_providin_2023_e5a2fb,SP-dan43009_thediffe_2024_06451f,SP-hakimmudas_masterin_2025_22b6f4,SP-uthexdroid_whatisth_2024_a14eae,SP-lgabs_usesyste_2025_b78ffb} \\
  \textbf{Reasoning Effort Not Explicitly Set (RENES)}  & ~\citeSP{SP-wen2025budgetthinker,SP-wang2024reasoningbudget,SP-chen2025robustinference,SP-themesLLM2025cain,SP-han2025tokenbudget} & \\
   \textbf{Raw Vision Payload (Section \ref{subsubsec:RVP})}  & ~\citeSP{SP-lee2025ergo,SP-vasu2024fastvlm,SP-qian2025zoomer}  &  \\
   \textbf{Overspecified Sampling Parameters (Section \ref{subsubsec:OSP})}  &     &\citeSP{SP-azure_openai_temperature_topp,SP-anthropic_claude_temperature_topp,SP-willison_llm_anthropic_2025,SP-openai_temp0_forum_2024,SP-n8n_anthropic_temp_topp_issue_2025,SP-novitaai_whatarel_2024_914f73,SP-malhar_understa_2024_00e99a,SP-medium_settingt_2024_5a6dd8,SP-albert_largelan_2024_7da749,SP-vectorshif_largelan_2024_9903eb,SP-intel_defaultv_2025_d4213a,SP-huggingfac_thedocum_2025_b77968,SP-repowise_ensurede_2025_3759a4,SP-mpfaffenbe_featurer_2025_639588,SP-haktanceti_llmrando_2025_ec24a3}  \\
  \textbf{Anonymous Inference Call (Section \ref{subsubsec:AIC})}  &   & ~\citeSP{SP-openai_safety_best_practices,SP-google_gemini_safety_guidance,SP-anthropic_claude_messages_create_python,SP-berriai_litellm_issue_10106,SP-openai_needhelp_2024_349ad1,SP-openaideve_apibanfr_2022_6e7ce6,SP-openaicomm_anysugge_2024_873212,SP-openai_lessonsl_2022_863fc6,SP-openai_apipolic_2024_325242,SP-anthropic_detectin_2025_f8b326,SP-nerdfortec_safeguar_2023_1af5bd,SP-promptengineering_temperature_top_p_2024} \\

  \textbf{Others} &\citeSP{SP-deOliveira2025trust,SP-IjasJoRaj2024XAI_LLM_TransparencyTrust,SP-na2024llmcpus} & ~\citeSP{SP-karthikeya_openaiap_2025_4fab36,SP-novitaai_whatarel_2024_914f73,SP-malhar_understa_2024_00e99a,SP-vectorshif_largelan_2024_9903eb,SP-runllama_featurer_2025_07be8e,SP-ollama_ollamaol_2025_6af93e,SP-owainlewis_owainlew_2025_c9fdf0,SP-humanlayer_12factor_2025_66d41b,SP-deepseekai_deepseek_2025_767b4a,SP-openai_gpt41cha_2025_3482c8,SP-openai_howtopas_2024_f80a3f,SP-quic_llmoutpu_2025_6021e8,SP-alltuner_adderror_2025_1161a0,SP-skyvernai_alwaysca_2025_835ac2,SP-jerseychee_advanced_2025_c58498,SP-doclingpro_vlmpipel_2025_d71e58,SP-blaizzy_commanda_2025_bc5bc9,SP-aidynamo_featgene_2025_304aac,SP-sglproject_docsaddt_2025_13d3d3,SP-m5stack_sampling_2025_aed1a0,SP-mistralai_featurer_2025_719ae5,SP-houseofbet_implemen_2025_795c25,SP-ollama_ollamafr_2025_ad8643,SP-springproj_chatmemo_2025_6ce4a7,SP-alexyang08_inferenc_2025_518dd3,SP-praveen76_llmsapiu_2025_4f93e7,SP-feibaobob_llmbased_2025_b51625,SP-andrewyng_aisuite_2025_1b4871,SP-0x6f677548_unicodei_2025_1e6bb6,SP-drpwner_promptsn_2025_d00848,SP-pablochaco_adversar_2025_ca6c98,SP-ridpath_llmvulne_2025_371dfc,SP-grafbase_grafbase_2025_9f0e32,SP-librai_libraiop_2025_13948c,SP-hqwuhitcs_awesomel_2025_abbaf2,SP-gongyichen_figstepj_2025_5e6a3d,SP-0xaidr_aidrbast_2025_bcfdcc} \\
   
\bottomrule
\end{tabularx}

\begin{tablenotes}\footnotesize
\item[*] \textit{Sources} - Literature: peer-reviewed literature; Grey: grey literature (official docs, tech reports, blogs); and Empirical: primary empirical data (e.g., commits, repos).
\item[*] \textit{Others} - Refers to general sources and other relevant results that did not lead to new LLM code smells.
\end{tablenotes}
\end{threeparttable}
\end{table*}

\subsection{Systematic review of research literature}
\label{subsec:systematic-review}

We followed the updated PRISMA guidelines \ccite{page2021prisma, page2021prismae} and Kitchenham et al.  \ccite{kitchenham2022segress} guidelines to review and report our findings.
We used three main phases: planning, conducting, and reporting the review. During the planning phase, we defined the objective of SLR and reviewed the protocol. The objective of this SLR is to answer this question: which kinds of integration problems, misconfigurations, and code level bad practices are documented when LLMs are embedded into software systems, tools, or pipelines?

It consists of four steps: 

\begin{itemize}
\item[\ding{202}] Definition of the research question
\item[\ding{203}] Formulation of the search query
\item[\ding{204}] Selection of the papers
\item[\ding{205}] Snowballing
\end{itemize}

In the following subsections, we explain each step of our review protocol.

\subsubsection{Definition of the research intent}

This section answers the following question: \textit{How can we derive a catalog of recurrent LLM code smells by
consolidating evidence from scientific literature, grey literature, and practitioner
artifacts?}

Answering this question yields a consolidated catalog of LLM integration defects that is grounded both in prior research and in the concrete practice of software teams. The resulting catalog provides a common vocabulary to describe recurrent integration problems, supports practitioners in recognizing these defects in their own systems, and serves as a foundation for automated detection approaches such as \textit{SpecDetect4LLM}. In the remainder of this section, we explain how we operationalise the RQ into concrete search, screening, and coding procedures to capture evidence about LLM code smells across heterogeneous sources.

\subsubsection{Formulation of the Search Query}
\label{subsec:searchstrategy}

We formulated our search query by applying the PICO (Population, Intervention, Comparison, Outcome) framework \ccite{pico}. We followed the following steps:
\begin{itemize}
\item[\ding{202}] Obtaining the main terms from our research question (RQ1). 
\item[\ding{203}] Identifying the possible synonyms of the main terms.
\item[\ding{204}] Applying the Boolean OR to combine possible synonyms of the main terms.
\item[\ding{205}] Applying the Boolean AND to combine expressions in the previous step.
\end{itemize}
As a result of PICO framework, we formulated the following search query:

\definecolor{qpink}{RGB}{180,0,60}
\definecolor{qblue}{RGB}{0,70,160}
\definecolor{qbg}{RGB}{245,245,245}

\newcommand{\qop}[1]{\textcolor{qblue}{#1}}
\newcommand{\qterm}[1]{\textcolor{qpink}{"#1"}}

\begin{tcolorbox}[colback=qbg, colframe=black!40, arc=3mm, boxrule=0.3pt]
\small
(\qterm{large language model*} \qop{OR} \qterm{LLM*} )
\qop{AND}
(\qterm{integrat*} \qop{OR} \qterm{API*} \qop{OR} \qterm{software system*})
\qop{AND}
(\qterm{misuse*} \qop{OR} \qterm{defect*} \qop{OR} \qterm{bug*} \qop{OR} \qterm{smell*} \qop{OR} \qterm{pitfall*})
\end{tcolorbox}

To obtain more comprehensive results, we used the asterisk (*) in search queries as a wildcard to match any sequence of characters.

\subsubsection{Selection of the Papers}
\begin{figure}[t]
  \centering
  \includegraphics[height=303pt]{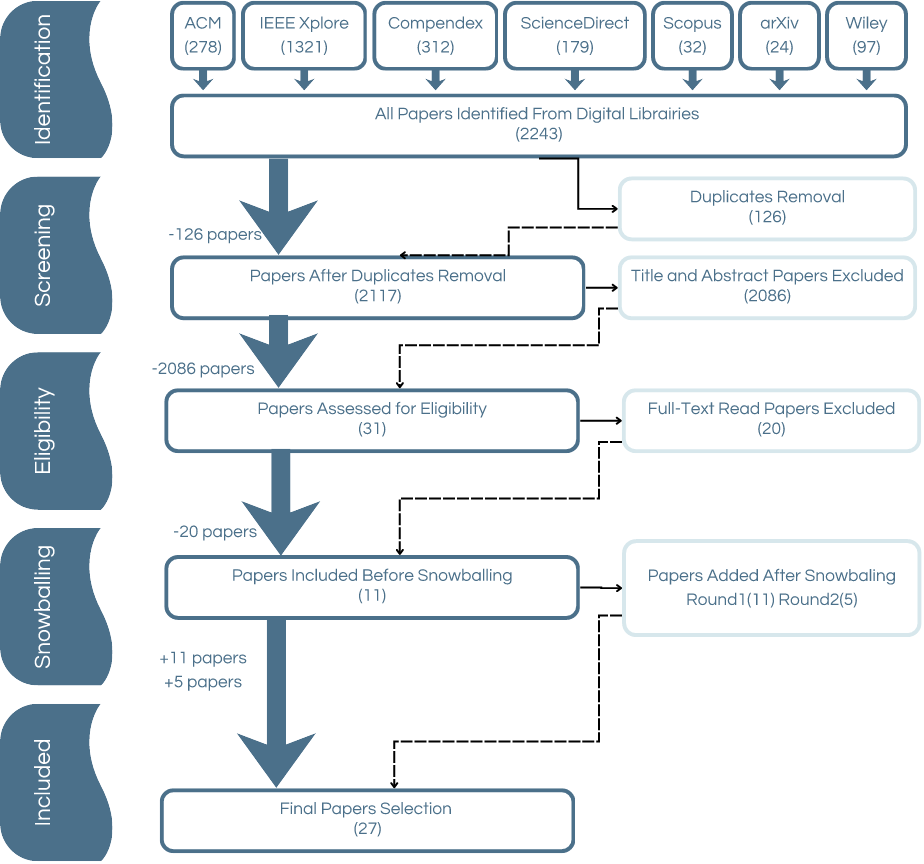}
  \caption{PRISMA flow for paper selection. Values in parentheses indicate the number of papers at each stage.}
  \label{fig:prisma_overview}
\end{figure}

We applied the PRISMA steps to select the literature papers. The main steps include database identification, removal of duplicates, screening, eligibility assessment, two rounds of both backward and forward snowballing, and quality assessment of each study. Figure \ref{fig:prisma_overview} summarizes 
those steps.

\noindent{\textit{\textbf{Databases Identification}}}
\label{sec:db_identification}

We selected seven online digital libraries: ACM Digital Library, Compendex, IEEE Xplore, ScienceDirect, SpringerLink, Scopus, arXiv, and Wiley. These libraries are widely used for literature reviews in software engineering, as recommended by Dyba et al.~\ccite{dyba2007applying}. 

We applied our search query to each of these digital libraries. However, some libraries impose restrictions when performing queries. For instance, ScienceDirect limits queries to a maximum of eight connectors, while the ACM Digital Library does not allow wildcards. We adjusted the search query to meet the specific requirements of each library.

Our search was confined to English-language, peer-reviewed scholarly articles published in journals, conferences, and workshops between 2017 and 2025. This time frame was chosen due to the increase in LLM-related literature since 2017 \ccite{XXX}. We initially retrieved a total of 2243 papers from seven libraries.

\noindent{\textit{\textbf{Duplicates Removal}}}
\label{sec:de_duplication}

Duplicates were identified based on an exact match of the study's title, first author, and venue (conference or journal). This reduced the total from  papers to 2117.

\noindent{\textit{\textbf{Screening}}}
\label{sec:screening}

We defined inclusion and exclusion criteria and applied them to select relevant primary papers while excluding irrelevant ones.

\noindent\textbf{Inclusion Criteria:}
\label{sec:inclCrit}

We considered the following inclusion criteria for paper selection:
\begin{itemize}
\item \textbf{IC1:} The study is written in English.
\item \textbf{IC2:} The study is published between 2017 and 2025.
\item \textbf{IC3:} The study presents an LLM based software system or an engineering practice involving LLM API or SDK integration.
\item \textbf{IC4:} The study reports at least one concrete integration issue, misuse, defect, or failure related to LLM based components.
\item \textbf{IC5:} The study provides technical detail about the LLM invocation, API configuration, request structure, or downstream processing.
\item \textbf{IC6:} The study provides enough information to characterise at least one LLM code smell.
\item \textbf{IC7:} The study has its full text available online.
\item \textbf{IC8:} The study is peer reviewed for scientific literature.
\end{itemize}

\textbf{Exclusion Criteria:} We considered the following exclusion criteria:
\begin{itemize}
\item \textbf{EC1:} The study is a secondary source (e.g., literature review, survey, opinion piece, or conceptual commentary).
\item \textbf{EC2:} The study does not describe any interaction with an LLM API, SDK, or integration mechanism.
\item \textbf{EC3:} The study provides no implementation detail allowing us to identify or analyse integration defects.
\item \textbf{EC4:} The study focuses exclusively on model training, dataset creation, benchmarking, or evaluation without addressing integration aspects.
\item \textbf{EC5:} The study's full text is not available online.
\item \textbf{EC6:} The study does not provide any evidence of LLM integration behavior relevant to code smell extraction.
\item \textbf{EC7:} The study is not written in English.
\end{itemize}

We applied our inclusion and exclusion criteria in two steps, first using the titles and abstracts, then the full texts. We acknowledge that some inclusion and exclusion criteria are logically complementary, for instance IC7 and EC5 describe opposing conditions. We explicitly state both inclusion and exclusion criteria separately to comply with the PRISMA guidelines, which recommend clearly articulating both positive and negative selection criteria to enhance transparency and reproducibility.

During the initial screening, we assessed the titles, abstracts, and page counts, and determined whether the papers qualified as primary. Two authors independently conducted the screening using predefined inclusion and exclusion criteria. To ensure consistency, we compared the screening results for a randomly selected subset of papers using Cohen's Kappa. We obtained near-perfect agreement (k=0.85), demonstrating the reliability of the screening process. For the remainder of the screening, the authors met regularly to review results and resolved disagreements through discussion and consensus.
After completing the initial screening, we selected the set of papers for full-text review as potential primary sources.

In the second round of screening, two authors independently examined papers in depth to verify whether it provided sufficient technical detail about LLM integration, including the invocation mechanism, the surrounding architectural or code-level context, and the observable integration behavior. To ensure a shared understanding of the criteria, we assessed the agreement between the two authors on 15 randomly selected papers using Cohen's Kappa. We obtained a near-perfect agreement (k=0.93), demonstrating strong consistency in the interpretation and application of the criteria. This level of agreement allowed us to confidently proceed with the eligibility assessment for the remaining papers. Following this stage, we identified the set of primary papers that satisfied all eligibility requirements for inclusion in our evidence base.

\subsubsection{Snowballing}
\label{sec:snowballing}

We performed two iterations of backward and forward snowballing to identify additional relevant papers. This process led to the inclusion of 16 additional studies in the final selection. Overall, after completing all phases of PRISMA flow, we obtained a final selection of 27 papers.

\subsection{Grey and Empirical literature Mining}
\label{subsec:grey-literature}

\begin{figure}[t]
  \centering
\includegraphics[height=303pt]{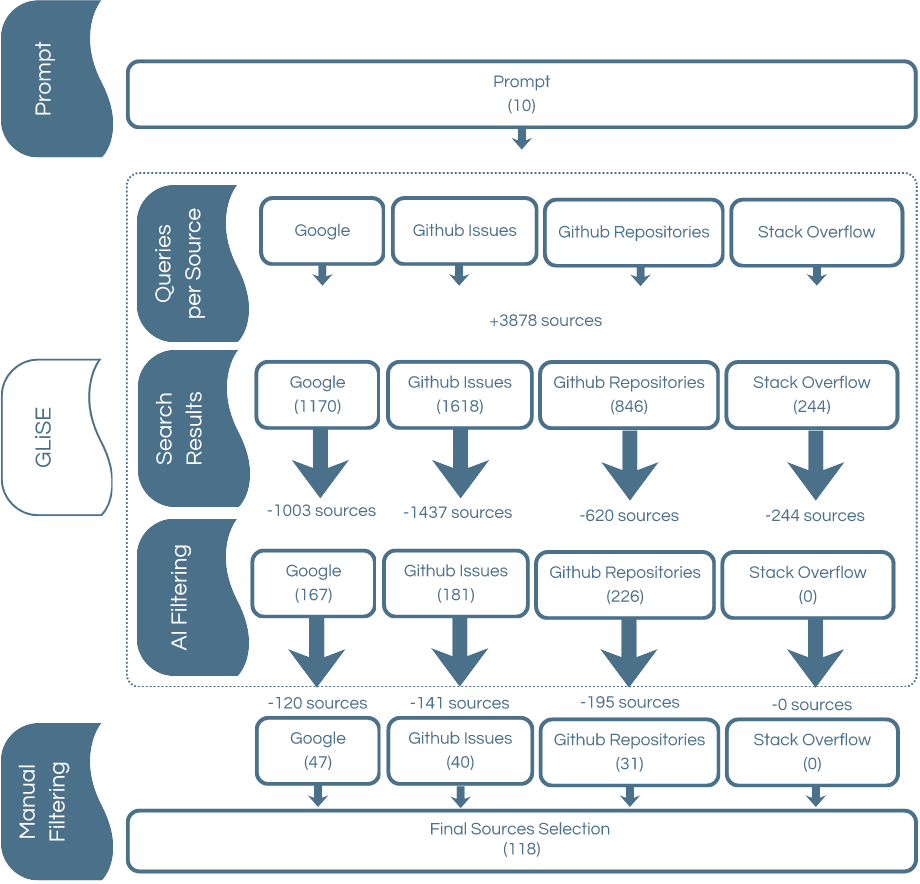}
\caption{Grey Literature Mining Flow.} 
\label{fig:grey_lit}
\end{figure}


Research literature provides a delayed and partial view of emerging integration practices. In contrast, grey and empirical literature such as engineering blogs, GitHub issues and repositories, programming forums, and official documentation captures real-world practices in near real time that are often overlooked in academic venues \ccite{garousi2021grey,alves2020grey}. Accounting for these sources is essential to software engineering research. Therefore, to capture real-world LLM integration practices in software systems, we complemented the systematic review with a structured mining of grey and empirical literature.

This allows us to explore new perspectives and to enrich the catalog by providing additional examples and counter-examples, as well as by documenting emerging integration issues not yet studied in academic venues, such as subtle misuses of provider features and configuration options. For each identified code smell, we associate relevant grey literature sources, as presented in the evidence table (Table \ref{tab:sources-code-smells}).

We followed a strict methodology, comparable to the one applied for scientific papers. We relied on GLiSE \ccite{cherief2025automatedgreyliteratureextraction}, a three-step, prompt-driven and ML-powered tool for the extraction and filtering software engineering grey literature sources from Google, GitHub issues, GitHub repositories, and Stack Overflow posts. We used GLiSE because, unlike general AI-based search tools such as Perplexity \ccite{PerplexityAI2025}, it is specialized for software engineering grey literature and is fully traceable and reproducible, allowing the extraction of intermediate traces and artifacts at each step of the search process.

Our methodology is composed of three steps: (i) Prompts Creation, (ii) Extraction with GLiSE, and (iii) Manual Filtering. We first defined ten textual prompts, which were then used with GLiSE. Finally, we manually filtered the resulting corpus to refine the selection and ensure relevance. After this process, we completed the grey literature corpus with targeted manual searches using search engines and navigating the official documentation of major LLM and VLM providers, including OpenAI, Gemini, Anthropic, Hugging Face, and Ollama \ccite{OpenAIDocs2025, GoogleGeminiDocs2025, AnthropicDocs2025, HuggingFaceHome2025, ollama_ollamaol_2025_6af93e}. This allowed us to identify complementary evidence for potential code smells, assess their relevance, and consider providers' guidance and practices. We focused primarily on official documentation, as these sources are both reliable and directly aligned with the subject of this study.

The complete grey literature mining workflow, along with the number of results and intermediate artefacts produced at each step, is shown in Figure \ref{fig:grey_lit}. All artefacts and traces generated throughout this process are included in the replication package \ccite{Replication_package} to support reproducibility.

\subsubsection{Prompts Creation}
To extract relevant and useful grey and empirical literature, we created ten distinct textual prompts designed for use with GLiSE. Some prompts were intentionally generic, while others targeted specific features and functionalities of LLM and VLM APIs and SDKs. This allowed us to ensure broad coverage of LLM and VLM inference in software systems, while increasing the likelihood of identifying precise and relevant documentation related to suspected integration pitfalls. Using multiple prompts with varying levels of granularity mitigates the risk associated with relying on a single prompt that could be either too generic to find specific issues or too narrow to cover all relevant integration aspects. Therefore, by using multiple prompts, we aimed to maximize the relevance and quality of the retrieved sources. 

All ten prompts are provided in the replication package \ccite{Replication_package}.

\subsubsection{Extraction with GLiSE}
To search for grey and empirical literature, we used GLiSE \ccite{cherief2025automatedgreyliteratureextraction} independently for each of the ten prompts created in the previous step, extracting and filtering sources from Google, GitHub, and Stack Overflow for each prompt. The final results obtained for each prompt were then grouped into a single corpus, while ensuring the absence of duplicates using a Python script.

GLiSE operates by first generating source-specific search queries from a textual prompt. These queries are then executed using the respective APIs of each source. Finally, retrieved results are screened based on their semantic relevance to the original prompt using the embeddings (vectorial representation of the semantics) of their metadata and machine learning classifiers. All generated queries, retrieved results, and settings can be extracted to support reproducibility. We provide ours in the replication package \ccite{Replication_package}.

After running GLiSE on the ten prompts and combining their final results, we obtained a total of 574 results, including 167 from Google Search, 181 from GitHub issues, 226 from GitHub repositories, and none from Stack Overflow.

\subsubsection{Manual Filtering}
After obtaining results using GliSE, we performed an additional manual screening step because the documentation required for this study is highly specific, and our relevance criteria are stricter than general topical relevance. Our subject can also be confused with related but distinct concepts, such as code smells in code generated by LLMs, which can result in sources being erroneously filtered as relevant. We therefore combined GLiSE's automated filtering, to speed up the process and enable broader searches, with manual screening to ensure the quality and reliability of the final grey literature corpus. 

At the end of this step, 118 sources remained, including 47 from Google search, 40 from GitHub issues, 31 from GitHub repositories, and none from Stack Overflow.


\section{LLM Code Smells}
\label{sec:catalog_taxo}


\textcolor{blue}{This section presents our catalog of LLM code smells. We first introduce the taxonomy that structures the catalog and clarifies the nature of the underlying smells. We then present the catalog of LLM code smells according to this taxonomy. Table~\ref{tab:defects-llm} summarizes, for these nine code smells, their category in the taxonomy and their main effects on software quality. Extended examples, further discussion, and additional instances are available in our replication package~\ccite{Replication_package}. New LLM code smells introduced in this paper are marked with *.}

\subsection{Taxonomy of LLM code smells}

\begin{figure}[ht]
    \centering
    \includegraphics[width=0.6\linewidth]{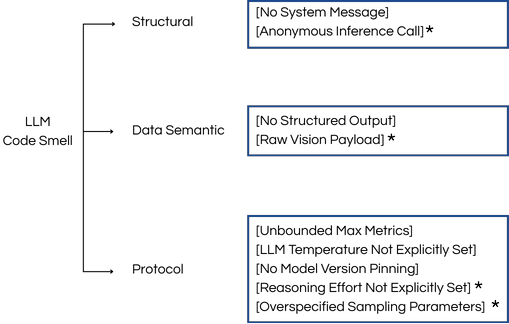}
    \caption{Classification of LLM code smells}
    \label{fig:NatureCS}
\end{figure}

\textcolor{blue}{We organize the catalog into a taxonomy composed of three categories of LLM code smells, defined according to the nature of the underlying smell. More specifically, an LLM code smell may be primarily related to \textit{Structural} or API usage, \textit{Data semantics}, or \textit{Protocol}. The categories were derived by grouping the identified smells according to the dimension of LLM integration they primarily concern, namely \textit{Structural or API usage}, \textit{Data semantics}, or \textit{Protocol}. These categories also differ in how strongly they are coupled to provider-specific API designs, which may evolve over time. This taxonomy, illustrated in Figure~\ref{fig:NatureCS}, provides a conceptual structure for the catalog by clarifying why each smell arises during development and by indicating the kind of reasoning typically required for its detection.}

\medskip
\noindent\textbf{Structural or API usage}
This category covers LLM code smells that arise from how developers invoke the LLM API, configure parameters, and compose requests. These source code defects are visible directly in the structure of the invocation, such as in parameter choices, the construction of the messages array, or the instantiation of helper objects. As a result, these smells can usually be identified through static analysis of the call site, without requiring reconstruction of full data-flow or execution semantics.

\textcolor{blue}{As this category is directly tied to the composition of requests and the way the API is invoked, it is sensitive to API evolution. However, the aspects covered by the current smells in this category appear relatively established across the majority of providers, reflecting a certain degree of standardization that has started to take place.}

\medskip
\noindent\textbf{Data semantics}
This category covers defects that arise from incorrect assumptions about the structure, validity, or stability of LLM inputs and outputs, including multimodal data such as images or typed JSON payloads. These LLM code smells manifest when downstream components operate on non-validated or misinterpreted data, for example when free text is consumed as a structured JSON or when raw screenshots are sent without cropping or downscaling. Detecting these smells typically requires relating the LLM inference to how its inputs and output are parsed, stored, or executed.

\textcolor{blue}{The concepts underlying this category are relatively stable, as they mainly concern the structure and handling of input and output data of the LLM rather than provider-specific API conventions. These smells are therefore less subject to change as APIs and models evolve.}

\medskip
\noindent\textbf{Protocol}
This category covers LLM code smells that concern the overall behavior of the LLM inference rather than a single invocation. These defects arise when defaults or unstable system-wide settings are used. Typical examples include implicit temperature, missing reasoning effort configuration, unbounded token budgets, or moving model aliases. Such defects affect reliability, reproducibility, and robustness across runs, as default values may evolve over time or differ between LLM providers.

\textcolor{blue}{The coupling to provider-specific behavioural parameters makes \textit{Protocol} smells the most subject to change as APIs and models evolve among the three categories, as these parameters are more likely to change as the field matures. For some smells in this category, such variation across model types and versions is already observable.}

\subsection{\noindent\textbf{LLM Code Smells Catalog}}

This section presents the nine LLM code smells identified in our catalog. Each smell is described using a common structure composed of its context, problem, solution, example, taxonomy category, and evolutionary stability.

\subsubsection{\noindent\textbf{No Structured Output (NSO)}}
\label{subsubsec:NSO}

\noindent\textbf{Context:} LLM-Integrating systems often expect typed fields (e.g., JSON) but rely on LLM inference outputs which may not respect the format and produce raw free-form text. This smell applies when the output is later parsed, indexed, or executed assuming structure.

\noindent\textbf{Problem:} Without an enforced output schema, the system may receive free-form text where structured fields are expected~\ccite{NeedStructuredOutput2024}. This increases error-proneness via schema drift, missing or renamed fields, type mismatches, and silent truncation that passes as success, breaking parsers and downstream steps. It degrades reliability as runs become inconsistent, data stores accumulate corrupted or hallucinated values, and execution/storage/display paths face injection~\ccite{DeveloperService2025PydanticLLM}. 

\noindent\textbf{Solution:} Enforce structured output at the API boundary. With OpenAI, declare a JSON Schema via \textit{response\_format} (chat completions) or \textit{text.format} (responses). With the Python SDK, you may instead bind the format directly to classes~\ccite{OpenAIStructuredOutput}. Always validate results to handle refusals or other errors~\ccite{Kharitonov2024EnforcingJSON, WymanBarber2024ValidateOutputs, OpenAIStructuredOutput}.

\noindent\textbf{Example:} In Listing~\ref{fig:NSO}, the faulty version consumes free-form text, whereas the corrected version enables strict parsing and safer downstream use by enforcing a JSON schema via \texttt{response\_format}.

\begin{lstlisting}[style=mystyle,basicstyle=\ttfamily\footnotesize, language=python,alsoletter=_,caption=No Structured Output (NSO),keywordstyle=\color{magenta},morekeywords={openai,create,ChatCompletion},deletekeywords={and,not,set,in},label=fig:NSO,moredelim={**[il][\DiffRemoved]{-}},
  moredelim={**[il][\DiffAdded]{+}}]
# Define a JSON schema e.g. result_schema = ...
- response = openai.chat.completions.create(
-     model="gpt-4o-2024-11-20", messages=messages)
+ response = openai.chat.completions.create(
+     model="gpt-4o-2024-11-20", response_format={
+         "type": "json_schema",
+         "json_schema": {"name": "Result", "schema": 
+         result_schema}}, messages=messages)
\end{lstlisting}
\noindent\textcolor{blue}{\textbf{Taxonomy Category:} \textit{Data semantics}. NSO belongs to this category because its core defect is semantic rather than purely structural. The system assumes structured data, such as JSON objects or typed fields, while the LLM call only guarantees free-form natural language output. The smell therefore arises from a mismatch between the semantic expectations of downstream components and the actual output contract enforced at the inference boundary.}

\noindent\textcolor{blue}{\textbf{Evolutionary Stability:} The core idea behind NSO, which is enforcing data structure and its validation before parsing, should remain consistent as the field of LLMs evolves. Additionally, structured output functionalities are becoming increasingly common in LLM APIs and SDKs.}

\subsubsection{\noindent\textbf{Unbounded Max Metrics (UMM)}}
\label{subsubsec:UMM}

\noindent\textbf{Context:} Hosted LLM APIs (e.g., OpenAI, Anthropic) expose finite token windows, per-request output caps, and rate limits. Pipelines that ignore these constraints, or omit their own limits on concurrency and response size, are prone to throttling and partial outputs.

\noindent\textbf{Problem:} Leaving token budgets, timeouts, and retries unbounded undermines system robustness and performance. Unspecified token budgets may yield outputs that exceed context limits (truncating mid-structure or return 400), overload downstream parsers, or inflate token costs and memory usage~\ccite{chen2025adaptivelyrobustllminference}. Not specifying timeouts and retry limits can cause requests to hang indefinitely, leading to unpredictable latency, rising costs, and reduced throughput as resources are tied up by long-running calls~\ccite{aws2025timeouts}. Not specifying these values also hinders maintainability, as defaults may change over time and tracking settings is essential for reproducibility.

\noindent\textbf{Solution:} Always bound and adjust the \textit{max\_output\_tokens}, \textit{timeout}, and \textit{max\_retries} parameters~\ccite{openai_python2025}. Monitoring the number of input tokens is also recommended.


\noindent \textbf{Example: }In Listing~\ref{fig:UMM}, the faulty version leaves metrics unbounded, whereas the corrected enforces bounded metrics.

\begin{lstlisting}[style=mystyle,basicstyle=\ttfamily\footnotesize,language=python,alsoletter=_,caption=Unbounded Max Metrics (UMM),keywordstyle=\color{magenta},morekeywords={OpenAI,OpenAIError,RateLimitError,responses,create,with_options},deletekeywords={and,not,set,in},label=fig:UMM, moredelim={**[il][\DiffRemoved]{-}},
  moredelim={**[il][\DiffAdded]{+}}]
- client = OpenAI()
- response = client.responses.create(model=
      "gpt-4o-2024-11-20", input=prompt)
+ client = OpenAI(timeout=20, max_retries=3)
+ response = client.responses.create(
+     model="gpt-4o-2024-11-20",
+     input=prompt, max_output_tokens=256)  
\end{lstlisting}
\noindent\textcolor{blue}{\textbf{Taxonomy Category:} \textit{Protocol}. UMM belongs to this category because its core issue concerns the operational contract that governs LLM inference rather than the local structure of a single call. The smell arises when essential execution bounds, such as \textit{max\_output\_tokens}, \textit{timeout}, and \textit{max\_retries} parameters, are left implicit or unspecified. It therefore affects the stability, robustness, and predictability of inference behaviour across executions, which makes it a protocol-level smell rather than a purely structural one.}

\noindent\textcolor{blue}{\textbf{Evolutionary Stability:} The core idea behind UMM, that the resources given to the inference should be specified to avoid unwanted repercussions, is not new to LLMs and should remain consistent as the field changes. The details of how this is applied may change and improve, but the principle of accordingly limiting request retries and token budget should remain.}

\subsubsection{\noindent\textbf{LLM Temperature Not Explicitly Set (TNES)}}
\label{subsubsec:TNES}


\noindent\textbf{Context:} In LLM APIs, SDKs, and runtimes such as OpenAI, Hugging Face, or Ollama, temperature controls sampling stochasticity~\ccite{VellumTemperature2025} and is a key decoding parameter, together with nucleus sampling (\texttt{top\_p}) and top k sampling (\texttt{top\_k}).

\noindent\textbf{Problem:} Relying on an implicit temperature reduces maintainability and reliability. Defaults temperature differ across providers/models~\ccite{OpenAIDocs2025, GoogleGeminiDocs2025, OllamaModelfile2025} and may change over time~\ccite{MontandonEtAl2025DABC}, which harms reproducibility, portability and can silently alter behavior. 

\noindent\textbf{Solution:} Always specify explicitly the temperature while minimizing node coupling. If temperature is used, always specify it explicitly and justify its value: low (0--0.3) for precise, repeatable automation, higher (0.7--1.0) for creative generation, and avoid extremes~\ccite{MinhEtAl2025MinpSampling}. If \texttt{top\_p} or \texttt{top\_k} is explicitly set, do not additionally set temperature, and instead document that randomness is governed by \texttt{top\_p} or \texttt{top\_k} to avoid overspecification and erratic behaviors \ccite{promptengineering_temperature_top_p_2024}.




\noindent\textbf{Example:} In Listing~\ref{fig:TNES}, the faulty version omits temperature, while the corrected version makes it explicit.

\begin{lstlisting}[style=mystyle,basicstyle=\ttfamily\footnotesize, language=python,alsoletter=_,caption=LLM Temperature Not Explicitly Set (TNES),keywordstyle=\color{magenta},morekeywords={openai, create, action,isMLMethodCall,hasExplicitHyperparameters},deletekeywords={and,not,set,in}, label=fig:TNES, moredelim={**[il][\DiffRemoved]{-}},
  moredelim={**[il][\DiffAdded]{+}}]
response = openai.chat.completions.create(
-  model = "gpt-4o-2024-11-20", messages = messages)
+  model = "gpt-4o-2024-11-20", messages = messages,temperature = 1.0)
\end{lstlisting}

\noindent\textcolor{blue}{\textbf{Taxonomy Category:} \textit{Protocol}. TNES belongs to this category because its core defect concerns the behavioural configuration of LLM inference rather than the local structure of the invocation. Temperature is a primary sampling parameter that directly affects reproducibility, output stability, and behavioural consistency. When it is left implicit, the system silently relies on provider- or model-specific defaults, which may vary across APIs, environments, or model updates. TNES is therefore a protocol-level defect because it weakens the explicit inference contract that governs system behaviour over time.}

\noindent\textcolor{blue}{\textbf{Evolutionary Stability:} The temperature parameter has been a relatively established standard across LLM APIs for some time. However, its use has recently been removed for some models, such as certain OpenAI reasoning models \cite{OpenAIDocs2025}. This suggests that this smell may become obsolete for some specific model families as the field evolves.}

\subsubsection{\noindent\textbf{No Model Version Pinning (NMVP)}}
\label{subsubsec:NMVP}


\noindent\textbf{Context:} In LLM APIs, runtimes, and hubs (e.g. OpenAI, Ollama, Hugging Face), models can be called via moving aliases (e.g., \textit{gpt-4o}) or immutable versions/snapshots (e.g., \textit{gpt-4o-2024-11-20}). Aliases may advance as providers update models~\ccite{Microsoft2025FoundationLifecycle}.

\noindent\textbf{Problem:} Using only a provider alias removes explicit versioning, so weights, prompts, and safety filters can change without notice and shift behavior~\ccite{Morishige}. It reduces maintainability by eroding traceability and reproducibility. Runs cannot be tied to a stable model build and portability across environments degrades as the same alias may yields different behavior.

\noindent\textbf{Solution:} Always specify an immutable identifier and record it with other run metadata. Update versions via change control~\ccite{HFTransformersDocs2025, OpenAIDocs2025, WilsonEtAl2014BestPractices}.


\noindent\textbf{Example:} In Listing~\ref{fig:NMVP}, the faulty version uses a moving alias, whereas the corrected version pins an immutable version/snapshot to ensure reproducibility and traceability.

\begin{lstlisting}[style=mystyle,basicstyle=\ttfamily\footnotesize, language=python,alsoletter=_,caption=No Model Version Pinning (NMVP),keywordstyle=\color{magenta},morekeywords={openai, create, action,isMLMethodCall,hasExplicitHyperparameters},deletekeywords={and,not,set,in}, label=fig:NMVP, moredelim={**[il][\DiffRemoved]{-}},
  moredelim={**[il][\DiffAdded]{+}}]
response = openai.chat.completions.create(
  - model="gpt-4o", messages=messages )
  + model="gpt-4o-2024-11-20", messages=messages)
\end{lstlisting}
\noindent\textcolor{blue}{\textbf{Taxonomy Category:} \textit{Protocol}. NMVP belongs to this category because its core defect concerns the stability of the inference contract over time rather than the immediate structure of a single invocation. When a system relies on a moving model alias instead of an immutable version or snapshot, its effective behaviour may change without any modification to the source code. NMVP therefore weakens reproducibility, traceability, and behavioural stability, which makes it a protocol-level defect.}

\noindent\textcolor{blue}{\textbf{Evolutionary Stability:} The principle behind NMVP is well rooted in established SE practices, where explicit versioning has long been a fundamental requirement for reproducibility and traceability. The aliasing versus pinning distinction remains consistent across major LLM providers, making this smell among the least subject to change as the field evolves.}

\subsubsection{\noindent\textbf{No System Message (NSM)}}
\label{subsubsec:NSM}


\noindent\textbf{Context:} In role-based chat APIs and runtimes, the system message sets global behavior, constraints, and tone for the assistant.

\noindent\textbf{Problem:} Without a system message, the model lacks high-level guidance, which reduces consistency and adherence to constraints. Outputs become more generic and harder to control. This degrades reliability, as additional iterations or longer prompts are often required to achieve adequate results~\ccite{PromptHub2025SystemMessages}.

\noindent\textbf{Solution:} Always include a clear system message that defines roles, goals, and constraints. Keep task specifics in the user message~\ccite{jeong2025messagegenerationuserpreferences}. 

\noindent\textbf{Example:} In Listing~\ref{fig:NSM}, the faulty version omits a \textit{system} message, whereas the corrected version adds a concise system instruction to anchor behavior, improve consistency and response quality.

\begin{lstlisting}[style=mystyle,basicstyle=\ttfamily\footnotesize, language=python,alsoletter=_,caption=No System Message (NSM),keywordstyle=\color{magenta},morekeywords={openai, create},deletekeywords={and,not,set,in}, label=fig:NSM, moredelim={**[il][\DiffRemoved]{-}},
  moredelim={**[il][\DiffAdded]{+}}]
response = openai.chat.completions.create(
  model="gpt-4o-2024-11-20",
- messages=[{"role": "user", "content": "Explain recursion with an example"}])
+ messages=[{"role": "system", "content": "You are a Computer Science tutor. Answer 
+ "clearly."},{"role":"user", "content": "Explain recursion with an example"}])
\end{lstlisting}
\noindent\textcolor{blue}{\textbf{Taxonomy Category:} \textit{Structural or API usage}. NSM belongs to this category because its core defect lies in how the LLM request is structurally constructed at the API level. In role-based chat interfaces, the system message is an explicit element of the message array that defines global behaviour, constraints, and tone. Omitting it is therefore a defect in request composition rather than a semantic mismatch in the exchanged data or a protocol-level inference policy issue.}

\noindent\textcolor{blue}{\textbf{Evolutionary Stability:} The concept of system messages has been more and more standardised accross models and providers. Even if the concrete implementation at the API level may vary slightly between APIs or SDKs, the core concept remains the same. This seems to indicate that this practice is there to stay and shouldn't change much as the field of LLMs evolve.}

\subsubsection{\noindent\textbf{Reasoning Effort Not Explicitly Set (RENES) *}}
\label{subsubsec:RENES}

\noindent\textbf{Context:} Modern RLMs providers expose controls on the depth or intensity of internal reasoning. Depending on the provider, these controls can be expressed as a parameter such as \textit{reasoning effort} or \textit{reasoning depth}, as a budget on internal reasoning tokens or steps, or as a dedicated reasoning or thinking mode that can be switched on or off. These influence how much structured internal reasoning the model performs before producing user visible tokens, which directly affects accuracy, robustness for complex tasks, latency, and cost \ccite{wen2025budgetthinker}. In many APIs, the default value of these controls is model specific and may change over time as models and deployments evolve \ccite{wen2025budgetthinker,wang2024reasoningbudget}.

\noindent\textbf{Problem:} When a reasoning capable model is invoked without explicitly configuring its reasoning control, the system silently relies on provider specific defaults. Since these defaults differ across models and can be updated without any change in the client code, migrating to a new model or endpoint can alter the effective depth of reasoning, as well as runtime and cost, in non transparent ways. This harms reproducibility because executions cannot be tied to a clearly specified reasoning configuration. It also harms maintainability because small changes in the model name or deployment can change behavior without a visible code change. Finally, it harms reliability, since developers and stakeholders may believe that the system systematically uses deep reasoning while, in practice, it may run with a shallow or even disabled reasoning mode \ccite{,chen2025robustinference}.

\noindent\textbf{Solution:} Use the reasoning control as part of the public contract of any call to a reasoning capable model. Always set the corresponding parameter explicitly in code and, when relevant, document the chosen level in configuration files and experimental protocols. For interactive or high throughput scenarios, select a lower reasoning effort or shallower reasoning depth to keep latency and cost under control and validate the choice empirically. For complex problem solving, high risk, or safety critical tasks, explicitly request higher reasoning effort. When providers introduce new reasoning levels or change their defaults, update code and documentation through explicit change management instead of relying on new defaults. Conceptually, the reasoning control should be handled as the primary parameter for depth of reasoning for these models, in the same way that temperature is treated as a primary parameter for sampling behavior in classical LLMs \ccite{,themesLLM2025cain,han2025tokenbudget}.

\noindent \textbf{Example: }In the following example Listing \ref{fig:RENES} , the faulty version calls a reasoning model without specifying any reasoning configuration, while the corrected version pins \textit{reasoning\_effort} to a value that reflects an explicit trade off between latency, cost and accuracy. The example is adapted from the GPT 5 cookbook from OpenAI.

\begin{lstlisting}[style=mystyle,basicstyle=\ttfamily\footnotesize,language=python,alsoletter=_,caption=Reasoning Effort Not Explicitly Set (RENES),keywordstyle=\color{magenta},morekeywords={OpenAI,OpenAIError,RateLimitError,responses,create,with_options},deletekeywords={and,not,set,in},label=fig:RENES, moredelim={**[il][\DiffRemoved]{-}},
  moredelim={**[il][\DiffAdded]{+}}]
 client = OpenAI()
 response = client.responses.create(
     model="gpt-5.1-mini-2025-02-01",
     input=[
         {
             "role": "system",
             "content": "You are a careful assistant that explains your reasoning for complex software engineering decisions."
         },
         {
             "role": "user",
             "content": (
                 "We plan to integrate a large language model into a safety critical pipeline for code review. Explain the main tradeoffs, risks, and mitigations in a structured way."
             ),
         },
     ],
-     temperature=0.2 )
+     temperature=0.2,
+     reasoning={"effort": "minimal"},
+ )
\end{lstlisting}

\noindent\textcolor{blue}{\textbf{Taxonomy Category:} \textit{Protocol}. RENES belongs to this category because its core defect concerns the explicit control of reasoning behaviour during inference rather than the local structure of the invocation. In reasoning models, parameters such as \textit{reasoning\_effort}, \textit{reasoning\_depth}, or \textit{thinking\_mode} govern the trade-off between accuracy, latency, and cost. When these controls are left implicit, the system silently depends on provider-specific defaults that may vary across models, endpoints, or updates. RENES is therefore a protocol-level defect because it weakens the explicit inference contract that should govern reasoning behaviour over time.}

\noindent\textcolor{blue}{\textbf{Evolutionary Stability:} The way reasoning effort is constrained in reasoning models currently lacks standardization across providers and models. Depending on the provider, the core mechanism used differs or may not even exist \cite{OpenAIChatAPI2025, AnthropicMessagesAPI2025, GoogleGeminiDocs2025}. This suggests that further changes are likely to come that may impact the concrete manifestation of this smell. However, the core idea of explicitly controlling the amount of reasoning allocated to a task should remain relevant as the field evolves.}

\subsubsection{\noindent\textbf{Raw Vision Payload (RVP) *}}
\label{subsubsec:RVP}

\noindent\textbf{Context:} Vision-enabled LLM APIs allow sending images such as UI screenshots, diagrams, or camera photos together with text. It is straightforward for developers to pass any bytes or encoded image directly as input. In practice, many systems send full resolution screenshots or raw images captured from large displays or high resolution devices, even when only a small region is relevant. Since vision inputs are internally converted to tokens, image size, resolution, detail level, and the number of images have a strong impact on latency, cost, and how much of the multimodal context window is consumed\ccite{lee2025ergo,qian2025zoomer,vasu2024fastvlm}.

\noindent\textbf{Problem:} When systems systematically send raw, unbounded vision payloads, the effective behavior of the application depends on how the provider handles very large images or many images in a single request. This increases latency and token usage, which can create cost spikes and rate limit issues. It also harms reliability, since the model may fail to focus on a small but critical detail in a large screenshot, and truncation of multimodal context may occur in ways that are opaque to the developer. Finally, unfiltered screenshots can leak unnecessary or sensitive interface elements, which complicates compliance and privacy reviews\ccite{lee2025ergo,vasu2024fastvlm}.

\noindent\textbf{Solution:} Treat vision input as a limited resource and introduce explicit budgets for image size and content. Crop images to the region that is relevant for the task, downscale large screenshots to a reasonable maximum resolution, and avoid sending multiple redundant images in the same call. When the API exposes configuration for image detail, explicitly select an appropriate level rather than relying on defaults, for example using a low detail setting for coarse inspection and reserving a high detail setting for cases that require fine grained visual analysis. For tasks that primarily depend on text, prefer extracting text or structured data from the source instead of sending a screenshot of that text. Implement preprocessing in a dedicated helper rather than inlining raw image bytes in the call, and document the chosen image budgets alongside text context and output limits so that multimodal behavior is reproducible and auditable \ccite{qian2025zoomer}.

\noindent\textbf{Example:}
Listing~\ref{fig:RVP} illustrates Raw Vision Payload (RVP), where the faulty version sends full resolution screenshots directly to the vision model while the corrected version first crops and resizes the image and configures a low detail setting to respect an explicit input budget.

\begin{lstlisting}[style=mystyle,basicstyle=\ttfamily\footnotesize,language=python,alsoletter=_,caption=Raw Vision Payload (RVP),keywordstyle=\color{magenta},morekeywords={OpenAI,responses,create},deletekeywords={and,not,set,in},label=fig:RVP, moredelim={**[il][\DiffRemoved]{-}},
  moredelim={**[il][\DiffAdded]{+}}]
  from openai import OpenAI

  client = OpenAI()

 def describe_bug(screenshot_bytes):
+     small_image = resize_and_crop(screenshot_bytes, max_side=1024)
     response = client.responses.create(
         model="gpt-4.1-mini",
         input=[
             {
                 "role": "user",
                 "content": [
                     {   "type": "input_text",
                         "text": "Describe the bug shown in this screenshot."   
                     },
-                    {   "type": "input_image",
-                        "image": screenshot_bytes,
-                    },
-                         
+                    {   "type": "input_image",
+                        "image": small_image,
+                        "detail": "low",     
+                    },
                 ],
             }
         ],
         temperature=0.2,
     )
     return response.output[0].content[0].text

\end{lstlisting}

\noindent\textcolor{blue}{\textbf{Taxonomy Category:} \textit{Data semantics}. RVP belongs to this category because its core defect concerns the semantic suitability of multimodal inputs rather than the structural form of the API call. Raw screenshots or full-resolution images are passed to the model as if their content, granularity, and size were already appropriate for the intended task. The smell therefore arises from a mismatch between the visual payload that is actually provided and the semantic assumptions made by the system about what the model should attend to and process reliably.}

\noindent\textcolor{blue}{\textbf{Evolutionary Stability:} The core principle behind RVP does not depend on any single API parameter, making the concept of this smell relatively stable. However, the specific mechanisms available to control image detail and resolution vary between providers and may evolve over time, meaning the concrete instances of this smell could change as the field evolves \cite{openai2025vision, anthropic2025vision, openai_usingthe_2024_dba98b}.}

\subsubsection{\noindent\textbf{Overspecified Sampling Parameters (OSP) *}}
\label{subsubsec:OSP}

\noindent\textbf{Context:} Modern LLM runtimes expose multiple sampling controls that shape the stochasticity of decoding, most commonly temperature, nucleus sampling (\texttt{top\_p}), and top-k sampling (\texttt{top\_k}). These parameters are not independent, since they all restrict or reweight the next token distribution, and their exact semantics and supported combinations vary across providers and model families \ccite{azure_openai_temperature_topp,anthropic_claude_temperature_topp,willison_llm_anthropic_2025,openai_temp0_forum_2024}.

\noindent\textbf{Problem:} OSP occurs when a call pins several randomness node at once, for example setting temperature together with \texttt{top\_p} or \texttt{top\_k}. This overspecification makes the effective decoding policy hard to reason about and difficult to reproduce across providers, because small API changes, different default behaviors, or different implementations of the same parameter can alter which constraint dominates. In practice, OSP reduces maintainability and portability, complicates debugging and tuning, and can introduce silent quality regressions where output diversity or determinism shifts without a clear causal explanation \ccite{huggingfac_thedocum_2025_b77968, repowise_ensurede_2025_3759a4}.

\noindent\textbf{Solution:} Prefer a minimal, explicit sampling contract that uses a single primary control for randomness and documents the rationale. If temperature is the chosen control, set it explicitly and tune by task (e.g. Listing \ref{fig:TNES}) for creative generation, while avoiding extremes~\ccite{MinhEtAl2025MinpSampling}. If \texttt{top\_p} or \texttt{top\_k} is used, keep temperature unset and document that randomness is governed by \texttt{top\_p} or \texttt{top\_k}. Centralize these choices and validate changes with small regression suites to ensure that decoding behavior remains stable across model and provider updates\ccite{ mpfaffenbe_featurer_2025_639588, haktanceti_llmrando_2025_ec24a3}.

\noindent\textbf{Example:}
Listing~\ref{fig:OSP} illustrates Overspecified Sampling Parameters (OSP), where the faulty version explicitly sets multiple sampling controls (e.g., \texttt{temperature} and \texttt{top\_p} or \texttt{top\_k}) even though they are alternative mechanisms for shaping randomness. The corrected version keeps a single primary control (\texttt{temperature}) to avoid unintended interactions and to make the generation behaviour easier to reason about and reproduce.

\begin{lstlisting}[style=mystyle,basicstyle=\ttfamily\footnotesize,language=python,alsoletter=_,caption=Overspecified Sampling Parameters (OSP),keywordstyle=\color{magenta},morekeywords={anthropic,messages,create},deletekeywords={and,not,set,in},label=fig:OSP, moredelim={**[il][\DiffRemoved]{-}},
  moredelim={**[il][\DiffAdded]{+}}]
  import anthropic

  client = anthropic.Anthropic()

  def generate_creative_story(topic):
      message = client.messages.create(
          model="claude-3-5-sonnet-20241022",
          max_tokens=1024,
          messages=[
              {"role": "user", "content": f"Write a sci-fi story about {topic}"}
          ],
-         temperature=0.9,
-         top_p=0.95, #or top_k
+         temperature=0.9, #or top_p=0.95,
      return message.content[0].text

\end{lstlisting}
\noindent\textcolor{blue}{\textbf{Taxonomy Category:} \textit{Protocol}. OSP belongs to this category because its core defect concerns the sampling policy that governs inference behaviour rather than the local structure of a single invocation. Parameters such as \texttt{temperature}, \texttt{top\_p}, and \texttt{top\_k} act as competing controls over decoding stochasticity, and combining several of them without a clear rationale weakens the clarity and stability of the inference contract. OSP is therefore a protocol-level defect because it makes model behaviour harder to reason about, reproduce, and tune consistently across providers and updates.}

\noindent\textcolor{blue}{\textbf{Evolutionary Stability:} While the underlying principle of OSP, keeping the sampling policy explicit and minimal, should remain relevant as the field evolves, its concrete application may change over time. As the interactions between parameters such as temperature and top_p become better understood, what constitutes overspecification may be redefined. Additionally, as noted for TNES, temperature may not be available for all model families. OSP is therefore among the smells more subject to change as the field matures.}

\subsubsection{\noindent\textbf{Anonymous Inference Call (AIC) *}}
\label{subsubsec:AIC}

\noindent\textbf{Context:} Many LLM providers expose optional request fields that allow applications to attach a stable end-user identifier or equivalent metadata to each inference call, enabling traceability, abuse monitoring, and accountable auditing across sessions and tenants. In practice, such identifiers are also useful for correlating logs, reproducing incidents, and performing user scoped analytics in multi user systems \ccite{ openai_apipolic_2024_325242, anthropic_detectin_2025_f8b326, nerdfortec_safeguar_2023_1af5bd}.

\noindent\textbf{Problem:} AIC arises when an application performs inference without attaching any user level identifier, even though the surrounding code clearly operates in a multi user context (e.g., a web request, session id, or authenticated principal). This omission reduces maintainability and reliability because failures, hallucination reports, cost spikes, and rate limit events cannot be attributed to a specific user or session, which weakens incident response and hampers debugging and evaluation. From a governance perspective, anonymous calls also reduce accountability and complicate safety workflows such as detecting misuse patterns and enforcing user specific throttling \ccite{openai_safety_best_practices,google_gemini_safety_guidance,anthropic_claude_messages_create_python}.

\noindent\textbf{Solution:} Propagate an explicit user identifier from the application boundary to every LLM inference and document the traceability contract. Concretely, when a request is associated with a session or authenticated user, include a stable identifier in the inference call, as illustrated in Listing~\ref{fig:AIC}. In practice, this identifier is typically a pseudonymous token such as a hash of the user email (or an internal account id), rather than raw personal data. Keep the mapping in application storage, and ensure consistency across providers by routing the identifier through a single client wrapper. Finally, log the identifier alongside model name, sampling parameters, and request ids so that debugging, auditing, and abuse monitoring remain feasible under provider changes and evolving deployment configurations \ccite{berriai_litellm_issue_10106,openai_needhelp_2024_349ad1, openaideve_apibanfr_2022_6e7ce6, openaicomm_anysugge_2024_873212, openai_lessonsl_2022_863fc6}.

\noindent\textbf{Example:}
Listing~\ref{fig:AIC} illustrates Anonymous Inference Call (AIC), where the faulty version issues an LLM request without attaching any user level identifier, which weakens traceability and makes it harder to audit usage, investigate incidents, or enforce per user policies. The corrected version provides an explicit \texttt{user} identifier to associate the request with the originating session and enable accountability and monitoring.

\begin{lstlisting}[style=mystyle,basicstyle=\ttfamily\footnotesize,language=python,alsoletter=_,caption=Anonymous Inference Call (AIC),keywordstyle=\color{magenta},morekeywords={create},deletekeywords={and,not,set,in},label=fig:AIC, moredelim={**[il][\DiffRemoved]{-}},
  moredelim={**[il][\DiffAdded]{+}}]
  from openai import OpenAI

  client = OpenAI()
  
  def get_response(user_session_id, prompt):
      response = client.chat.completions.create(
          model="gpt-4o",
          messages=[{"role": "user", "content": prompt}],
          temperature=0.7,
+          user=user_session_id # Traceability: Link execution to the specific user
      )
      return response.choices[0].message.content
\end{lstlisting}

\noindent\textcolor{blue}{\textbf{Taxonomy Category:} \textit{Structural or API usage}. AIC belongs to this category because its core defect lies in how the inference request is constructed at the API level. Many providers expose explicit attribution fields such as \texttt{user} or \texttt{user\_id} to associate a request with an application-level principal or session. Omitting this information, despite its availability in the surrounding code, is therefore a defect in request composition rather than a semantic mismatch in the exchanged data or a protocol-level inference policy issue.}

\noindent\textcolor{blue}{\textbf{Evolutionary Stability:} While the principle of tracking users to ensure proper auditing of usage and abuse is there to stay, the way it is concretely implemented across APIs is still actively changing. For example, the \textit{user} parameter as described in this smell is being deprecated in favor of \textit{safety_identifier} and \textit{prompt_cache_key} in OpenAI's Chat Completions API \cite{OpenAIChatAPI2025}. This smell is therefore currently at high risk of obsolescence, even if the concrete mechanisms may eventually stabilize and standardize across providers.}

\medskip
The nine instrumented LLM code smells have been introduced above with their context, problem, solution, example, and taxonomy category. To complement these individual descriptions, Table~\ref{tab:defects-llm} summarizes their main software quality effects. More specifically, it shows whether each smell primarily impacts robustness, performance, maintainability, and reliability, thereby providing a compact cross-smell view of the catalog.

\newcolumntype{E}{>{\hsize=.25\hsize\centering\arraybackslash}X}

\begingroup
\renewcommand{\arraystretch}{1}
\setlength{\extrarowheight}{0pt}

\setlength{\aboverulesep}{0.2ex}
\setlength{\belowrulesep}{0.2ex}
\setlength{\cmidrulesep}{0.3ex}

\renewcommand{\cmark}{\smash{\raisebox{0.06ex}{\scalebox{0.78}{\ding{51}}}}}
\renewcommand{\tabcolsep}{5pt}

\makeatletter
\g@addto@macro\TPT@defaults{\setlength{\itemsep}{0pt}\setlength{\parsep}{0pt}}
\makeatother

\begin{table*}[t]
\centering
\footnotesize
\begin{threeparttable}
\caption{Overview of the taxonomy and effects of LLM code smells.}
\label{tab:defects-llm}
\renewcommand{\arraystretch}{1}

\begin{tabularx}{\textwidth}{@{} >{\raggedright\arraybackslash}X >{\centering\arraybackslash}m{2.5cm} *{4}{E} @{}}
\toprule

\multirow{2}{*}{\textbf{Code Smell}} &
\multirow{2}{*}{\textbf{Taxonomy}} &
\multicolumn{4}{c}{\textbf{Effect}} \\

\cmidrule(lr){3-6}
& & \textbf{RO} & \textbf{P} & \textbf{M} & \textbf{R} \\
\midrule

\textbf{No Structured Output (NSO)}                   & Data semantics           & \cmark &        &        & \cmark \\
\textbf{Unbounded Max Metrics (UMM)}                 & Protocol                 & \cmark & \cmark & \cmark &        \\
\textbf{LLM Temperature Not Explicitly Set (TNES)}   & Protocol                 &        &        & \cmark & \cmark \\
\textbf{No Model Version Pinning (NMVP)}             & Protocol                 &        &        & \cmark & \cmark \\
\textbf{No System Message (NSM)}                     & Structural or API usage  &        &        & \cmark & \cmark \\
\textbf{Reasoning Effort Not Explicitly Set (RENES) *} & Protocol                 & \cmark & \cmark & \cmark & \cmark \\
\textbf{Raw Vision Payload (RVP) *}                    & Data semantics           &        & \cmark & \cmark &        \\
\textbf{Overspecified Sampling Parameters (OSP) *}     & Protocol                 &        &        & \cmark & \cmark \\
\textbf{Anonymous Inference Call (AIC) *}              & Structural or API usage  & \cmark &        & \cmark &        \\

\bottomrule
\end{tabularx}

\begin{tablenotes}\footnotesize
\item[*] \textit{RO: Robustness} operational resilience, error-prone behavior
\item[*] \textit{P: Performance} cost, latency, memory
\item[*] \textit{M: Maintainability} reproducibility, portability, observability
\item[*] \textit{R: Reliability} response quality, consistency, stability over time
\end{tablenotes}
\end{threeparttable}
\end{table*}

\endgroup


\section{Detection Approach with SpecDetect4LLM}
\label{sec:detection_approach}

This section explains how we built our LLM code smell detection tool, \textit{SpecDetect4LLM}, based on an existing tool for AI-Specific code smells, \textit{SpecDetect4AI}\ccite{mahmoudi2025ai}. In this section, we briefly explain the underlying DSL, describe how we instantiated rules for the smells in the taxonomy, and discuss the limits of the current tool, with a focus on intra file occurrences that can be captured by static analysis.

\subsection{From the \textit{SpecDetect4AI} DSL to \textit{SpecDetect4LLM}}

\textit{SpecDetect4AI} models each code smell as a rule over the abstract syntax tree (AST) of a single Python file across the system. Rules are written in a small domain specific language (DSL) where variables range over AST nodes and where primitive predicates express syntactic or local control flow properties. The compiler translates a specification into a Python visitor that traverses one AST, enriches it with parent links and line numbers, and reports a violation whenever the rule body holds for some node.

\textit{SpecDetect4LLM} keeps this overall architecture intact. The main design effort consists in adding a library of LLM-oriented predicates so that each smell in the taxonomy can be expressed as one rule in the DSL. We adopted a pragmatic strategy. Whenever the existing predicates from \textit{SpecDetect4AI} were sufficient, we reused them unchanged. New predicates were added only when needed to encode a concrete LLM smell. This keeps the rule language compact and facilitates maintenance.

The new predicates fall into three intuitive groups. First, classification predicates identify LLM related calls and distinguish specialised families. For example, \texttt{isLLMCall} and \texttt{isTextGeneratingCall} recognise provider-specific text generation calls through their call path and simple alias analysis inside the file. Predicates such as \texttt{isVisionModelCall} and \texttt{isReasoningModelCall} refine this classification by checking model identifiers and constructor patterns for vision capable and reasoning capable models.

Second, configuration predicates test whether specific parameters are set at the call site. \texttt{hasNoTemperatureParameter} checks for missing temperature, \texttt{hasNoModelVersionPinning} checks that model identifiers are not version pinned, \texttt{hasNoReasoningEffort} detects calls to reasoning models without effort or thinking depth, \texttt{hasNoBoundedMetrics} focuses on resource limits such as output length or timeout, and \texttt{hasNoStructuredOutput} indicates that no structured response format or parser is attached. These predicates inspect both direct keyword arguments and simple dictionary based configuration, using a backward scan to recover the latest assignment to a configuration dictionary in the same file.

Third, prompt and payload predicates characterise chat like prompts and multimodal content. \texttt{isRoleBasedLLMChat} identifies role based chat invocations, while \texttt{hasNoSystemMessage} checks whether any system message or instruction can be recovered from the call or from a local configuration dictionary. For multimodal prompts, \texttt{hasImageContent} detects the presence of image payloads, \texttt{hasImagePreprocessing} searches for evidence of preprocessing pipelines, and \texttt{hasExplicitDetailLevel} checks whether a detail or quality level is configured. All these predicates operate within the AST and rely on simple local reasoning such as last assignment and pattern based recognition of image processing calls.

\subsection{Rule specification and alignment with the taxonomy}

Each smell in the taxonomy is encoded as a single declarative rule that combines a small set of predicates. As in \textit{SpecDetect4AI}, rules are written as quantified formulas over the AST. Variables range over calls or small local control-flow fragments inside the file, and the rule body is expressed as a conjunction of predicates that makes the symptoms of the smell explicit. Our implementation extends the predicate library to capture the diversity of LLM SDK surfaces and orchestration patterns observed in repositories. This design keeps the rules local, auditable, and directly aligned with the taxonomy introduced above.

\paragraph{No Structured Output.}
The NSO rule detects generation calls whose outputs are likely to be consumed as structured artifacts without an explicit output contract at the inference boundary. It relies on LLM-call classification predicates to identify text-generation calls and then applies \texttt{hasNoStructuredOutput} to determine whether the call lacks a structured response format, output parser, JSON schema, or equivalent local validation mechanism. The predicate inspects direct keyword arguments, locally assigned configuration dictionaries, and common SDK-specific fields used to request structured outputs. NSO is aligned with the Data semantics category because the rule captures a mismatch between the expected structure of downstream data and the absence of an explicit structure enforced at the LLM call.

\paragraph{Unbounded Max Metrics.}
The UMM rule detects LLM calls for which no local resource or execution bound can be recovered. It first identifies LLM inference calls using \texttt{isLLMCall}, then reports a violation when \texttt{hasNoBoundedMetrics} holds. This predicate checks whether the call or its locally recoverable configuration specifies limits such as output length, timeout, retry budget, or provider-specific generation bounds. It also accounts for nested configuration styles, including Gemini generation dictionaries and wrapper-level options when they are visible in the same file. UMM is aligned with the Protocol category because it concerns the explicit operational contract that governs latency, cost, and robustness during inference.

\paragraph{LLM Temperature Not Explicitly Set.}
The TNES rule targets text-generation calls for which temperature is required but not explicitly configured. It restricts attention to generative endpoints using \texttt{isTextGeneratingCall}, which recognises OpenAI-style \texttt{chat.completions.create} and \texttt{responses.create}, Gemini \texttt{generate\_content}, common orchestration entrypoints such as \texttt{invoke} and \texttt{predict}, and variable calls originating from a Hugging Face \texttt{pipeline("text-generation")}. The rule then reports a violation when \texttt{isLLMCallRequiringTemperature} holds and \texttt{hasNoTemperatureParameter} is true. Constructor-like calls and pipeline constructors are explicitly guarded against to avoid spurious alerts. TNES is aligned with the Protocol category because temperature defines part of the sampling contract and affects reproducibility, stability, and behavioral consistency across executions.

\paragraph{No Model Version Pinning.}
The NMVP rule detects LLM calls that rely on moving model aliases instead of stable model versions or snapshots. The rule first restricts the search space to calls where model versioning is meaningful, then applies \texttt{hasNoModelVersionPinning}. This predicate checks direct model arguments, locally assigned model variables, and simple configuration dictionaries when the model identifier can be recovered from the same file. The rule reports a violation when the model identifier corresponds to an alias or non-versioned family name rather than an immutable version. NMVP is aligned with the Protocol category because model identifiers define part of the inference contract and directly affect traceability, reproducibility, and behavioral stability over time.

\paragraph{No System Message.}
The NSM rule detects role-based chat calls that omit a system-level instruction. It combines \texttt{isRoleBasedLLMChat}, which recognises chat-style message arrays, with \texttt{hasNoSystemMessage}, which searches the local call context for a \texttt{system} role or an equivalent provider-specific instruction field. The rule reports a violation when the request is composed only from user or assistant messages and no recoverable global instruction is present. NSM is aligned with the Structural or API usage category because the relevant evidence lies in the request structure itself, especially in the construction of the message array passed to the LLM call.

\paragraph{Reasoning Effort Not Explicitly Set.}
The RENES rule detects reasoning-capable model calls that do not explicitly configure the amount or depth of reasoning. It first applies \texttt{isReasoningModelCall}, which recognises multiple provider-specific naming conventions, including OpenAI reasoning-oriented model families and pinned GPT-5 variants, Anthropic thinking-capable models, and Gemini thinking variants. The rule then applies \texttt{hasNoReasoningEffort}, which searches for direct reasoning arguments and nested configuration objects such as \texttt{generation\_config}. RENES is aligned with the Protocol category because reasoning effort controls the trade-off between accuracy, latency, and cost, and leaving it implicit weakens the inference contract of reasoning-capable systems.

\paragraph{Raw Vision Payload.}
The RVP rule detects vision-capable LLM calls that receive image payloads without visible preprocessing or explicit detail control. It combines \texttt{isVisionModelCall} with \texttt{hasImageContent}, then reports a violation when neither \texttt{hasImagePreprocessing} nor \texttt{hasExplicitDetailLevel} holds. The \texttt{isVisionModelCall} predicate recognises vision-capable providers and wrappers, while \texttt{hasImageContent} performs structural search over nested dictionaries, lists, common image encodings, and URL-based payload patterns. RVP is aligned with the Data semantics category because it concerns the semantic suitability of multimodal inputs before they are sent to the model, including whether the image content is bounded, selected, and appropriate for the intended task.

\paragraph{Overspecified Sampling Parameters.}
The OSP rule detects generation calls that configure several randomness controls simultaneously. It is implemented through \texttt{hasOverspecifiedSampling}, which identifies calls that set \texttt{temperature} together with \texttt{top\_p} or \texttt{top\_k}, either directly on the call or through an enclosing configuration context such as \texttt{with\_options}. The rule focuses on cases where the local decoding policy becomes harder to interpret because multiple sampling mechanisms are pinned at the same time. OSP is aligned with the Protocol category because it concerns the sampling policy that governs inference behavior and affects reproducibility, portability, and tuning consistency across providers.

\paragraph{Anonymous Inference Call.}
The AIC rule detects LLM calls executed in an apparent multi-user context without local user attribution. It combines \texttt{hasMultiUserContext}, which approximates the presence of request-scoped or session-scoped identifiers by searching enclosing scopes for patterns such as \texttt{request.user.id} and \texttt{session["user\_id"]}, with \texttt{hasUserAttribution}, which checks whether the call propagates a user identifier through explicit \texttt{user} parameters, metadata fields, safety identifiers, or keys passed through \texttt{**kwargs}. The rule reports a violation when an LLM call occurs in a multi-user context while lacking any recoverable attribution field. AIC is aligned with the Structural or API usage category because the defect concerns how the inference request is constructed at the API level, especially whether available attribution information is propagated as part of the call.

Overall, this rule specification follows the same structure for all nine LLM code smells. Each rule identifies a relevant LLM inference context, checks for the absence or problematic combination of local evidence, and reports a violation only when the smell can be justified from the analysed file. This design preserves the lightweight nature of SpecDetect4LLM while maintaining a direct correspondence between the implementation and the taxonomy. Structural or API usage smells are detected from the shape of the request and its immediate arguments. Data semantics smells are detected by relating LLM calls to the structure and constraints of their inputs or outputs. Protocol smells are detected by checking whether configuration choices that govern behavior across executions are made explicit at the call boundary.

\subsection{Scope of \textit{SpecDetect4LLM} }

\textcolor{blue}{As in \textit{SpecDetect4AI}, the current \textit{SpecDetect4LLM} tool is intentionally scoped as a lightweight AST-based static analyser. Each rule is evaluated on the AST of a single file and relies only on locally recoverable evidence, such as call structure, argument values, local control constructs, node relations, line numbers, and limited last-assignment recovery. This observation model is well suited to LLM code smell occurrences that are locally explicit at the call site or in its immediate context. This design choice preserves scalability, keeps reported smells easy to localise and inspect, and supports large-scale analysis over open-source systems. It also makes detection explainable, since each reported smell can be traced back to source-level evidence visible in the analysed file. Accordingly, the current prototype targets statically observable local occurrences rather than full system-level behaviour. occurrences that depend on information distributed across files, propagated through interprocedural control-flow or data-flow relations, or materialised through external configuration artefacts remain outside its scope. Despite this restricted observation model, the tool already achieves encouraging precision and recall on the smells targeted in this study. Richer forms of static reasoning could further improve coverage for such non-local occurrences, but incorporating them would have substantially enlarged the implementation and evaluation scope of the present work and was therefore left outside the scope of this paper.
}

\section{Validation Design and Results}
\label{sec:resuls_and_eval}
In this section, we present the study that aims to validate our tool-based approach. We follow a mixed-method methodology through quantitative and qualitative data collection and analysis.

\subsection{Research Questions}
\label{sec:research_questions}
We aim to respond to the following research questions:

\begin{itemize}

    \item \textbf{\textit{RQ$_{1}$:} How precise is \textit{SpecDetect4LLM} to detect LLM code smells?} This question assesses whether \textit{SpecDetect4LLM} can accurately detect LLM code smells.

    \item \textbf{\textit{RQ$_{2}$:} Does \textit{SpecDetect4LLM} scale efficiently across diverse LLM-integrating systems, regardless of system size?}  
    This question measures the runtime performance and scalability of the tool across a large set of LLM-integrating systems.

    \item \textcolor{blue}{\textbf{\textit{RQ$_{3}$:} How prevalent are LLM code smells in open-source software systems that integrate LLMs, and which LLM code smells occur most frequently?} This question quantifies the relative prevalence of each smell in our corpus.}
    
\end{itemize}


\subsection{Objects of the Validation}
\label{sub:dataset_selection}

We apply \textit{SpecDetect4LLM} to a large-scale corpus of LLM-integrating systems through a two-source construction that combines targeted GitHub mining with reuse of a previously curated corpus. For the GitHub mining phase, we implement an automated collector that queries the GitHub Search API and extracts a target sample of 500 repositories. Candidates are retrieved using a set of LLM and multimodal keywords and topics, including \textit{openai}, \textit{anthropic}, \textit{gemini}, \textit{ollama}, \textit{vllm}, \textit{llava}, \textit{vlm}, \textit{lvlm}, and phrases such as \textit{vision language} and \textit{reasoning model}, expressed as multiple search queries over repository metadata and README content. To avoid low-signal projects and reduce noise, we filter out forks and archived repositories and enforce a minimum popularity threshold \texttt{stars:$\geq$3}. We further prioritize recently maintained repositories by retaining projects with at least one push in the last 365 days, and by sorting candidates by the most recent push timestamp in descending order. To mitigate false positives due to keyword-only matches, we verify likely integration by fetching the repository README and probing common dependency manifests and lockfiles when available.

To strengthen external validity and mitigate sampling bias induced by GitHub search and ranking heuristics, we complement our crawl with 100 repositories from Shao et al.~\ccite{shao2025llmscorrectlyintegratedsoftware}. Their corpus captures widely used and previously studied systems, while our crawl emphasizes more recent and actively maintained projects, thereby increasing diversity in both project maturity and development activity. We then merge the two sources and remove duplicates (8 repositories). Duplicates include identical \texttt{owner/repo} identifiers and near-duplicates introduced by forks or mirrors of the same codebase. Next, we harmonize repository metadata across both subsets. Concretely, we standardize fields such as stars, forks, last update timestamps, the default branch, and the reference commit used for analysis. After fork removal, duplicate filtering, and metadata harmonization, the final dataset contains 692 repositories spanning both established and emerging LLM-integrating systems and supporting consistent downstream validation.

\subsection{Sampling Design and Ground-Truth Construction}
\label{subsec:gt}

As shown in Section \ref{sub:dataset_selection}, our full evaluation corpus contains 692 systems, 171,194 source files, and 40,919,625 LOC, targeting the functional Python components of the repository. Producing a complete line-level ground truth for the entire corpus would require well over 5,000 human-hours when extrapolating from the annotation throughput reported in~\ccite{Nandani2023DACOS}, which is infeasible within a single study.

To keep the workload tractable while preserving statistical validity, we construct a stratified random sample across repositories. This choice explicitly mitigates sampling bias. A global simple random draw would be dominated by large repositories and could systematically under-represent smaller ones. In contrast, stratified sampling ensures that each repository contributes to the sample proportionally to its size, improving representativeness across heterogeneous projects.

We set the target sample size using Cochran's formula for proportions~\ccite{cochran1977sampling}. With a 95\% confidence level, $Z = 1.96$, a margin of error $e = 0.05$, and the conservative choice $p = 0.5$, the large-population estimate is
\[
n_0 = \frac{Z^2 \, p \, \left(1 - p\right)}{e^2} \approx  384.16.
\]
Because our population is finite, we apply the finite population correction and obtain
\[
n = \frac{n_0}{1 + \frac{n_0 - 1}{N}} \approx 381.
\]
Accordingly, we annotate a stratified random sample of $n = 381$ files.

Strata correspond to files containing LLM inference call. Let $N_i$ denote the number of eligible files in repository $i$. The allocation for repository $i$ is computed proportionally as
\[
n_i = n \cdot \frac{N_i}{N},
\]
followed by standard rounding and a small adjustment to ensure $\sum_i n_i = n$. Within each repository, files are selected uniformly at random. Post-hoc checks show that per-smell frequencies deviate by at most 2.3 percentage points from the population, supporting the representativeness of the reduced labeling effort.

Because no public ground truth covering these LLM code smells currently exists, we constructed an adjudicated dataset through manual labeling by seven annotators. The annotator pool comprised the two first authors, both PhD students in software engineering whose research focuses on AI-based software systems, LLM-integrating systems, code smells, and static analysis, one Master's-level student with two years of hands-on AI development experience and prior familiarity with the AI and ML code smells of Zhang et al.~\ccite{zhang2022codesmellsmachinelearning}, and five additional contributors working in software engineering, including two with direct AI industry experience and two with general software engineering experience. Overall, the annotators had, on average, 3.4 years of professional experience. To avoid confirmation bias, none of the four annotators had prior knowledge of our detection tool \textit{SpecDetect4LLM} or its outputs, and the labeling was conducted blind to tool predictions.

All annotators followed a structured protocol adapted from prior work~\ccite{passi2018,Nandani2023DACOS}. The protocol was documented in the replication package and included, for each smell, a definition, operational cues describing how the smell manifests in code, and illustrative examples clarifying borderline cases. To standardize the process and reduce transcription errors, annotators performed labeling through a dedicated web interface that supports file browsing and line-level marking of smell occurrences. The workload was distributed evenly, with each annotator labeling 55 files, for a total of 385 labeling assignments, which covers the full sample and supports overlap for agreement estimation.


Blind multi-annotator labeling followed by adjudication yielded strong inter-rater reliability, with Fleiss'~$\kappa = 0.79$ and per-smell agreement ranging from 0.72 to 0.92, indicating near-perfect agreement. Throughout the process, annotators were blind to \textit{SpecDetect4LLM}'s implementation, internal smells, and any tool-produced outputs, which reduces the risk of confirmation bias. To support auditability and reuse, we release the full guidelines, code templates, and raw labels in the replication package~\cite{Replication_package}.

\subsection{Evaluation Metrics}
\label{subsec:metrics}

We compare the approach’s detections ($D$) with the ground-truth set ($G$).
Table~\ref{tab:confusion} summarises the four outcomes; $V$ denotes
validated positives, $V^{c}$ validated negatives.

\begin{table}[ht]
\scriptsize\centering
\setlength{\tabcolsep}{4pt} 
\caption{Outcome categories (TP: True Positives, FP: False Positives, FN: False Negatives, TN: True Negatives). 
$V$: smell instances, $V^c$: clean instances.}
\label{tab:confusion}
\begin{tabular}{c|cc}
\toprule
\textbf{Reported} & \textbf{$V$ (smell)} & \textbf{$V^{c}$ (clean)} \\ \midrule
$D$ (detected) & $\textit{TP}=D\cap V$ & $\textit{FP}=D\cap V^{c}$ \\
$D^{c}$ (missed) & $\textit{FN}=D^{c}\cap V$ & $\textit{TN}=D^{c}\cap V^{c}$ \\
\bottomrule
\end{tabular}
\vspace{-2mm}
\end{table}

As summarized in Table~\ref{tab:confusion},  
\emph{precision} is the proportion of validated positives among all detections
$\text{P}=\frac{\text{TP}}{\text{D}}$,  
\emph{recall} is the proportion of validated positives among all ground-truth
instances $\text{R}=\frac{\text{TP}}{\text{TP}+\text{FN}}$,  
and \emph{$F_{1}$-score} is the harmonic mean of the two $\text{$F_{1}$-score =}\frac{2\!\times\!\text{P}\times\!\ \text{R}}{\text{P}+\text{R}}$.

\subsection{Results}
\label{sub:results}
In this section, we study and answer the research questions.

\subsubsection{\textit{RQ$_{1}$:} How precise is \textit{SpecDetect4LLM} to detect LLM code smells?}

\begin{table}[t]
\centering
\caption{RQ1 per smell confusion outcomes and accuracy metrics on the annotated set. For RENES and AIC, recall is not reported because the annotated set contains too few positives to support a stable estimate. In these cases, TP and FP are obtained through a manual audit of flagged instances.}
\label{tab:rq1_detection_accuracy}
\small
\begin{tabular}{llrrrrrrr}
\toprule
 Smell & TP & FP & FN & TN & Precision & Recall & F1 \\
\midrule
 NMVP  & 84 & 13 & 14 & 121 & 0.866 & 0.857 & 0.862 \\
 TNES  & 33 &  9 & 15 & 175 & 0.786 & 0.688 & 0.733 \\
 NSO   & 26 &  4 &  8 & 194 & 0.867 & 0.765 & 0.812 \\
 UMM   & 18 &  5 & 14 & 195 & 0.783 & 0.562 & 0.655 \\
 NSM   & 16 &  0 &  1 & 215 & 1.000 & 0.941 & 0.970 \\
 RVP   &  3 &  0 &  2 & 227 & 1.000 & 0.600 & 0.750 \\
 OSP   &  3 &  0 &  1 & 228 & 1.000 & 0.750 & 0.857 \\
 RENES &  0 &  0 & N/A & N/A & 0.000 & N/A   & N/A \\
 AIC   &  0 &  0 & N/A & N/A & 0.000 & N/A   & N/A \\
\bottomrule
\end{tabular}
\end{table}

\begin{table}[t]
\centering
\caption{RQ1 aggregated metrics across smells on the annotated set.}
\label{tab:rq1_aggregates}
\small
\begin{tabular}{lrrr}
\toprule
Aggregation & Precision & Recall & F1 \\
\midrule
Micro average & 0.855 & 0.769 & 0.810 \\
Macro average & 0.900 & 0.738 & 0.806 \\
\bottomrule
\end{tabular}
\end{table}

This section evaluates the correctness of \textit{SpecDetect4LLM} when detecting LLM code smells in real world projects. We compare tool predictions against our manually labeled ground truth on the annotated subset. For each smell, we compute standard confusion outcomes. True positives correspond to smell instances correctly reported by the tool. False positives correspond to reported instances that are not confirmed by manual inspection. False negatives correspond to manually confirmed instances that are missed by the tool. True negatives correspond to non smell instances correctly left unflagged. From these counts, we derive precision, recall, and the $F_{1}$-score per smell.

Table~\ref{tab:confusion} summarises the smell-level results. We report both micro- and macro-averaged metrics to account for the imbalance between frequent and rare smells. Micro averaging pools all true positives, false positives, and false negatives across smells before computing precision, recall, and $F_{1}$-score. It is therefore mainly influenced by frequent smells. Macro averaging first computes the metric independently for each smell and then averages the smell-level values, giving the same weight to each smell regardless of its frequency. Across smells, \textit{SpecDetect4LLM} achieves a micro-averaged precision of 0.855, a micro-averaged recall of 0.769, and a micro-averaged $F_{1}$-score of 0.810 (Table~\ref{tab:rq1_aggregates}). The macro-averaged precision is higher at 0.900, while macro-averaged recall is slightly lower at 0.738, yielding a macro-averaged $F_{1}$-score of 0.806. This divergence is consistent with a performance profile where some less frequent smells exhibit very high alert correctness, which increases macro precision, while their incomplete coverage depresses macro recall. Conversely, micro averaging is more influenced by the frequent smells in the annotated set, particularly NMVP, TNES, UMM, and NSO, whose precision and recall are high but not perfect.

At the smell level, NMVP and NSO show strong and relatively balanced behavior, with $F_{1}$-score values of 0.862 and 0.812 respectively, indicating that most tool alerts correspond to true issues and that a large fraction of manually confirmed instances are recovered. TNES remains effective but exhibits residual error, with 9 false positives and 15 false negatives. UMM follows a similar pattern, with 5 false positives and 14 false negatives. NSM displays perfect precision but lower recall (0.941), suggesting that when the analyser flags a missing system message it is reliable, yet many true instances are missed. RVP and OSP achieve perfect precision with moderate recall, which is consistent with locally explicit cues that are easy to validate when present, but not always exposed in the syntactic forms captured by an intra file smell.

The error surface is largely explained by the observability limits of intra file static detection, a limitation inherited from \textit{SpecDetect4AI} and deliberately preserved in \textit{SpecDetect4LLM} to maintain scalability over large corpora. Concretely, the analyser reasons over local AST level evidence and does not reconstruct inter file data flow, interprocedural control flow, or runtime configuration resolution. As a result, false positives may arise when a call site appears underspecified in isolation although the constraint is enforced elsewhere, for instance through a shared wrapper that injects defaults or validates outputs. False negatives may arise when the relevant evidence is implemented through indirection, encapsulated in another module, or expressed through framework abstractions that hide the underlying provider call.

The following listings illustrate representative patterns that were encountered during manual adjudication and that explain how false positives and false negatives can be produced by an intra file analyser. These cases are not artefacts of our implementation but stem from the same design constraints and tradeoffs as \textit{SpecDetect4AI}, on which \textit{SpecDetect4LLM} is directly based.

Listing~\ref{fig:TNES_FP_wrapper} exemplifies a typical cross module configuration pattern that challenges intra file reasoning. The call site in \texttt{app/chat.py} does not expose any \texttt{temperature} argument and is therefore indistinguishable, locally, from a truly underspecified call. However, the project centralises the policy in \texttt{app/llm\_client.py}, where the client is constructed with defaults that enforce the sampling configuration globally. Under an intra file AST model, this yields a natural false positive for TNES because the evidence needed to validate compliance is external to the analysed unit. Eliminating such cases requires inter file data flow reconstruction and interprocedural reasoning about how configuration and client instances propagate to call sites, which is deliberately out of scope for a scalable static analyser.

Listing~\ref{fig:UMM_FP_factory_runtime} pushes this limitation further by introducing runtime selection of the concrete client implementation. The handler in \texttt{app/handler.py} invokes \texttt{get\_llm()} and then performs a completion without explicitly setting \texttt{timeout} and \texttt{max\_tokens}. Yet the factory in \texttt{app/factory.py} can return either a bounded wrapper or a raw client depending on a runtime flag derived from the environment. This means that the boundedness property is not a syntactic attribute of the call site, nor even a fixed attribute of the program under a single static interpretation. It is a property of a particular execution configuration. Consequently, an intra file static analyser cannot soundly decide whether the call will be bounded at runtime, and must approximate. In practice, this approximation can materialise as a false positive when the bounded client is selected, and as a true positive when the unbounded path is active.

\textcolor{blue}{To further explain this interpretation, we manually classified the residual FPs and FNs according to the type of information that is not available under the current intra file observation model. Among the 86 errors, 60 are related to inter file data flow or local value propagation that the analyser cannot fully reconstruct, including cases where relevant arguments are passed through variables, dictionaries, wrappers, helper abstractions, or client instances. Another 14 errors are related to runtime configuration resolution or cross module configuration layers, where the analysed call site appears underspecified although the missing constraint is introduced elsewhere in the project. Finally, 12 errors are related to interprocedural control flow, where the correctness of the detection depends on execution paths, conditional construction patterns, dynamic dispatch, or dependency injection. This distribution supports our interpretation that most remaining errors stem from the restricted observation model of the analyser rather than from inconsistencies in the smell definitions.}

These examples motivate the evaluation stance adopted in RQ1. \textit{SpecDetect4LLM} is a static, intra file analyser whose primary objective is to provide scalable and reliable warnings over large corpora. In this setting, maximising precision is often preferable to aggressively pursuing recall because warnings are meant to be actionable and triaged by developers, and excessive noise can undermine adoption. We therefore design smell detection rules to be conservative with respect to locally observable evidence, aiming to be \emph{soundy}~\cite{liblit2005soundy} within the intra file observation model, in the sense that reported violations are supported by explicit syntactic evidence at the analysed call site. Achieving system level soundness, in the strict sense, would require whole program reasoning that accounts for cross module configuration, dynamic dispatch, dependency injection, and runtime inputs such as environment variables and remote policies. Such an approach would substantially increase analysis cost and complexity, and it would still face undecidability and modelling gaps in dynamic Python ecosystems.

\begin{lstlisting}[style=mystyle,basicstyle=\ttfamily\footnotesize,language=python,alsoletter=_,caption=TNES false positive through cross module configuration injection,keywordstyle=\color{magenta},morekeywords={create},deletekeywords={and,not,set,in},label=fig:TNES_FP_wrapper, moredelim={**[il][\DiffRemoved]{-}},
  moredelim={**[il][\DiffAdded]{+}}]
# file: app/chat.py
from app.llm_client import client

def reply(prompt):
    return client.chat.completions.create(
        model="gpt-4o",
        messages=[{"role": "user", "content": prompt}],
-        # flagged as TNES because temperature is absent at the call site
    )

# file: app/llm_client.py
from openai import OpenAI
from app.config import settings

client = OpenAI()

+ client = OpenAI(
+     # temperature is enforced centrally and not visible at the call site
+     default_headers={"x-temperature": str(settings.temperature)}
+ )
\end{lstlisting}

\begin{lstlisting}[style=mystyle,basicstyle=\ttfamily\footnotesize,language=python,alsoletter=_,caption=UMM false positive because bounds are injected through a runtime selected client implementation in another module,keywordstyle=\color{magenta},morekeywords={create},deletekeywords={and,not,set,in},label=fig:UMM_FP_factory_runtime,moredelim={**[il][\DiffRemoved]{-}},moredelim={**[il][\DiffAdded]{+}}]
# file: app/handler.py
from app.factory import get_llm

def answer(prompt):
    llm = get_llm()
-    return llm.chat.completions.create(
-        model="gpt-4o",
-        messages=[{"role": "user", "content": prompt}],
        # flagged as UMM because timeout and max_tokens are not visible here
    )

# file: app/factory.py
import os
from app.bounded_client import BoundedOpenAI
from openai import OpenAI

def get_llm():
    # runtime switch, cannot be decided reliably intra file
    if os.getenv("LLM_BOUNDED_MODE", "1") == "1":
        return BoundedOpenAI()
    return OpenAI()

# file: app/bounded_client.py
import os
from openai import OpenAI

class BoundedOpenAI(OpenAI):
    def __init__(self):
        super().__init__()
        self._timeout_s = float(os.getenv("LLM_TIMEOUT_S", "15.0"))
        self._max_tokens = int(os.getenv("LLM_MAX_TOKENS", "512"))

    class chat:
    class completions:
        @staticmethod
        def create(*args, **kwargs):
            timeout_s = float(os.getenv("LLM_TIMEOUT_S", "15.0"))
            max_tokens = int(os.getenv("LLM_MAX_TOKENS", "512"))

            kwargs.setdefault("timeout", timeout_s)
            kwargs.setdefault("max_tokens", max_tokens)

            client = OpenAI()
            return client.chat.completions.create(*args, **kwargs)

\end{lstlisting}

\noindent{\textbf{Precision oriented validation for low prevalence smells (RENES and AIC)}}

In our annotated subset, \textit{RENES}  and \textit{AIC} occur too rarely to support a statistically meaningful recall estimate under random file sampling. In such sparse regimes, recall becomes either undefined or dominated by sampling noise, which would make any confusion matrix analysis for these smells uninformative. We therefore adopted a precision oriented protocol that directly evaluates the correctness of alerts conditional on the analyser emitting a warning.

Concretely, we manually audited the complete set of instances reported by \textit{SpecDetect4LLM} for RENES and AIC and adjudicated each alert as a true positive or false positive using the same operational decision sheet as in the main manual study. All audited alerts for these low prevalence smells were confirmed as true positives, with no false positives observed in the flagged set. This result supports the practical utility of these smells for practitioners, because when prevalence is low the primary question is whether emitted warnings can be trusted, rather than whether the analyser can recover an unobserved population of positives.

To mitigate the threat that an apparent absence of false positives is an artefact of narrow corpus patterns, we complemented the audit with smell level regression tests that explicitly encode both positive and negative scenarios. For RENES, the tests cover multiple providers and model families where reasoning controls are exposed through different API surfaces, including OpenAI reasoning oriented models, Anthropic models with explicit thinking configuration, and Gemini models with thinking related generation settings. For AIC, the tests capture anonymized or dynamically routed inference calls in which the effective model identity is not statically recoverable, as well as representative true negative controls in which the model is explicitly pinned to a concrete version. Importantly, both suites include false positive guards, such as non reasoning model calls that omit any reasoning configuration for RENES, and non LLM objects exposing a \texttt{create} method or wrapper based clients that hardcode a version pinned model for AIC. In combination, the manual audit provides external validity on real project instances, while the regression tests strengthen internal validity by constraining the smell semantics and preventing future changes to the analyser from reintroducing spurious matches.

\noindent This precision oriented validation does not claim that RENES and AIC have high recall in the wild. Instead, it establishes that when the analyser reports these low prevalence smells, the warning corresponds to a genuine integration issue under our specification. Further, by explicitly linking the regression tests to concrete positive and negative code patterns, we ensure that future tool revisions preserve the intended semantics and avoid reintroducing known sources of false positives.

\begin{tcolorbox}[
    enhanced,
    boxrule=0.5pt,
    colback=gray!05,
    colframe=black,
    arc=4pt,
    boxsep=4pt,
    left=4pt,
    right=4pt,
    top=0pt,
    bottom=0pt,
]
\textbf{Answer to \textit{RQ$_{1}$}:}\\
\textbf{\textit{SpecDetect4LLM} is precise at detecting LLM code smells}. Aggregating all smells, the analyser achieves a micro averaged precision of 0.855 and a macro averaged precision of 0.900, indicating that reported warnings are correct in the large majority of cases. At the smell level, precision remains consistently high across the main smells, reaching 0.866 for NMVP, 0.867 for NSO, 0.783 for UMM, and 0.786 for TNES, while NSM, RVP, and OSP exhibit perfect precision in this sample. The remaining false positives are largely explained by the intra file static observation model inherited from \textit{SpecDetect4AI}, where call sites can appear underspecified even though constraints are enforced through cross module wrappers, factories, or runtime configuration. In line with a soundy design objective, \textit{SpecDetect4LLM} prioritises conservative, locally justified findings to keep warnings actionable and to limit noise during developer triage. For low prevalence smells such as RENES and AIC, the annotated set contains too few positives to support stable recall estimation, so RQ1 conclusions focus on the \textbf{high precision of the smells that appear with sufficient frequency} in the adjudicated ground truth.
\end{tcolorbox}

\subsubsection{\textit{RQ$_{2}$:} Does \textit{SpecDetect4LLM} scale efficiently across diverse LLM-integrating systems, regardless of system size?}  

We evaluate the runtime performance and scalability of \textit{SpecDetect4LLM} by applying it to our full corpus of \textbf{171{,}194} Python files mined from Github repositories. Because the analyser is static, its computational cost is primarily driven by parsing and AST traversal within each module, rather than by global program structure. We therefore report both corpus level throughput and the distribution of per file runtimes to characterise scalability across diverse systems and file sizes.

\noindent
\begin{table}[ht]
\centering
\caption{General Statistics of the Analysis}
\label{tab:general_stats}
\scriptsize
\begin{tabular}{@{}lr@{}}
\toprule
\textbf{Metric} & \textbf{Value} \\ \midrule
Number of \texttt{.py} files analysed & 171{,}194 \\
Total analysis time & 36{,}746.9~s (10h 12min 27s) \\
Throughput & 4.66 files/s \\
Mean time per file & 0.215~s \\
Median time per file & 0.0222~s \\
95$^{\text{th}}$-percentile time per file & 0.681~s \\
Minimum time per file & 7.77$\times$10$^{-5}$~s \\
Maximum time per file & 1{,}607.1~s \\
\bottomrule
\end{tabular}
\end{table}

The complete run required \textbf{36{,}746.9~s} (10h 12min 27s), corresponding to an overall throughput of \textbf{4.66 files/s}. This throughput indicates that the analysis remains feasible at corpus scale, and it also provides a practical upper bound for expected wall clock time when processing new snapshots of similar magnitude. All experiments were executed on a MacBook Pro equipped with an Apple M4 Pro processor and \textbf{24\,GB} of RAM. Since analysis is embarrassingly parallel across files, this throughput can be further increased through parallel execution on multi core machines, while preserving identical detection semantics.

\begin{figure}[t]
\centering
\includegraphics[width=0.85\linewidth]{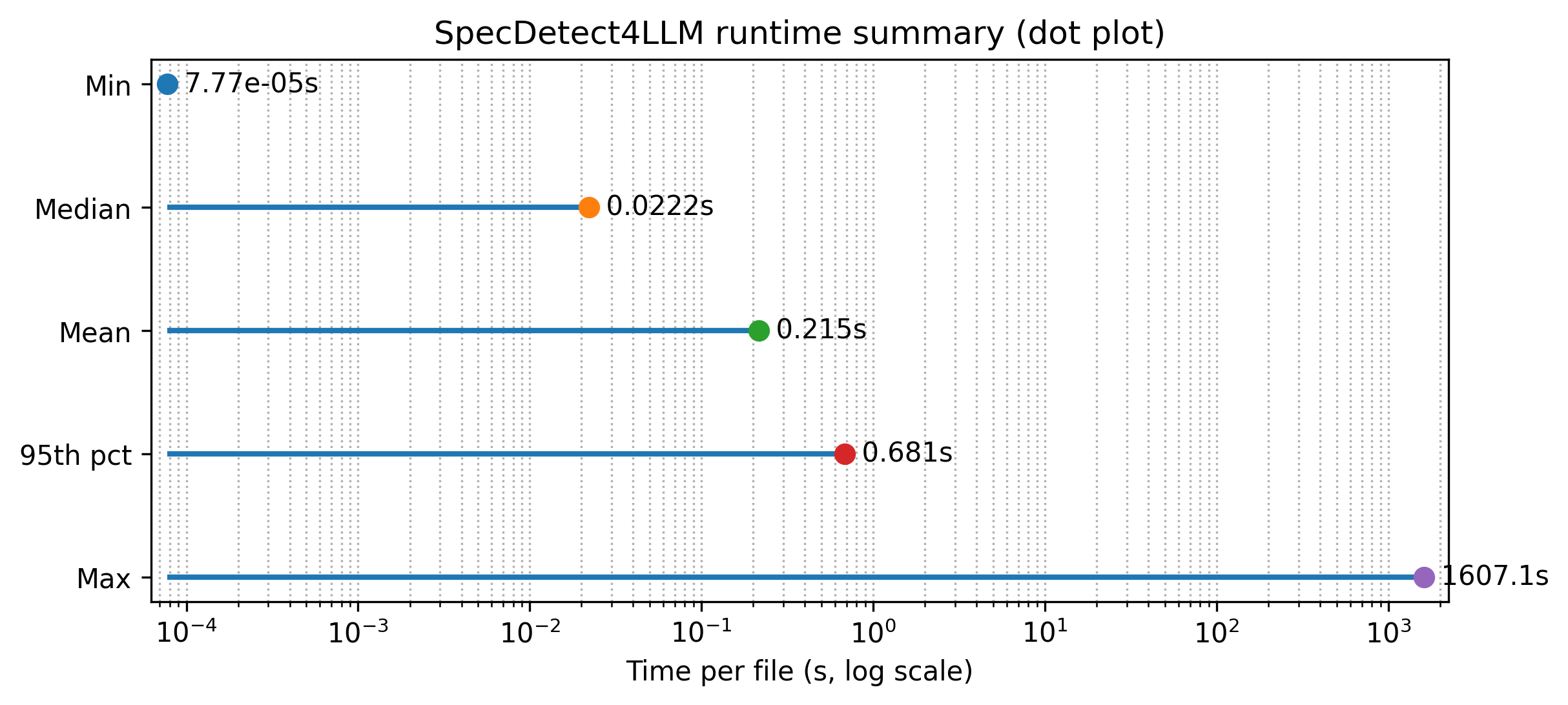}
\caption{Runtime summary per file on a log scale using a dot plot with stems (min, median, mean, 95th percentile, max).}
\label{fig:runtime-summary-dot}
\end{figure}

Runtime is highly concentrated on a small tail of difficult inputs. The median file is processed in \textbf{0.0222s}, and \textbf{95\%} of files finish within \textbf{0.681s}, showing that for most modules the detector runs in the millisecond range. The mean is substantially higher at \textbf{0.215s}, which is consistent with a heavy tailed distribution where a small fraction of files dominates the total cost. Figure~\ref{fig:runtime-summary-dot} visualises this skew using a log scaled summary plot.

This distributional profile is important for interpreting scalability with respect to system size. In large repositories, most files are typically small to medium sized, and their cost remains close to the median. Therefore, total runtime tends to grow approximately linearly with the number of files, with limited sensitivity to the overall architectural complexity of the system. In other words, repository scale affects wall clock time mainly through file count and the presence of exceptionally large modules, rather than through cross module interactions, which the tool intentionally does not model.

To better understand the sources of extreme runtimes, we manually inspected a sample of the slowest files. The maximum observed runtime was \textbf{1{,}607.1s}, and many other outliers correspond to very large test suites or vendored third party code, including substantial upstream libraries embedded in repositories. These cases are characterised by unusually large ASTs, deep nesting, and extensive use of metaprogramming patterns that increase parse and traversal cost. This evidence suggests that runtime variability is driven primarily by AST complexity and Python parsing overhead rather than by typical application code paths. From an engineering standpoint, these outliers motivate practical mitigations when the tool is used operationally, such as excluding vendor and generated code directories, limiting analysis to first party modules, or enforcing per file timeouts to prevent a small number of extreme inputs from delaying a full run.

Overall, the measured runtime behavior aligns with the scalability expectations of an intra file static detector. The strong right skew indicates that \textit{SpecDetect4LLM} is fast on the vast majority of files encountered in heterogeneous LLM integrating projects, while a small set of exceptionally large or complex modules dominates the tail. This profile supports efficient large scale analysis across systems of varying size and structure, and it provides clear operational guidance for controlling worst case costs through common repository.

\begin{tcolorbox}[
    enhanced,
    boxrule=0.5pt,
    colback=gray!05,
    colframe=black,
    arc=4pt,
    boxsep=4pt,
    left=4pt,
    right=4pt,
    top=0pt,
    bottom=0pt,
]
\textbf{Answer to \textit{RQ$_2$}:}\\
\textbf{\textit{SpecDetect4LLM} scales efficiently across diverse LLM integrating systems.} On 171{,}194 Python files, it sustains a throughput of 4.66 files/s and processes the typical module in 0.0222s, with 95\% of files completing within 0.681s. The remaining cost is concentrated in a small heavy tail of exceptionally large or syntactically complex files, most often originating from test suites or vendored third party code, indicating that runtime is governed by file level AST complexity and parsing overhead rather than by system level architectural size or cross module dependencies.
\end{tcolorbox}

\subsection{\textit{RQ$_{3}$:} How prevalent are LLM code smells in open-source software systems that integrate LLMs, and which LLM code smells occur most frequently?}

Having established the accuracy and scalability of \textit{SpecDetect4LLM}, we next examine how LLM code smells manifest in practice across the studied corpus.

To answer this question, we analyze the prevalence of each LLM code smell using complementary frequency indicators, namely the number of occurrences, affected files, and affected projects. This analysis allows us to identify the smells that are most prominent in practice and to derive empirical insights into their distribution across open source systems.

\subsection{Dataset and prevalence metrics}
We conducted a large scale prevalence study by running \textit{SpecDetect4LLM} on our entire dataset of 692 LLM-integrating systems comprising 171,194 source files. We report prevalence at two complementary levels of granularity. File level prevalence measures the proportion of files that contain at least one instance of a smell. Project level prevalence measures the proportion of projects in which at least one file contains an instance of a smell. In addition, we report the number of occurrences, which corresponds to the number of alerts emitted by the detector and captures intensity beyond simple presence.

\subsection{Overall prevalence}
Across the full dataset, 7,388 files are flagged by at least one rule, which corresponds to a file level prevalence of 4.32\% (7,388 out of 171,194). At the project level, 509 projects contain at least one alert, yielding a project level prevalence of 73.55\% (509 out of 692). These results highlight a clear contrast between breadth and locality. Smells are widespread across systems, yet typically concentrated in a small subset of files per project, consistent with the fact that LLM integration logic is usually implemented in dedicated modules or pipeline components rather than scattered uniformly throughout the codebase.

From an intensity perspective, we observe 28,617 occurrences overall, with an average of 3.87 occurrences per flagged file (28,617/7,388) and 56.22 occurrences per affected project (28,617/ 509). The median affected project, however, has 12 occurrences and only 4 flagged files, indicating a strongly right skewed distribution in which a minority of projects concentrate a large share of alerts.

\subsection{Per rule prevalence and dominant patterns}
\begin{figure}[t]
\centering
\includegraphics[width=0.82\linewidth]{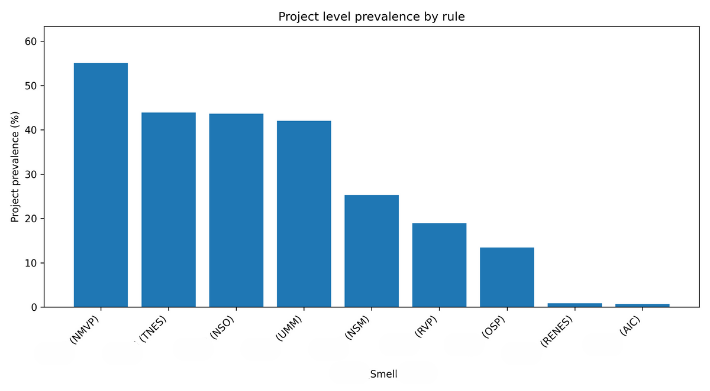}
\caption{Project level prevalence by rule. Values correspond to the fraction of projects (out of 692) in which each smell is triggered at least once.}
\label{fig:prev_project_by_rule}
\end{figure}

Table~\ref{tab:prevalence_by_rule} summarises prevalence per rule and Figure~\ref{fig:prev_project_by_rule} ranks rules by project level prevalence. Beyond raw frequency, each smell exhibits a characteristic mechanism that explains why it tends to appear, where it concentrates in code, and what risks it creates for LLM-integrating systems.

\begin{table}[t]
\centering
\caption{Prevalence per rule across 171,194 files and 692 projects. File and project prevalence are computed against the full dataset totals.}
\label{tab:prevalence_by_rule}
\small
\begin{tabular}{llrrrrr}
\toprule
 Smell & Files & File prev. (\%) & Projects & Project prev. (\%) & Occurrences \\
\midrule
 NMVP& 5,543 & 3.24 & 381 & 55.06 & 15,100 \\
 TNES& 1,859 & 1.09 & 304 & 43.93 & 4,219 \\
 NSO & 1,524 & 0.89 & 302 & 43.64 & 3,495 \\
 UMM & 1,425 & 0.83 & 291 & 42.05 & 3,372 \\
 NSM & 761 & 0.44 & 175 & 25.29 & 1,665 \\
 RVP & 302 & 0.18 & 131 & 18.93 & 441 \\
 OSP & 183 & 0.11 & 93 & 13.44 & 302 \\
 RENES& 6 & 0.00 & 6 & 0.87 & 15 \\
 AIC  & 7 & 0.00 & 5 & 0.72 & 8 \\
\bottomrule
\end{tabular}
\end{table}

\paragraph{NMVP, LLM Version Pinning Not Explicitly Set.}
NMVP is the dominant smell, affecting 55.06\% of projects (381 out of 692) and 3.24\% of files (5,543 out of 171,194), and accounting for 52.77\% of all occurrences (15,100 out of 28,617). This prevalence is consistent with an engineering practice where developers pass a high level model alias (for example a family name) rather than a pinned model identifier or revision. The key particularity of NMVP is that it directly threatens reproducibility and auditability. When the provider updates or reroutes an alias, the same code can yield different outputs without any change in the repository history, which complicates debugging, regression testing, and safety reviews. The high occurrence density (2.72 occurrences per touched file on average) suggests that once a project adopts non pinned calls, the pattern propagates across wrappers and utilities, producing repeated instantiations of the same fragility. A practical implication is that pinning should be treated as part of the call contract and enforced at the boundary layer, typically through a central client factory that accepts only pinned identifiers and rejects aliases during CI.

\paragraph{TNES, LLM Temperature Not Explicitly Set.}
TNES affects 43.93\% of projects (304 out of 692) but only 1.09\% of files (1,859 out of 171,194), with 4,219 occurrences. The defining trait of TNES is that it silently delegates a core stochasticity control to provider defaults, which are both model specific and subject to change. This smell is particularly likely to occur in early prototypes and research code, where teams focus on prompt engineering and treat sampling as an implementation detail. In production settings, TNES undermines stability because output variance can change across environments, across SDK versions, or across provider side updates. It also weakens evaluation because experimental comparisons can become partially confounded by implicit sampling. The observed pattern, high project prevalence with low file prevalence, indicates that temperature omission is concentrated in a small number of integration files, often a shared helper or wrapper. This makes remediation tractable. Introducing an explicit temperature policy, including a deterministic default for automation and an exploratory default for creative tasks, can reduce variance while preserving intentional diversity when needed.

\paragraph{UMM, LLM Calls Without Bounded Metrics.}
UMM appears in 42.05\% of projects (291 out of 692) with 3,372 occurrences. UMM is distinctive because it is an operational risk smell rather than a purely functional one. Unbounded calls typically omit constraints such as maximum tokens, timeouts, retry budgets, or cost ceilings, thereby exposing systems to latency spikes, runaway costs, and cascading failures under load. In practice, this smell often emerges when the call site is written as a direct SDK invocation without passing guardrail parameters, assuming that infrastructure will handle stability. The prevalence suggests that many projects still do not encode resource governance at the LLM boundary. A key recommendation is to centralise budget and latency constraints in a single gateway wrapper that applies strict defaults and logs budgets per request, since these constraints are difficult to enforce reliably when scattered across call sites.

\paragraph{NSO, No Structured Output in Pipeline.}
NSO affects 43.64\% of projects (302 out of 692) with 3,495 occurrences. NSO is characterised by treating LLM responses as free form text even when the downstream pipeline expects a parseable or schema compliant artefact. This smell is especially salient in LLM based automation, where unstructured outputs create brittle parsing heuristics, increase error handling complexity, and amplify silent failure modes. Its co prevalence with TNES and UMM is not accidental. Systems that treat the LLM call as an informal component often also omit explicit sampling and resource contracts. The most effective mitigation is to adopt a typed interface at the integration boundary, such as JSON schema constraints, response models, or constrained decoding, which reduces ambiguity and improves testability by enabling deterministic validators for output shape.

\paragraph{NSM, LLM With No System Message.}
NSM affects 25.29\% of projects (175 out of 692) with 1,665 occurrences. NSM is conceptually different from the configuration oriented smells above because it targets instruction hierarchy and behavioral alignment. Omitting a system message can lead to weak global constraints, inconsistent tone, and increased susceptibility to prompt injection when user content is untrusted or externally sourced. This smell is also sensitive to API style. Some frameworks encourage system prompts, while others embed global instructions in templates or middleware. Hence, a portion of the observed prevalence likely reflects genuine omissions at the call site, while another portion may correspond to projects that rely on hidden system layers that intra file analysis cannot observe. Practically, teams should treat system level policy as a first class artefact, store it in version control, and ensure that all user driven interactions are mediated by a stable instruction scaffold.

\paragraph{RVP, Raw Vision Payload.}
RVP affects 18.93\% of projects (131 out of 692) with 441 occurrences. RVP typically arises when vision requests embed raw images, often as large base64 payloads or unprocessed originals, directly into prompts. The particularity of RVP is that it couples security, privacy, and cost risks. Raw payloads can leak sensitive metadata or unnecessary content, and large images inflate tokenisation and inference cost while increasing latency. The non trivial prevalence indicates that multimodal usage is sufficiently common that payload hygiene is now a practical concern. Mitigation is largely engineering oriented. Preprocess images through resizing, compression, redaction, and optional feature extraction, then pass the minimal representation needed for the task, while enforcing explicit size limits at the gateway.

\paragraph{OSP, Overspecified Sampling Parameters.}
OSP affects 13.44\% of projects (93 out of 692) with 302 occurrences. OSP is characterised by specifying multiple sampling controls simultaneously, such as temperature together with top\_p and additional penalties, without a clear rationale or calibration. The risk is not merely redundancy. Over specification can yield unstable or unintuitive distributions, complicate reproducibility, and make behavior difficult to debug because small parameter shifts interact non linearly. The moderate prevalence suggests that some teams adopt parameter bundles copied from examples or tutorials rather than tuning parameters for a task. A principled mitigation is to standardise a minimal sampling policy, specify only the parameters necessary for the intended behavior, and document the rationale for each non default choice in the wrapper layer.

\paragraph{RENES, Reasoning Effort Not Explicitly Set and AIC, Anonymous Inference Call.}
RENES and AIC are rare in this corpus, affecting 0.87\% (6 out of 692) and 0.72\% (5 out of 692) of projects, respectively. For RENES, rarity can reflect limited adoption of provider specific reasoning effort controls, or the fact that such controls are often configured outside the call site, for example via environment driven configuration, which reduces detector observability. For AIC, low prevalence is consistent with the widespread use of wrappers and SDK clients that encapsulate provider identity, model routing, or authentication, meaning that anonymity may occur at a higher abstraction layer than the immediate call site. These long tail results should therefore be interpreted cautiously. Low measured prevalence may indicate genuine rarity, reduced static observability, or both.

\paragraph{Relating prevalence to the taxonomy.}
To connect the prevalence results to the taxonomy (Section~\ref{sec:catalog_taxo}), we interpret each smell through its defect category, namely Structural or API usage, Data semantics, and Protocol. This perspective reveals a clear skew toward protocol-related defects, which capture missing explicitness in system-wide LLM invocation contracts such as model stability, sampling control, and resource bounds. Concretely, when aggregating occurrences by category, protocol smells (NMVP, TNES, UMM, RENES) account for 22{,}706 out of 28{,}617 alerts (79.3\%), while data-semantics smells (NSO, RVP) account for 3{,}936 (13.8\%) and structural or API-usage smells (NSM, OSP, AIC) account for 1{,}975 (6.9\%). This distribution suggests that, in open-source LLM integrations, the dominant risks arise less from exotic API misuse and more from implicit defaults and unstable configurations that affect reproducibility and operational reliability across runs. It also provides a parsimonious explanation for why a small subset of smells dominates prevalence, as protocol-related choices tend to be shared across many call sites once adopted in a project-level wrapper or client factory.

\subsection{Implications and actionable guidance}
The prevalence results yield three actionable implications for engineering practice. First, the combination of high project level prevalence and low file level prevalence indicates that LLM related integration issues are widespread across systems yet structurally localised within codebases. In other words, smells tend to concentrate in a small set of dedicated modules that encapsulate model access, such as client factories, wrappers, and orchestration pipelines. This concentration suggests that remediation can be cost effective. Addressing a limited number of integration files can provide a disproportionate reduction in system wide risk, without requiring broad refactoring across unrelated components.

Second, the dominance of NMVP provides evidence that reproducibility and traceability remain systemic weaknesses in current LLM integration practice. When calls rely on non pinned model identifiers, system behavior can drift over time due to provider side updates, even when the application code remains unchanged. This undermines regression testing, incident analysis, and compliance oriented audits. Consequently, model identification should be treated as part of the stable interface of an LLM call, at the same level as prompt specification and output validation. Practically, teams can operationalise this requirement by enforcing pinned model identifiers through configuration schemas, code review checklists, and continuous integration guards, so that model upgrades become explicit, reviewable, and reversible changes.

Third, the high prevalence of the second tier smells, notably TNES, UMM, and NSO, indicates that reliability and operational robustness are recurring challenges once LLM functionality is introduced. These smells reflect missing explicitness at the call boundary, covering sampling control, resource and cost bounds, and the structure of returned outputs. A robust mitigation strategy is to standardise an integration contract that makes such parameters explicit and enforceable. In practice, a lightweight yet effective design is to centralise all provider calls within a single wrapper or client abstraction that applies explicit defaults, enforces bounded resources, logs critical parameters, and returns validated structured outputs when required. Restricting direct provider calls outside this abstraction reduces both defect introduction and maintenance overhead by consolidating integration policy in one auditable location.

\begin{tcolorbox}[
    enhanced,
    boxrule=0.5pt,
    colback=gray!05,
    colframe=black,
    arc=4pt,
    boxsep=4pt,
    left=4pt,
    right=4pt,
    top=0pt,
    bottom=0pt,
]
\textbf{Answer to \textit{RQ$_{3}$}:}\\
\textbf{LLM code smells are widespread in open source systems, with a small set of smells dominating prevalence}. Across 692 LLM integrating projects and 171{,}194 Python files, \textit{SpecDetect4LLM} reports that the most prevalent smells are concentrated in underspecified call site practices, where configuration and interaction contracts are not made explicit at the invocation point. The prevalence distribution is strongly skewed: a few smells account for most affected files and projects, while the remaining smells appear only sporadically. This pattern suggests that open source LLM integrations are primarily challenged by missing explicitness and incomplete invocation level specification, rather than by niche provider specific features. At the same time, the long tail of rare smells should not be interpreted as irrelevance, as their low observed prevalence can reflect both genuinely uncommon practices and limited local observability when constraints are implemented through cross module wrappers, factories, or runtime configuration. Overall, RQ3 indicates that prevalence is not uniform across smells, and that practitioners are most likely to encounter a recurring subset of high frequency smells that are broadly shared across projects, complemented by a sparse tail of context dependent smells.
\end{tcolorbox}

\label{sec:prevalence_study}

\section{Limitations and threats to validity}
\label{sec:limit_threats}

This section summarizes the main threats to validity. For clarity, we structure each validity dimension around three parts of the study, namely the catalog construction, the detection approach, and the prevalence study.

\subsection{Construct validity}

\subsubsection{Catalog construction}
A key threat is whether the proposed smells match real recurring problems in LLM integration. Because the catalog is built from mixed evidence sources, similar recommendations can sometimes refer to different issues depending on context. We mitigate this by keeping only smells that are observable in code and linked to a concrete remediation. Still, some practices are context dependent, so the same pattern may not always be harmful.

\subsubsection{Detection}
Some smells describe properties that may be enforced outside the local file, for example via wrappers or configuration layers. A static analyser can therefore flag a call that looks incomplete even if constraints are applied elsewhere, or miss cases where evidence is split across modules. We address this by defining conservative rules and by interpreting matches as local indicators rather than full system level guarantees. In addition, our tool is builds on \textit{SpecDetect4AI}, a state of the art static analyser for AI-specific code smells that was designed to be more precise and extensible, which provides a strong and stable foundation for implementing and evolving LLM specific rules.

\subsubsection{Prevalence}
Prevalence is measured from what the rules can observe, which can bias counts up or down when a rule relies on a local proxy. We therefore interpret prevalence as observable prevalence under local static analysis, not as the true frequency of violations in the whole system.

\subsection{Internal validity}

\subsubsection{Catalog construction}
The catalog may be influenced by what is most documented, for example popular providers or frameworks, which could bias the set of smells. We reduce this by requiring converging evidence across source types and by defining smells in provider agnostic terms when possible.

\subsubsection{Detection}
Detection results can be affected by implementation choices and by code patterns that hide calls, such as aliasing or abstraction layers. We mitigate this through regression tests covering typical and tricky cases, and by manual auditing of flagged instances when needed. Some residual errors may remain due to the limits of lightweight static analysis.

\subsubsection{Prevalence}
Prevalence can be confounded by corpus composition, for instance large repositories, vendored code, or test suites that concentrate certain patterns. We mitigate this by reporting prevalence at multiple levels such as occurrences, files, and projects, and by interpreting results with the observed skew in mind.

\subsection{Conclusion validity}

\subsubsection{Catalog construction}
Evidence does not imply the same impact in all settings. We therefore present smells as risk factors and avoid framing them as universally harmful defects.

\subsubsection{Detection}
For rare smells, recall estimates from random sampling are unstable. We address this by adapting the evaluation, focusing on precision for low prevalence smells, and avoiding strong claims when the number of positives is too small.

\subsubsection{Prevalence}
Prevalence distributions are often highly skewed, so comparisons across smells can be sensitive to a few large projects. We report complementary indicators and draw cautious conclusions, especially for rare smells.

\subsection{External validity}

\subsubsection{Catalog construction}
The main external threat to validity is that results may not fully generalize beyond open source Python systems. Other languages and environments may implement LLM integration through different mechanisms. We keep definitions conceptual, but extensions will be needed in other ecosystems.

\textcolor{blue}{There is also the threat of temporal instability and potential obsolescence of individual smells. This threat is a direct consequence of the nature of the field, which is still being established and is in rapid evolution, creating instability in standards and practices. However, we argue that it is precisely because of this that such work is essential, so that guidelines can be established and standards may follow. This is also a recurring limitation in smell catalogs in software engineering more broadly, where catalogs are often tied to API versions and require periodic revision as ecosystems evolve. Furthermore, even when a newer API version or model deprecates a given LLM Code Smell, older API versions and models still continue to be used in existing systems, meaning the smells remain applicable on these codebases.}

\textcolor{blue}{Additionally, the possible future obsolescence of an LLM code smell does not imply that the principle behind it becomes obsolete with it. We argue that the core ideas and principles motivating the smells remain relevant as the field of LLMs matures, even if the way some concepts are applied may evolve. Therefore, this work positions itself as an initial foundation to be built upon as the field matures, rather than a fixed specification. }

\textcolor{blue}{It is also important to note that not all smells are equally exposed to this risk. In general, Structural and Data Semantic LLM code smells are less at risk compared to Protocol ones, which are more directly dependent on provider API designs. Even within Protocol smells, the risk is not uniform. Smells such as TNES and RENES, for instance, are already subject to some variation across providers and models, and are therefore more susceptible than others.}

\subsubsection{Detection}

\textcolor{blue}{A threat to validity comes from the fact that the current \textit{SpecDetect4LLM} tool is based on lightweight AST-only static analysis. This design is effective for locally explicit manifestations, but it becomes more limited when the relevant evidence is distributed across files, procedures, wrappers, or configuration artefacts. LLM code smells may therefore depend on non-local relations that cannot be reconstructed within the current observation model, including inter-file, control-flow, and data-flow dependencies. As a result, the analyser may produce false positives when a call appears underspecified in isolation although the relevant constraint is enforced elsewhere, and false negatives when the decisive occurences are implemented through indirection or resolved only at runtime. We mitigate this threat by defining conservative rules and by interpreting detections as locally observable indicators rather than full system-level guarantees. Future work will explore richer forms of static reasoning to improve coverage for such non-local manifestations.}


\textcolor{blue}{Our detection tool SpecDetect4LLM also inherits the obsolescence risks of the smells it detects, as their concrete manifestations may change with API versions. However, SpecDetect4LLM's DSL and rule-based nature, inherited from SpecDetect4AI, not only makes it extensible but also easily maintainable and adjustable over time, substantially reducing obsolescence concerns at the detection level.}

\subsubsection{Prevalence}
Observed prevalence in open source may differ from industrial or regulated settings, where practices and encapsulation differ. We therefore interpret prevalence as trends in our corpus and encourage replications in other contexts.

\section{Related Work}
\label{sec:related_works}

\subsection{Domain-Specific Code Smells}

The concept of \textbf{code smells} is well established in software engineering~\ccite{10.5555/311424}. Over time, researchers have extended the notion by developing catalogs and taxonomies for specific application domains, such as for Android applications ~\ccite{Carvalho2019AndroidSmells} and Machine Learning (ML) ~\ccite{zhang2022codesmellsmachinelearning}. We essentially follow the same approach, but for LLM-integrating systems. Therefore, our LLM code smells distinguish themselves by their application domain specific to LLM inference inside source code. 

Mahhmoudi et al. introduced \textit{SpecDetect4AI}~\ccite{mahmoudi2025ai}, a static source code analysis tool for the detection of AI-specific code smells~\ccite{zhang2022codesmellsmachinelearning}. As mentioned in section~\ref{sec:detection_approach}, \textit{SpecDetect4LLM} is derived from this foundation by incorporating a new set of DSL-defined detection rules for LLM code smells. We selected \textit{SpecDetect4AI} as the base analyser because, in prior evaluations, it outperformed existing tools from the literature in terms of accuracy, and because its DSL-based rule layer is explicitly designed for extensibility, enabling the systematic addition of new detection rules without reengineering the underlying analysis pipeline. While both tools share the same architecture and workflow, they address distinct application domains.

\subsection{LLM Integration Defects}

In our previous work~\ccite{Mahmoudi2026LLMCodeSmells}, we introduced the concept of LLM code smells, presented an initial catalog of five defects, and provided the first version of \textit{SpecDetect4LLM}. However, that study was limited to preliminary precision measurements and a prevalence study on 200 open-source projects. This paper consolidates and extends that work by expanding the catalog to nine defects, introducing a lightweight taxonomy, extending \textit{SpecDetect4LLM}, and presenting a comprehensive evaluation of detection effectiveness (precision and recall on 288 source files) and prevalence (on a corpus of 692 projects).

The closest existing works to our study, with the exception of our previous work, are the taxonomy of code defects for LLM-based autonomous agents and the taxonomy of defects for LLM-enabled and RAG systems, introduced by \textbf{Ning et al.}~\ccite{ning2024definingdetectingdefectslarge} and \textbf{Shao et al.}~\ccite{shao2025llmscorrectlyintegratedsoftware} respectively.



While some defects introduced by \textbf{Ning et al.}~\ccite{ning2024definingdetectingdefectslarge}, such as \textit{LLM API-related Defect}, relate to similar integration concerns, their taxonomy differs from ours in scope and granularity. Ning et al. focus on broader agent-level defects, whereas our work targets recurrent source-code practices at LLM inference call sites, which may appear both in simple integrations and agentic systems~\ccite{ke2025surveyfrontiersllmreasoning}. Similarly, the defects studied by \textbf{Shao et al.}~\ccite{shao2025llmscorrectlyintegratedsoftware} describe higher-level symptoms in LLM-enabled and RAG systems, rather than concrete coding practices. In contrast, our LLM code smells identify implementation-level practices that undermine maintainability, reliability, performance, or robustness without necessarily constituting direct failures.

\textcolor{blue}{\textbf{Shao et al.}'s~\ccite{shao2025llmscorrectlyintegratedsoftware} catalog of defects is based on general failure modes, \textbf{Ning et al.}'s~\ccite{ning2024definingdetectingdefectslarge} catalog contains mainly agent workflow issues, while our LLM code smells present concrete implementation practices at the inference call site that should be avoided, like an implicit contract that should be respected. While there is some overlap between the integration defects found in both \textbf{Ning et al.} and \textbf{Shao et al.}, it remains limited.} 

\textcolor{blue}{The biggest overlap for both is with NSO. In \textbf{Shao et al.}, it relates to \textit{Incompatible LLM output format}, which describes downstream components receiving LLM responses that do not match the expected structure. In \textbf{Ning et al.}, NSO overlaps with \textit{LLM Output Parsing Error}, which describes the absence of error-handling or flexible parsing mechanisms when parsing model outputs. Both target the same risk as NSO, a mismatch between the expected structure and the free-form aspect of LLM outputs. The difference is that these two defects describe the resulting failure, while NSO targets the concrete coding practice that allows the failure to occur in the first place.}

\textcolor{blue}{For \textbf{Shao et al.}~\ccite{shao2025llmscorrectlyintegratedsoftware} there is also partial overlap with UMM and NSM, UMM partially overlaps with Exceeding LLM context limit, which discusses prompts that surpass the maximum context window and cause truncation. The overlap is with the token-budget dimension of UMM, since Exceeding LLM context limit does not address timeout or retry configuration. It is also taken from a different angle, Shao et al.'s defect focuses on how truncation may happen when the context limit is reached, while we argue that an unspecified token budget may result in additional costs and delays. We therefore recommend for practitioners to choose a token budget tuned to the needs of the task given to the LLM. As for NSM, it overlaps very slightly with Unclear context in prompt and Lacking restrictions in prompt. While these integration defects relate more to prompt smells, which are out of the scope of this work, we recognize some overlap with the fact of missing a system message, as it is a useful mechanism to properly specify instructions given to an LLM.}

\textcolor{blue}{The remaining LLM Code Smells, have no direct counterpart in either taxonomy. The full mapping is summarized in Table \ref{tab:table-overlap}.}

\begin{table}[t]
\centering
\caption{LLM code smell catalog overlap with Shao et al. and Ning et al.}
\label{tab:table-overlap}
\footnotesize
\begin{tabular}{p{2cm}|p{4cm}p{3.5cm}}
\toprule
\textbf{LLM Code Smell} & \textbf{Shao et al.} & \textbf{Ning et al.} \\
\midrule
\textbf{NMVP} & - & - \\ \hline
\textbf{TNES} & - & - \\ \hline
\textbf{NSO} & Incompatible LLM output format & LLM Output Parsing Error \\ \hline
\textbf{UMM} & Exceeding LLM context limit & - \\ \hline
\textbf{NSM} & Unclear context in prompt; Lacking restriction in prompt & - \\ \hline
\textbf{RVP} & - & - \\ \hline
\textbf{OSP} & - & - \\ \hline
\textbf{RENES} & - & - \\ \hline
\textbf{AIC} & - & - \\ 
\bottomrule
\end{tabular}
\end{table}

\subsection{Related Concepts}

Multiple similar, but fundamentally different concepts and works exist, these includes:
\begin{itemize}
    \item \textbf{Taxonomies of failures} in LLM-Integrating systems that emphasize high-level symptoms rather than code-level practices~\ccite{11081716,cemri2025multiagentllmsystemsfail,le-jeune-etal-2025-realharm};
    \item \textbf{Prompt-smell taxonomies} that are complementary to our study, but that operate on prompts, not code \ccite{ronanki2024promptsmellsomenundesirable,tian2025taxonomypromptdefectsllm};
    \item \textbf{Anti-patterns} for LLM inference \textbf{benchmarking} rather than integration~\ccite{agrawal2025evaluatingperformancellminference};
    \item \textbf{Analyses of defects, anti-patterns, and code smells in LLM-generated code}, whereas we study human-written source code that integrate LLMs~\ccite{zhuo2025identifyingmitigatingapimisuse,esfahani2024understandingdefectsgeneratedcodes}.
    \item \textcolor{blue}{\textbf{Large Language Model Operations (LLMOps)}, a subtype of MLOps, addresses the operational lifecycle of LLM models (deployment, serving, monitoring) \cite{10612341, 10779344, google_cloud_llmops, ibm_llmops}. Our work is distinct, addressing the quality of source code that integrates LLMs, and therefore positioning itself within the software engineering practices related to the lifecycle of LLM-integrating systems.}
\end{itemize}

\section{Conclusion}
\label{sec:conclusion}

In this work, we refine the concept of LLM code smells by extending our earlier preliminary work~\ccite{Mahmoudi2026LLMCodeSmells}. We construct a catalog of nine LLM code smells through a systematic literature review combined with grey literature mining, and organize this catalog into a taxonomy composed of three categories: Structural, Data-Semantic, and Protocol-Related. We also present \textit{SpecDetect4LLM}, a static analysis tool for detecting LLM code smells, and evaluate its detection effectiveness in terms of precision and recall on 381 source code files. Finally, we complement this study with an empirical assessment of the prevalence of LLM code smells across 692 open-source software projects, totaling 171,194 source files. 

Our results show that \textit{SpecDetect4LLM} achieves a macro-averaged precision of 0.913 and a macro-averaged recall of 0.718, indicating effective detection. They further reveal that 73.55\% of the analyzed projects exhibit at least one LLM code smell, suggesting that such defects are widespread in LLM-integrating systems and reinforcing the relevance of a dedicated catalog and taxonomy.

\section{Appendix}
\begin{appendices}

\begingroup
\makeatletter
\let\@auxout\@auxoutSP
\renewcommand{\refname}{Selected Papers}
\renewcommand{\bibname}{Selected Papers}
\renewcommand{\@biblabel}[1]{[SP#1]}

\makeatother
\endgroup

\end{appendices}


\begin{thebibliography}{103}
\ifx \bisbn   \undefined \def \bisbn  #1{ISBN #1}\fi
\ifx \binits  \undefined \def \binits#1{#1}\fi
\ifx \bauthor  \undefined \def \bauthor#1{#1}\fi
\ifx \batitle  \undefined \def \batitle#1{#1}\fi
\ifx \bjtitle  \undefined \def \bjtitle#1{#1}\fi
\ifx \bvolume  \undefined \def \bvolume#1{\textbf{#1}}\fi
\ifx \byear  \undefined \def \byear#1{#1}\fi
\ifx \bissue  \undefined \def \bissue#1{#1}\fi
\ifx \bfpage  \undefined \def \bfpage#1{#1}\fi
\ifx \blpage  \undefined \def \blpage #1{#1}\fi
\ifx \burl  \undefined \def \burl#1{\textsf{#1}}\fi
\ifx \doiurl  \undefined \def \doiurl#1{\url{https://doi.org/#1}}\fi
\ifx \betal  \undefined \def \betal{\textit{et al.}}\fi
\ifx \binstitute  \undefined \def \binstitute#1{#1}\fi
\ifx \binstitutionaled  \undefined \def \binstitutionaled#1{#1}\fi
\ifx \bctitle  \undefined \def \bctitle#1{#1}\fi
\ifx \beditor  \undefined \def \beditor#1{#1}\fi
\ifx \bpublisher  \undefined \def \bpublisher#1{#1}\fi
\ifx \bbtitle  \undefined \def \bbtitle#1{#1}\fi
\ifx \bedition  \undefined \def \bedition#1{#1}\fi
\ifx \bseriesno  \undefined \def \bseriesno#1{#1}\fi
\ifx \blocation  \undefined \def \blocation#1{#1}\fi
\ifx \bsertitle  \undefined \def \bsertitle#1{#1}\fi
\ifx \bsnm \undefined \def \bsnm#1{#1}\fi
\ifx \bsuffix \undefined \def \bsuffix#1{#1}\fi
\ifx \bparticle \undefined \def \bparticle#1{#1}\fi
\ifx \barticle \undefined \def \barticle#1{#1}\fi
\bibcommenthead
\ifx \bconfdate \undefined \def \bconfdate #1{#1}\fi
\ifx \botherref \undefined \def \botherref #1{#1}\fi
\ifx \url \undefined \def \url#1{\textsf{#1}}\fi
\ifx \bchapter \undefined \def \bchapter#1{#1}\fi
\ifx \bbook \undefined \def \bbook#1{#1}\fi
\ifx \bcomment \undefined \def \bcomment#1{#1}\fi
\ifx \oauthor \undefined \def \oauthor#1{#1}\fi
\ifx \citeauthoryear \undefined \def \citeauthoryear#1{#1}\fi
\ifx \endbibitem  \undefined \def \endbibitem {}\fi
\ifx \bconflocation  \undefined \def \bconflocation#1{#1}\fi
\ifx \arxivurl  \undefined \def \arxivurl#1{\textsf{#1}}\fi
\csname PreBibitemsHook\endcsname

\bibitem[\protect\citeauthoryear{Xia
  et~al.}{2025}]{xia2025analyzing16193llmpapers}
\begin{botherref}
\oauthor{\bsnm{Xia}, \binits{Z.}},
\oauthor{\bsnm{Zhu}, \binits{L.}},
\oauthor{\bsnm{Li}, \binits{B.}},
\oauthor{\bsnm{Chen}, \binits{F.}},
\oauthor{\bsnm{Li}, \binits{Q.}},
\oauthor{\bsnm{Liao}, \binits{C.}},
\oauthor{\bsnm{Wang}, \binits{F.}},
\oauthor{\bsnm{Liu}, \binits{H.}}:
Analyzing 16,193 LLM Papers for Fun and Profits
(2025).
\url{https://arxiv.org/abs/2504.08619}
\end{botherref}
\endbibitem

\bibitem[\protect\citeauthoryear{Shao
  et~al.}{2025}]{shao2025llmscorrectlyintegratedsoftware}
\begin{botherref}
\oauthor{\bsnm{Shao}, \binits{Y.}},
\oauthor{\bsnm{Huang}, \binits{Y.}},
\oauthor{\bsnm{Shen}, \binits{J.}},
\oauthor{\bsnm{Ma}, \binits{L.}},
\oauthor{\bsnm{Su}, \binits{T.}},
\oauthor{\bsnm{Wan}, \binits{C.}}:
Are LLMs Correctly Integrated into Software Systems?
(2025).
\url{https://arxiv.org/abs/2407.05138}
\end{botherref}
\endbibitem

\bibitem[\protect\citeauthoryear{Khatun}{2024}]{khatun2024reliability}
\begin{botherref}
\oauthor{\bsnm{Khatun}, \binits{A.}}:
Uncovering the reliability and consistency of ai language models: A systematic
  study.
PhD thesis,
University of Waterloo
(August 2024).
\url{https://uwspace.uwaterloo.ca/items/e01e11a6-e033-4f6a-85c6-849fba74e039}
\end{botherref}
\endbibitem

\bibitem[\protect\citeauthoryear{Yang et~al.}{2025}]{YANG2025113503}
\begin{barticle}
\bauthor{\bsnm{Yang}, \binits{W.}},
\bauthor{\bsnm{Some}, \binits{L.}},
\bauthor{\bsnm{Bain}, \binits{M.}},
\bauthor{\bsnm{Kang}, \binits{B.}}:
\batitle{A comprehensive survey on integrating large language models with
  knowledge-based methods}.
\bjtitle{Knowledge-Based Systems}
\bvolume{318},
\bfpage{113503}
(\byear{2025})
\doiurl{10.1016/j.knosys.2025.113503}
\end{barticle}
\endbibitem

\bibitem[\protect\citeauthoryear{Bucaioni
  et~al.}{2025}]{bucaioni2025functionalsoftwarereferencearchitecture}
\begin{botherref}
\oauthor{\bsnm{Bucaioni}, \binits{A.}},
\oauthor{\bsnm{Weyssow}, \binits{M.}},
\oauthor{\bsnm{He}, \binits{J.}},
\oauthor{\bsnm{Lyu}, \binits{Y.}},
\oauthor{\bsnm{Lo}, \binits{D.}}:
A Functional Software Reference Architecture for LLM-Integrated Systems
(2025).
\url{https://arxiv.org/abs/2501.12904}
\end{botherref}
\endbibitem

\bibitem[\protect\citeauthoryear{Fowler et~al.}{1999}]{10.5555/311424}
\begin{bbook}
\bauthor{\bsnm{Fowler}, \binits{M.}},
\bauthor{\bsnm{Beck}, \binits{K.}},
\bauthor{\bsnm{Brant}, \binits{J.}},
\bauthor{\bsnm{Opdyke}, \binits{W.}},
\bauthor{\bsnm{Roberts}, \binits{D.}}:
\bbtitle{Refactoring: Improving the Design of Existing Code}.
\bpublisher{Addison-Wesley Longman Publishing Co., Inc.},
\blocation{USA}
(\byear{1999})
\end{bbook}
\endbibitem

\bibitem[\protect\citeauthoryear{Zhang
  et~al.}{2022}]{zhang2022codesmellsmachinelearning}
\begin{botherref}
\oauthor{\bsnm{Zhang}, \binits{H.}},
\oauthor{\bsnm{Cruz}, \binits{L.}},
\oauthor{\bsnm{Deursen}, \binits{A.}}:
Code Smells for Machine Learning Applications
(2022).
\url{https://arxiv.org/abs/2203.13746}
\end{botherref}
\endbibitem

\bibitem[\protect\citeauthoryear{Tian
  et~al.}{2025}]{tian2025taxonomypromptdefectsllm}
\begin{botherref}
\oauthor{\bsnm{Tian}, \binits{H.}},
\oauthor{\bsnm{Wang}, \binits{C.}},
\oauthor{\bsnm{Yang}, \binits{B.}},
\oauthor{\bsnm{Zhang}, \binits{L.}},
\oauthor{\bsnm{Liu}, \binits{Y.}}:
A Taxonomy of Prompt Defects in LLM Systems
(2025).
\url{https://arxiv.org/abs/2509.14404}
\end{botherref}
\endbibitem

\bibitem[\protect\citeauthoryear{Paul et~al.}{2025}]{paul2025smells}
\begin{botherref}
\oauthor{\bsnm{Paul}, \binits{D.G.}},
\oauthor{\bsnm{Zhu}, \binits{H.}},
\oauthor{\bsnm{Bayley}, \binits{I.}}:
Investigating the Smells of LLM Generated Code.
SSRN.
Available at SSRN
(2025).
\doiurl{10.2139/ssrn.5601126} .
\url{https://ssrn.com/abstract=5601126}
\end{botherref}
\endbibitem

\bibitem[\protect\citeauthoryear{Mahmoudi
  et~al.}{2026}]{Mahmoudi2026LLMCodeSmells}
\begin{bchapter}
\bauthor{\bsnm{Mahmoudi}, \binits{B.}},
\bauthor{\bsnm{Chenail-Larcher}, \binits{Z.}},
\bauthor{\bsnm{Moha}, \binits{N.}},
\bauthor{\bsnm{Sti{\'e}venart}, \binits{Q.}},
\bauthor{\bsnm{Avellaneda}, \binits{F.}}:
\bctitle{Specification and detection of {LLM} code smells}.
In: \bbtitle{Proceedings of the 2026 IEEE/ACM 48th International Conference on
  Software Engineering, New Ideas and Emerging Results (ICSE-NIER '26)}.
\bpublisher{Association for Computing Machinery},
\blocation{New York, NY, USA}
(\byear{2026}).
\doiurl{10.1145/3786582.3786835} .
\burl{https://doi.org/10.1145/3786582.3786835}
\end{bchapter}
\endbibitem

\bibitem[\protect\citeauthoryear{Mahmoudi et~al.}{2025}]{mahmoudi2025ai}
\begin{botherref}
\oauthor{\bsnm{Mahmoudi}, \binits{B.}},
\oauthor{\bsnm{Moha}, \binits{N.}},
\oauthor{\bsnm{Stievenert}, \binits{Q.}},
\oauthor{\bsnm{Avellaneda}, \binits{F.}}:
AI-Specific Code Smells: From Specification to Detection
(2025).
\doiurl{10.48550/arXiv.2509.20491}
\end{botherref}
\endbibitem

\bibitem[\protect\citeauthoryear{Vaswani
  et~al.}{2017}]{10.5555/3295222.3295349}
\begin{bchapter}
\bauthor{\bsnm{Vaswani}, \binits{A.}},
\bauthor{\bsnm{Shazeer}, \binits{N.}},
\bauthor{\bsnm{Parmar}, \binits{N.}},
\bauthor{\bsnm{Uszkoreit}, \binits{J.}},
\bauthor{\bsnm{Jones}, \binits{L.}},
\bauthor{\bsnm{Gomez}, \binits{A.N.}},
\bauthor{\bsnm{Kaiser}, \binits{L.}},
\bauthor{\bsnm{Polosukhin}, \binits{I.}}:
\bctitle{Attention is all you need}.
In: \bbtitle{Proceedings of the 31st International Conference on Neural
  Information Processing Systems}.
\bsertitle{NIPS'17},
pp. \bfpage{6000}--\blpage{6010}.
\bpublisher{Curran Associates Inc.},
\blocation{Red Hook, NY, USA}
(\byear{2017})
\end{bchapter}
\endbibitem

\bibitem[\protect\citeauthoryear{Zhang et~al.}{2024}]{zhang2024vlmsurvey}
\begin{barticle}
\bauthor{\bsnm{Zhang}, \binits{J.}},
\bauthor{\bsnm{Huang}, \binits{J.}},
\bauthor{\bsnm{Jin}, \binits{S.}},
\bauthor{\bsnm{Lu}, \binits{S.}}:
\batitle{Vision-language models for vision tasks: A survey}.
\bjtitle{IEEE Transactions on Pattern Analysis and Machine Intelligence}
\bvolume{46}(\bissue{8}),
\bfpage{5625}--\blpage{5644}
(\byear{2024})
\doiurl{10.1109/TPAMI.2024.3369699}
\end{barticle}
\endbibitem

\bibitem[\protect\citeauthoryear{{OpenAI}}{2024}]{openai2024learning}
\begin{botherref}
\oauthor{\bsnm{{OpenAI}}}:
Learning to Reason with LLMs.
Technical report
(2024).
\url{https://openai.com/index/learning-to-reason-with-llms/}
\end{botherref}
\endbibitem

\bibitem[\protect\citeauthoryear{{ISO/IEC/IEEE}}{2017}]{iso24765_2017}
\begin{botherref}
\oauthor{\bsnm{{ISO/IEC/IEEE}}}:
{ISO/IEC/IEEE 24765:2017 Systems and software engineering: Vocabulary}.
International standard,
International Organization for Standardization
(2017)
\end{botherref}
\endbibitem

\bibitem[\protect\citeauthoryear{{ISO/IEC}}{2023}]{iso25010_2023}
\begin{botherref}
\oauthor{\bsnm{{ISO/IEC}}}:
{ISO/IEC 25010:2023 Systems and software engineering: Systems and software
  Quality Requirements and Evaluation (SQuaRE): Product quality model}.
International standard,
International Organization for Standardization
(2023)
\end{botherref}
\endbibitem

\bibitem[\protect\citeauthoryear{{IEEE}}{1990}]{ieee61012_1990}
\begin{botherref}
\oauthor{\bsnm{{IEEE}}}:
{IEEE Standard Glossary of Software Engineering Terminology}.
Ieee std 610.12-1990,
Institute of Electrical and Electronics Engineers
(1990).
\doiurl{10.1109/IEEESTD.1990.101064}
\end{botherref}
\endbibitem

\bibitem[\protect\citeauthoryear{Kitchenham and
  Charters}{2007}]{kitchenham2007guidelines}
\begin{botherref}
\oauthor{\bsnm{Kitchenham}, \binits{B.}},
\oauthor{\bsnm{Charters}, \binits{S.}}:
Guidelines for performing systematic literature reviews in software
  engineering.
Technical Report EBSE-2007-01,
EBSE 2007
(2007).
\url{https://www.elsevier.com/__data/promis_misc/525444systematicreviewsguide.pdf}
\end{botherref}
\endbibitem

\bibitem[\protect\citeauthoryear{Cherief
  et~al.}{2025}]{cherief2025automatedgreyliteratureextraction}
\begin{botherref}
\oauthor{\bsnm{Cherief}, \binits{H.A.}},
\oauthor{\bsnm{Mahmoudi}, \binits{B.}},
\oauthor{\bsnm{Chenail-Larcher}, \binits{Z.}},
\oauthor{\bsnm{Moha}, \binits{N.}},
\oauthor{\bsnm{Sti'evenart}, \binits{Q.}},
\oauthor{\bsnm{Avellaneda}, \binits{F.}}:
An Automated Grey Literature Extraction Tool for Software Engineering
(2025).
\url{https://arxiv.org/abs/2512.23066}
\end{botherref}
\endbibitem

\bibitem[\protect\citeauthoryear{Page et~al.}{2021a}]{page2021prisma}
\begin{barticle}
\bauthor{\bsnm{Page}, \binits{M.J.}},
\bauthor{\bsnm{McKenzie}, \binits{J.E.}},
\bauthor{\bsnm{Bossuyt}, \binits{P.M.}},
\bauthor{\bsnm{Boutron}, \binits{I.}},
\bauthor{\bsnm{Hoffmann}, \binits{T.C.}},
\bauthor{\bsnm{Mulrow}, \binits{C.D.}},
\bauthor{\bsnm{Shamseer}, \binits{L.}},
\bauthor{\bsnm{Tetzlaff}, \binits{J.M.}},
\bauthor{\bsnm{Akl}, \binits{E.A.}},
\bauthor{\bsnm{Brennan}, \binits{S.E.}},
\bauthor{\bsnm{Chou}, \binits{R.}},
\bauthor{\bsnm{Glanville}, \binits{J.}},
\bauthor{\bsnm{Grimshaw}, \binits{J.M.}},
\bauthor{\bsnm{Hrobjartsson}, \binits{A.}},
\bauthor{\bsnm{Lalu}, \binits{M.M.}},
\bauthor{\bsnm{Li}, \binits{T.}},
\bauthor{\bsnm{Loder}, \binits{E.W.}},
\bauthor{\bsnm{Mayo-Wilson}, \binits{E.}},
\bauthor{\bsnm{McDonald}, \binits{S.}},
\bauthor{\bsnm{McGuinness}, \binits{L.A.}},
\bauthor{\bsnm{Stewart}, \binits{L.A.}},
\bauthor{\bsnm{Thomas}, \binits{J.}},
\bauthor{\bsnm{Tricco}, \binits{A.C.}},
\bauthor{\bsnm{Welch}, \binits{V.A.}},
\bauthor{\bsnm{Whiting}, \binits{P.}},
\bauthor{\bsnm{Moher}, \binits{D.}}:
\batitle{The prisma 2020 statement: An updated guideline for reporting
  systematic reviews}.
\bjtitle{BMJ}
\bvolume{372},
\bfpage{71}
(\byear{2021})
\doiurl{10.1136/bmj.n71}
\end{barticle}
\endbibitem

\bibitem[\protect\citeauthoryear{Page et~al.}{2021b}]{page2021prismae}
\begin{barticle}
\bauthor{\bsnm{Page}, \binits{M.J.}},
\bauthor{\bsnm{McKenzie}, \binits{J.E.}},
\bauthor{\bsnm{Bossuyt}, \binits{P.M.}},
\bauthor{\bsnm{Boutron}, \binits{I.}},
\bauthor{\bsnm{Hoffmann}, \binits{T.C.}},
\bauthor{\bsnm{Mulrow}, \binits{C.D.}},
\bauthor{\bsnm{Shamseer}, \binits{L.}},
\bauthor{\bsnm{Tetzlaff}, \binits{J.M.}},
\bauthor{\bsnm{Akl}, \binits{E.A.}},
\bauthor{\bsnm{Brennan}, \binits{S.E.}},
\bauthor{\bsnm{Chou}, \binits{R.}},
\bauthor{\bsnm{Glanville}, \binits{J.}},
\bauthor{\bsnm{Grimshaw}, \binits{J.M.}},
\bauthor{\bsnm{Hrobjartsson}, \binits{A.}},
\bauthor{\bsnm{Lalu}, \binits{M.M.}},
\bauthor{\bsnm{Li}, \binits{T.}},
\bauthor{\bsnm{Loder}, \binits{E.W.}},
\bauthor{\bsnm{Mayo-Wilson}, \binits{E.}},
\bauthor{\bsnm{McDonald}, \binits{S.}},
\bauthor{\bsnm{McGuinness}, \binits{L.A.}},
\bauthor{\bsnm{Stewart}, \binits{L.A.}},
\bauthor{\bsnm{Thomas}, \binits{J.}},
\bauthor{\bsnm{Tricco}, \binits{A.C.}},
\bauthor{\bsnm{Welch}, \binits{V.A.}},
\bauthor{\bsnm{Whiting}, \binits{P.}},
\bauthor{\bsnm{Moher}, \binits{D.}}:
\batitle{Prisma 2020 explanation and elaboration: Updated guidance and
  exemplars for reporting systematic reviews}.
\bjtitle{BMJ}
\bvolume{372},
\bfpage{160}
(\byear{2021})
\doiurl{10.1136/bmj.n160}
\end{barticle}
\endbibitem

\bibitem[\protect\citeauthoryear{Kitchenham
  et~al.}{2023}]{kitchenham2022segress}
\begin{barticle}
\bauthor{\bsnm{Kitchenham}, \binits{B.}},
\bauthor{\bsnm{Madeyski}, \binits{L.}},
\bauthor{\bsnm{Budgen}, \binits{D.}}:
\batitle{Segress: Software engineering guidelines for reporting secondary
  studies}.
\bjtitle{IEEE Transactions on Software Engineering}
\bvolume{49}(\bissue{3}),
\bfpage{1273}--\blpage{1298}
(\byear{2023})
\doiurl{10.1109/TSE.2022.3174092}
\end{barticle}
\endbibitem

\bibitem[\protect\citeauthoryear{Schardt et~al.}{2007}]{pico}
\begin{barticle}
\bauthor{\bsnm{Schardt}, \binits{C.}},
\bauthor{\bsnm{Adams}, \binits{M.B.}},
\bauthor{\bsnm{Owens}, \binits{T.}},
\bauthor{\bsnm{Keitz}, \binits{S.}},
\bauthor{\bsnm{Fontelo}, \binits{P.}}:
\batitle{Utilization of the pico framework to improve searching pubmed for
  clinical questions}.
\bjtitle{BMC Medical Informatics and Decision Making}
\bvolume{7},
\bfpage{16}
(\byear{2007})
\doiurl{10.1186/1472-6947-7-16}
\end{barticle}
\endbibitem

\bibitem[\protect\citeauthoryear{Dyb{\aa} et~al.}{2007}]{dyba2007applying}
\begin{bchapter}
\bauthor{\bsnm{Dyb{\aa}}, \binits{T.}},
\bauthor{\bsnm{Dings{\o}yr}, \binits{T.}},
\bauthor{\bsnm{Hanssen}, \binits{G.K.}}:
\bctitle{Applying systematic reviews to diverse study types: An experience
  report}.
In: \bbtitle{Proceedings of the First International Symposium on Empirical
  Software Engineering and Measurement (ESEM 2007)},
pp. \bfpage{225}--\blpage{234}.
\bpublisher{IEEE}
(\byear{2007}).
\doiurl{10.1109/ESEM.2007.59} .
\burl{https://doi.org/10.1109/ESEM.2007.59}
\end{bchapter}
\endbibitem

\bibitem[\protect\citeauthoryear{Zhao et~al.}{2023}]{XXX}
\begin{botherref}
\oauthor{\bsnm{Zhao}, \binits{W.X.}},
\oauthor{\bsnm{Zhou}, \binits{K.}},
\oauthor{\bsnm{Li}, \binits{J.}},
\oauthor{\bsnm{Tang}, \binits{T.}},
\oauthor{\bsnm{Wang}, \binits{X.}},
\oauthor{\bsnm{Hou}, \binits{Y.}},
\oauthor{\bsnm{Min}, \binits{Y.}},
\oauthor{\bsnm{Zhang}, \binits{B.}},
\oauthor{\bsnm{Zhang}, \binits{J.}},
\oauthor{\bsnm{Dong}, \binits{Z.}},
\oauthor{\bsnm{Du}, \binits{Y.}},
\oauthor{\bsnm{Yang}, \binits{C.}},
\oauthor{\bsnm{Chen}, \binits{Y.}},
\oauthor{\bsnm{Jiang}, \binits{J.}},
\oauthor{\bsnm{Ren}, \binits{R.}},
\oauthor{\bsnm{Li}, \binits{Y.}},
\oauthor{\bsnm{Tang}, \binits{X.}},
\oauthor{\bsnm{Liu}, \binits{Z.}},
\oauthor{\bsnm{Liu}, \binits{P.}},
\oauthor{\bsnm{Nie}, \binits{J.}},
\oauthor{\bsnm{Wen}, \binits{J.}}:
A survey of large language models.
arXiv preprint
(2023)
{\href{https://arxiv.org/abs/2303.18223}{{arXiv:2303.18223}}}
{[cs.CL]}
\end{botherref}
\endbibitem

\bibitem[\protect\citeauthoryear{Garousi et~al.}{2019}]{garousi2021grey}
\begin{barticle}
\bauthor{\bsnm{Garousi}, \binits{V.}},
\bauthor{\bsnm{Felderer}, \binits{M.}},
\bauthor{\bsnm{M{"a}ntyl{"a}}, \binits{M.V.}}:
\batitle{Guidelines for including grey literature and conducting multivocal
  literature reviews in software engineering}.
\bjtitle{Information and Software Technology}
\bvolume{106},
\bfpage{101}--\blpage{121}
(\byear{2019})
\doiurl{10.1016/j.infsof.2018.09.006}
\end{barticle}
\endbibitem

\bibitem[\protect\citeauthoryear{Kamei et~al.}{2020}]{alves2020grey}
\begin{bchapter}
\bauthor{\bsnm{Kamei}, \binits{F.}},
\bauthor{\bsnm{Wiese}, \binits{I.}},
\bauthor{\bsnm{Pinto}, \binits{G.}},
\bauthor{\bsnm{Ribeiro}, \binits{M.}},
\bauthor{\bsnm{Soares}, \binits{S.}}:
\bctitle{On the use of grey literature: A survey with the brazilian software
  engineering research community}.
In: \bbtitle{Proceedings of the 34th Brazilian Symposium on Software
  Engineering}.
\bsertitle{SBES '20}.
\bpublisher{Association for Computing Machinery}
(\byear{2020}).
\doiurl{10.1145/3422392.3422442}
\end{bchapter}
\endbibitem

\bibitem[\protect\citeauthoryear{AI}{2025}]{PerplexityAI2025}
\begin{botherref}
\oauthor{\bsnm{AI}, \binits{P.}}:
Perplexity \url{https://www.perplexity.ai/}.
\url{https://www.perplexity.ai/}
(2025)
\end{botherref}
\endbibitem

\bibitem[\protect\citeauthoryear{{OpenAI}}{2025}]{OpenAIDocs2025}
\begin{botherref}
\oauthor{\bsnm{{OpenAI}}}:
OpenAI Platform Documentation
(2025).
\url{https://platform.openai.com/docs}
Accessed 2025-09-25
\end{botherref}
\endbibitem

\bibitem[\protect\citeauthoryear{{Google}}{2025}]{GoogleGeminiDocs2025}
\begin{botherref}
\oauthor{\bsnm{{Google}}}:
Gemini API - Google AI for Developers
(2025).
\url{https://ai.google.dev/}
Accessed 2025-09-25
\end{botherref}
\endbibitem

\bibitem[\protect\citeauthoryear{{Anthropic}}{2025}]{AnthropicDocs2025}
\begin{botherref}
\oauthor{\bsnm{{Anthropic}}}:
Claude Documentation (API, Models)
(2025).
\url{https://docs.claude.com/}
Accessed 2025-09-25
\end{botherref}
\endbibitem

\bibitem[\protect\citeauthoryear{{Hugging Face}}{2025}]{HuggingFaceHome2025}
\begin{botherref}
\oauthor{\bsnm{{Hugging Face}}}:
Hugging Face - The AI Community Building the Future.
\url{https://huggingface.co/}
Accessed 2025-09-25
\end{botherref}
\endbibitem

\bibitem[\protect\citeauthoryear{ollama}{2025}]{ollama_ollamaol_2025_6af93e}
\begin{botherref}
\oauthor{\bsnm{ollama}}:
Ollama/ollama.
\url{https://github.com/ollama/ollama}
Accessed 2025-12-29
\end{botherref}
\endbibitem

\bibitem[\protect\citeauthoryear{Mahmoudi and
  Chenail~Larcher}{2025}]{Replication_package}
\begin{botherref}
\oauthor{\bsnm{Mahmoudi}, \binits{B.}},
\oauthor{\bsnm{Chenail~Larcher}, \binits{Z.}}:
Replication_Package_LLM-code_smells.
\url{https://github.com/Brahim-Mahmoudi/Code_Smell_LLM}
(2025)
\end{botherref}
\endbibitem

\bibitem[\protect\citeauthoryear{Liu et~al.}{2024}]{NeedStructuredOutput2024}
\begin{bchapter}
\bauthor{\bsnm{Liu}, \binits{M.X.}},
\bauthor{\bsnm{Liu}, \binits{F.}},
\bauthor{\bsnm{Fiannaca}, \binits{A.J.}},
\bauthor{\bsnm{Koo}, \binits{T.}},
\bauthor{\bsnm{Dixon}, \binits{L.}},
\bauthor{\bsnm{Terry}, \binits{M.}},
\bauthor{\bsnm{Cai}, \binits{C.J.}}:
\bctitle{{``We Need Structured Output'': Towards User-centered Constraints on
  Large Language Model Output}}.
In: \bbtitle{Extended Abstracts of the CHI Conference on Human Factors in
  Computing Systems}.
\bsertitle{CHI EA '24}.
\bpublisher{Association for Computing Machinery},
\blocation{New York, NY, USA}
(\byear{2024}).
\doiurl{10.1145/3613905.3650756} .
\burl{https://doi.org/10.1145/3613905.3650756}
\end{bchapter}
\endbibitem

\bibitem[\protect\citeauthoryear{{Developer
  Service}}{2025}]{DeveloperService2025PydanticLLM}
\begin{botherref}
\oauthor{\bsnm{{Developer Service}}}:
A Practical Guide on Structuring LLM Outputs with Pydantic.
\url{https://dev.to/devasservice/a-practical-guide-on-structuring-llm-outputs-with-pydantic-50b4}
Accessed 2025-09-25
\end{botherref}
\endbibitem

\bibitem[\protect\citeauthoryear{{OpenAI}}{2025}]{OpenAIStructuredOutput}
\begin{botherref}
\oauthor{\bsnm{{OpenAI}}}:
API Reference - Structured model outputs
(2025).
\url{https://platform.openai.com/docs/guides/structured-outputs}
Accessed 2025-09-25
\end{botherref}
\endbibitem

\bibitem[\protect\citeauthoryear{Kharitonov}{2024}]{Kharitonov2024EnforcingJSON}
\begin{botherref}
\oauthor{\bsnm{Kharitonov}, \binits{D.}}:
Enforcing JSON Outputs in Commercial LLMs.
Towards Data Science.
\url{https://medium.com/data-science/enforcing-json-outputs-in-commercial-llms-3db590b9b3c8}
Accessed 2025-09-25
\end{botherref}
\endbibitem

\bibitem[\protect\citeauthoryear{Wyman and
  Barber}{2024}]{WymanBarber2024ValidateOutputs}
\begin{botherref}
\oauthor{\bsnm{Wyman}, \binits{M.}},
\oauthor{\bsnm{Barber}, \binits{S.}}:
How to Validate the Output of LLM-Based Products.
\url{https://okareo.com/blog/posts/validate-llm-output}
Accessed 2025-09-25
\end{botherref}
\endbibitem

\bibitem[\protect\citeauthoryear{Chen
  et~al.}{2025}]{chen2025adaptivelyrobustllminference}
\begin{botherref}
\oauthor{\bsnm{Chen}, \binits{Z.}},
\oauthor{\bsnm{Ye}, \binits{Y.}},
\oauthor{\bsnm{Zhou}, \binits{Z.}}:
Adaptively Robust LLM Inference Optimization under Prediction Uncertainty
(2025).
\url{https://arxiv.org/abs/2508.14544}
\end{botherref}
\endbibitem

\bibitem[\protect\citeauthoryear{{Amazon Web Services}}{2025}]{aws2025timeouts}
\begin{botherref}
\oauthor{\bsnm{{Amazon Web Services}}}:
Timeouts, retries and backoff.
\url{https://aws.amazon.com/fr/builders-library/timeouts-retries-and-backoff-with-jitter}.
Accessed 2025-09-25
(2025)
\end{botherref}
\endbibitem

\bibitem[\protect\citeauthoryear{{OpenAI}}{2025}]{openai_python2025}
\begin{botherref}
\oauthor{\bsnm{{OpenAI}}}:
openai-python.
\url{https://github.com/openai/openai-python}.
Accessed 2025-09-25
(2025)
\end{botherref}
\endbibitem

\bibitem[\protect\citeauthoryear{{Vellum AI}}{2025}]{VellumTemperature2025}
\begin{botherref}
\oauthor{\bsnm{{Vellum AI}}}:
LLM Temperature: How It Works and When You Should Use It
(2025).
\url{https://www.vellum.ai/llm-parameters/temperature}
Accessed 2025-09-25
\end{botherref}
\endbibitem

\bibitem[\protect\citeauthoryear{{Ollama}}{2025}]{OllamaModelfile2025}
\begin{botherref}
\oauthor{\bsnm{{Ollama}}}:
Modelfile: Valid parameters and values
(2025).
\url{https://github.com/ollama/ollama/blob/main/docs/modelfile.md#valid-parameters-and-values}
Accessed 2025-09-25
\end{botherref}
\endbibitem

\bibitem[\protect\citeauthoryear{Montandon
  et~al.}{2025}]{MontandonEtAl2025DABC}
\begin{botherref}
\oauthor{\bsnm{Montandon}, \binits{J.E.}},
\oauthor{\bsnm{Silva}, \binits{L.L.}},
\oauthor{\bsnm{Politowski}, \binits{C.}},
\oauthor{\bsnm{Prates}, \binits{D.}},
\oauthor{\bsnm{Brito~Bonif{\'a}cio}, \binits{A.}},
\oauthor{\bsnm{Boussaidi}, \binits{G.E.}}:
Unboxing default argument breaking changes in data science libraries
(2025)
\doiurl{10.48550/arXiv.2408.05129}
{\href{https://arxiv.org/abs/2408.05129}{{arXiv:2408.05129}}}
{[cs.SE]}.
JSS
\end{botherref}
\endbibitem

\bibitem[\protect\citeauthoryear{Minh et~al.}{2025}]{MinhEtAl2025MinpSampling}
\begin{bchapter}
\bauthor{\bsnm{Minh}, \binits{N.N.}},
\bauthor{\bsnm{Baker}, \binits{A.}},
\bauthor{\bsnm{Neo}, \binits{C.}},
\bauthor{\bsnm{Roush}, \binits{A.G.}},
\bauthor{\bsnm{Kirsch}, \binits{A.}},
\bauthor{\bsnm{Shwartz{-}Ziv}, \binits{R.}}:
\bctitle{Turning up the heat: Min-p sampling for creative and coherent llm
  outputs}.
In: \bbtitle{ICLR 2025}
(\byear{2025})
\end{bchapter}
\endbibitem

\bibitem[\protect\citeauthoryear{Institute}{2024}]{promptengineering_temperature_top_p_2024}
\begin{botherref}
\oauthor{\bsnm{Institute}, \binits{P.E.}}:
Complete Guide to Prompt Engineering with Temperature and Top-p.
Accessed: 2025-12-31
(2024).
\url{https://promptengineering.org/prompt-engineering-with-temperature-and-top-p/}
\end{botherref}
\endbibitem

\bibitem[\protect\citeauthoryear{{Microsoft
  Learn}}{2025}]{Microsoft2025FoundationLifecycle}
\begin{botherref}
\oauthor{\bsnm{{Microsoft Learn}}}:
Design to Support Foundation Model Life Cycles
(2025).
\url{https://learn.microsoft.com/en-us/azure/architecture/ai-ml/guide/manage-foundation-models-lifecycle}
Accessed 2025-09-25
\end{botherref}
\endbibitem

\bibitem[\protect\citeauthoryear{Morishige and Koshihara}{2025}]{Morishige}
\begin{bchapter}
\bauthor{\bsnm{Morishige}, \binits{M.}},
\bauthor{\bsnm{Koshihara}, \binits{R.}}:
\bctitle{Ensuring reproducibility in generative ai systems for general use
  cases: A framework for regression testing and open datasets}.
(\byear{2025}).
\doiurl{10.48550/arXiv.2505.02854}
\end{bchapter}
\endbibitem

\bibitem[\protect\citeauthoryear{{Hugging Face}}{2025}]{HFTransformersDocs2025}
\begin{botherref}
\oauthor{\bsnm{{Hugging Face}}}:
Transformers Documentation (Generation, Chat Templates, Model Revisions)
(2025).
\url{https://huggingface.co/docs/transformers}
Accessed 2025-09-25
\end{botherref}
\endbibitem

\bibitem[\protect\citeauthoryear{Wilson
  et~al.}{2014}]{WilsonEtAl2014BestPractices}
\begin{barticle}
\bauthor{\bsnm{Wilson}, \binits{G.}},
\bauthor{\bsnm{Aruliah}, \binits{D.A.}},
\bauthor{\bsnm{Brown}, \binits{C.T.}},
\bauthor{\bsnm{Chue~Hong}, \binits{N.P.}},
\bauthor{\bsnm{Davis}, \binits{M.}},
\bauthor{\bsnm{Guy}, \binits{R.T.}},
\bauthor{\bsnm{Haddock}, \binits{S.H.D.}},
\bauthor{\bsnm{Huff}, \binits{K.D.}},
\bauthor{\bsnm{Mitchell}, \binits{I.M.}},
\bauthor{\bsnm{Plumbley}, \binits{M.D.}},
\bauthor{\bsnm{Waugh}, \binits{B.}},
\bauthor{\bsnm{White}, \binits{E.P.}},
\bauthor{\bsnm{Wilson}, \binits{P.}}:
\batitle{Best practices for scientific computing}.
\bjtitle{PLOS Biology}
\bvolume{12}(\bissue{1}),
\bfpage{1001745}
(\byear{2014})
\end{barticle}
\endbibitem

\bibitem[\protect\citeauthoryear{Cleary}{2025}]{PromptHub2025SystemMessages}
\begin{botherref}
\oauthor{\bsnm{Cleary}, \binits{D.}}:
System Messages: Best Practices, Real-world Experiments \& Prompt Injection
  Protectors.
PromptHub Blog.
\url{https://www.prompthub.us/blog/everything-system-messages-how-to-use-them-real-world-experiments-prompt-injection-protectors}
Accessed 2025-09-25
\end{botherref}
\endbibitem

\bibitem[\protect\citeauthoryear{Jeong
  et~al.}{2025}]{jeong2025messagegenerationuserpreferences}
\begin{botherref}
\oauthor{\bsnm{Jeong}, \binits{M.}},
\oauthor{\bsnm{Cho}, \binits{J.}},
\oauthor{\bsnm{Khang}, \binits{M.}},
\oauthor{\bsnm{Jung}, \binits{D.}},
\oauthor{\bsnm{Hong}, \binits{T.}}:
System Message Generation for User Preferences using Open-Source Models
(2025).
\url{https://arxiv.org/abs/2502.11330}
\end{botherref}
\endbibitem

\bibitem[\protect\citeauthoryear{Wen et~al.}{2025}]{wen2025budgetthinker}
\begin{botherref}
\oauthor{\bsnm{Wen}, \binits{H.}},
\oauthor{\bsnm{Wu}, \binits{X.}},
\oauthor{\bsnm{Sun}, \binits{Y.}},
\oauthor{\bsnm{Zhang}, \binits{F.}},
\oauthor{\bsnm{Chen}, \binits{L.}},
\oauthor{\bsnm{Wang}, \binits{J.}},
\oauthor{\bsnm{Liu}, \binits{Y.}},
\oauthor{\bsnm{Liu}, \binits{Y.}},
\oauthor{\bsnm{Zhang}, \binits{Y.}},
\oauthor{\bsnm{Li}, \binits{Y.}}:
Budgetthinker: Empowering budget-aware llm reasoning with control tokens
(2025)
\doiurl{10.48550/arXiv.2508.17196}
\end{botherref}
\endbibitem

\bibitem[\protect\citeauthoryear{Wang et~al.}{2024}]{wang2024reasoningbudget}
\begin{bchapter}
\bauthor{\bsnm{Wang}, \binits{J.}},
\bauthor{\bsnm{Jain}, \binits{S.}},
\bauthor{\bsnm{Zhang}, \binits{D.}},
\bauthor{\bsnm{Ray}, \binits{B.}},
\bauthor{\bsnm{Kumar}, \binits{V.}},
\bauthor{\bsnm{Athiwaratkun}, \binits{B.}}:
\bctitle{Reasoning in token economies: Budget-aware evaluation of llm reasoning
  strategies}.
(\byear{2024}).
\doiurl{10.18653/v1/2024.emnlp-main.1112} .
\burl{https://aclanthology.org/2024.emnlp-main.1112/}
\end{bchapter}
\endbibitem

\bibitem[\protect\citeauthoryear{Chen et~al.}{2025}]{chen2025robustinference}
\begin{botherref}
\oauthor{\bsnm{Chen}, \binits{Z.}},
\oauthor{\bsnm{Ye}, \binits{Y.}},
\oauthor{\bsnm{Zhou}, \binits{Z.}}:
Adaptively robust llm inference optimization under prediction uncertainty
(2025)
\end{botherref}
\endbibitem

\bibitem[\protect\citeauthoryear{Author et~al.}{2025}]{themesLLM2025cain}
\begin{bchapter}
\bauthor{\bsnm{Author}, \binits{F.}},
\bauthor{\bsnm{Author}, \binits{S.}},
\bauthor{\bsnm{Author}, \binits{T.}}:
\bctitle{Themes of building llm-based applications for production: A
  practitioner's view}.
In: \bbtitle{Proceedings of the 2025 IEEE/ACM 4th International Conference on
  AI Engineering (CAIN)}.
\bpublisher{IEEE},
\blocation{Ottawa, Canada}
(\byear{2025})
\end{bchapter}
\endbibitem

\bibitem[\protect\citeauthoryear{Han et~al.}{2025}]{han2025tokenbudget}
\begin{bchapter}
\bauthor{\bsnm{Han}, \binits{T.}},
\bauthor{\bsnm{Wang}, \binits{Z.}},
\bauthor{\bsnm{Fang}, \binits{C.}},
\bauthor{\bsnm{Zhao}, \binits{S.}},
\bauthor{\bsnm{Ma}, \binits{S.}},
\bauthor{\bsnm{Chen}, \binits{Z.}}:
\bctitle{Token-budget-aware llm reasoning}.
(\byear{2025}).
\doiurl{10.18653/v1/2025.findings-acl} .
\burl{https://doi.org/10.18653/v1/2025.findings-acl}
\end{bchapter}
\endbibitem

\bibitem[\protect\citeauthoryear{{OpenAI}}{2025}]{OpenAIChatAPI2025}
\begin{botherref}
\oauthor{\bsnm{{OpenAI}}}:
API Reference - Chat Completions
(2025).
\url{https://platform.openai.com/docs/api-reference/chat}
Accessed 2025-09-25
\end{botherref}
\endbibitem

\bibitem[\protect\citeauthoryear{{Anthropic}}{2025}]{AnthropicMessagesAPI2025}
\begin{botherref}
\oauthor{\bsnm{{Anthropic}}}:
Messages API - Claude Docs
(2025).
\url{https://docs.claude.com/en/api/messages}
Accessed 2025-09-25
\end{botherref}
\endbibitem

\bibitem[\protect\citeauthoryear{Lee et~al.}{2025}]{lee2025ergo}
\begin{botherref}
\oauthor{\bsnm{Lee}, \binits{J.}},
\oauthor{\bsnm{Shin}, \binits{W.}},
\oauthor{\bsnm{Yang}, \binits{S.}},
\oauthor{\bsnm{Song}, \binits{K.-U.}},
\oauthor{\bsnm{Lim}, \binits{D.}},
\oauthor{\bsnm{Kim}, \binits{J.}},
\oauthor{\bsnm{Kim}, \binits{T.-H.}},
\oauthor{\bsnm{Kim}, \binits{B.-K.}}:
Ergo: Efficient high-resolution visual understanding for vision-language models
(2025)
\doiurl{10.48550/arXiv.2509.21991}
\end{botherref}
\endbibitem

\bibitem[\protect\citeauthoryear{Qian et~al.}{2025}]{qian2025zoomer}
\begin{botherref}
\oauthor{\bsnm{Qian}, \binits{J.}},
\oauthor{\bsnm{Wang}, \binits{C.}},
\oauthor{\bsnm{Yang}, \binits{Y.}},
\oauthor{\bsnm{Zhang}, \binits{C.}},
\oauthor{\bsnm{Jiang}, \binits{H.}},
\oauthor{\bsnm{Luo}, \binits{X.}},
\oauthor{\bsnm{Kang}, \binits{Y.}},
\oauthor{\bsnm{Lin}, \binits{Q.}},
\oauthor{\bsnm{Zhang}, \binits{A.}},
\oauthor{\bsnm{Jiang}, \binits{S.}},
\oauthor{\bsnm{Cao}, \binits{T.}},
\oauthor{\bsnm{Mao}, \binits{T.}},
\oauthor{\bsnm{Banerjee}, \binits{S.}},
\oauthor{\bsnm{Liu}, \binits{G.}},
\oauthor{\bsnm{Rajmohan}, \binits{S.}},
\oauthor{\bsnm{Zhang}, \binits{D.}},
\oauthor{\bsnm{Yang}, \binits{Y.}},
\oauthor{\bsnm{Zhang}, \binits{Q.}},
\oauthor{\bsnm{Qiu}, \binits{L.}}:
Zoomer: Adaptive image focus optimization for black-box mllm
(2025)
\doiurl{10.48550/arXiv.2505.00742}
\end{botherref}
\endbibitem

\bibitem[\protect\citeauthoryear{Vasu et~al.}{2024}]{vasu2024fastvlm}
\begin{botherref}
\oauthor{\bsnm{Vasu}, \binits{P.K.A.}},
\oauthor{\bsnm{Faghri}, \binits{F.}},
\oauthor{\bsnm{Li}, \binits{C.-L.}},
\oauthor{\bsnm{Koc}, \binits{C.}},
\oauthor{\bsnm{True}, \binits{N.}},
\oauthor{\bsnm{Antony}, \binits{A.}},
\oauthor{\bsnm{Santhanam}, \binits{G.}},
\oauthor{\bsnm{Gabriel}, \binits{J.}},
\oauthor{\bsnm{Grasch}, \binits{P.}},
\oauthor{\bsnm{Tuzel}, \binits{O.}},
\oauthor{\bsnm{Pouransari}, \binits{H.}}:
Fastvlm: Efficient vision encoding for vision language models
(2024)
\doiurl{10.48550/arXiv.2412.13303}
\end{botherref}
\endbibitem

\bibitem[\protect\citeauthoryear{{OpenAI}}{2025}]{openai2025vision}
\begin{botherref}
\oauthor{\bsnm{{OpenAI}}}:
Images and Vision | OpenAI API Documentation
(2025).
\url{https://developers.openai.com/api/docs/guides/images-vision}
\end{botherref}
\endbibitem

\bibitem[\protect\citeauthoryear{{Anthropic}}{2025}]{anthropic2025vision}
\begin{botherref}
\oauthor{\bsnm{{Anthropic}}}:
Vision - Claude API Documentation
(2025).
\url{https://platform.claude.com/docs/en/build-with-claude/vision}
\end{botherref}
\endbibitem

\bibitem[\protect\citeauthoryear{OpenAI}{2024}]{openai_usingthe_2024_dba98b}
\begin{botherref}
\oauthor{\bsnm{OpenAI}}:
Using the Vision API: Best Practices - API - OpenAI Developer.
\url{https://community.openai.com/t/using-the-vision-api-best-practices/942342}
Accessed 2025-12-29
\end{botherref}
\endbibitem

\bibitem[\protect\citeauthoryear{{Microsoft
  Corporation}}{2025}]{azure_openai_temperature_topp}
\begin{botherref}
\oauthor{\bsnm{{Microsoft Corporation}}}:
Azure OpenAI in Microsoft Foundry Models REST API.
Accessed 2025-12-09.
\url{https://learn.microsoft.com/en-us/azure/ai-foundry/openai/latest?view=foundry-classic}
\end{botherref}
\endbibitem

\bibitem[\protect\citeauthoryear{{Anthropic}}{2025}]{anthropic_claude_temperature_topp}
\begin{botherref}
\oauthor{\bsnm{{Anthropic}}}:
Create a Text Completion Claude API Reference.
Accessed 2025-12-09.
\url{https://docs.anthropic.com/claude/reference/complete_post}
\end{botherref}
\endbibitem

\bibitem[\protect\citeauthoryear{Willison}{2025}]{willison_llm_anthropic_2025}
\begin{botherref}
\oauthor{\bsnm{Willison}, \binits{S.}}:
Llm-anthropic.
Python package version 0.23 accessed 2025-12-09.
\url{https://pypi.org/project/llm-anthropic/}
\end{botherref}
\endbibitem

\bibitem[\protect\citeauthoryear{{OpenAI Developer
  Community}}{2024}]{openai_temp0_forum_2024}
\begin{botherref}
\oauthor{\bsnm{{OpenAI Developer Community}}}:
Clarifications on setting temperature = 0.
Discussion thread accessed 2025-12-09
(2024).
\url{https://community.openai.com/t/clarifications-on-setting-temperature-0/886447}
\end{botherref}
\endbibitem

\bibitem[\protect\citeauthoryear{Face}{2025}]{huggingfac_thedocum_2025_b77968}
\begin{botherref}
\oauthor{\bsnm{Face}, \binits{H.}}:
The document of generation seems to wrongly describe the default value of
  top_p, top_k and temperature
(2025).
\url{https://github.com/huggingface/transformers/issues/35045}
Accessed 2025-12-29
\end{botherref}
\endbibitem

\bibitem[\protect\citeauthoryear{RepoWise}{2025}]{repowise_ensurede_2025_3759a4}
\begin{botherref}
\oauthor{\bsnm{RepoWise}}:
Ensure Deterministic \& Fully Consistent LLM Output
(2025).
\url{https://github.com/RepoWise/backend/issues/11}
Accessed 2025-12-29
\end{botherref}
\endbibitem

\bibitem[\protect\citeauthoryear{mpfaffenberger}{2025}]{mpfaffenbe_featurer_2025_639588}
\begin{botherref}
\oauthor{\bsnm{mpfaffenberger}}:
Feature Request: Add Recommended Sampling Parameters for GLM Model
(2025).
\url{https://github.com/mpfaffenberger/code_puppy/issues/70}
Accessed 2025-12-29
\end{botherref}
\endbibitem

\bibitem[\protect\citeauthoryear{haktancetin}{2025}]{haktanceti_llmrando_2025_ec24a3}
\begin{botherref}
\oauthor{\bsnm{haktancetin}}:
LLM Randomness
(2025).
\url{https://github.com/haktancetin/496_CookBuddyProject/issues/24}
Accessed 2025-12-29
\end{botherref}
\endbibitem

\bibitem[\protect\citeauthoryear{OpenAI}{2024}]{openai_apipolic_2024_325242}
\begin{botherref}
\oauthor{\bsnm{OpenAI}}:
API Policy Violation Warning - OpenAI Developer Community.
\url{https://community.openai.com/t/api-policy-violation-warning-advice-on-how-to-best-resolve/788354}
Accessed 2025-12-29
\end{botherref}
\endbibitem

\bibitem[\protect\citeauthoryear{Anthropic}{2025}]{anthropic_detectin_2025_f8b326}
\begin{botherref}
\oauthor{\bsnm{Anthropic}}:
Detecting and Countering Malicious Uses of Claude.
\url{https://www.anthropic.com/news/detecting-and-countering-malicious-uses-of-claude-march-2025}
Accessed 2025-12-29
\end{botherref}
\endbibitem

\bibitem[\protect\citeauthoryear{Tech}{2023}]{nerdfortec_safeguar_2023_1af5bd}
\begin{botherref}
\oauthor{\bsnm{Tech}, \binits{N.F.}}:
Safeguarding your ai: Best practices for securing your openai api
(2023).
Accessed 2025-12-29
\end{botherref}
\endbibitem

\bibitem[\protect\citeauthoryear{{OpenAI}}{2025}]{openai_safety_best_practices}
\begin{botherref}
\oauthor{\bsnm{{OpenAI}}}:
Safety best practices.
\url{https://platform.openai.com/docs/guides/safety-best-practices}.
Accessed 2025-12-10
(2025)
\end{botherref}
\endbibitem

\bibitem[\protect\citeauthoryear{{Google
  DeepMind}}{2025}]{google_gemini_safety_guidance}
\begin{botherref}
\oauthor{\bsnm{{Google DeepMind}}}:
Safety guidance.
\url{https://ai.google.dev/gemini-api/docs/safety-guidance}.
Accessed 2025-12-10
(2025)
\end{botherref}
\endbibitem

\bibitem[\protect\citeauthoryear{{Anthropic}}{2025}]{anthropic_claude_messages_create_python}
\begin{botherref}
\oauthor{\bsnm{{Anthropic}}}:
Create a Message.
\url{https://platform.claude.com/docs/en/api/python/messages/create}.
Accessed 2025-12-10
(2025)
\end{botherref}
\endbibitem

\bibitem[\protect\citeauthoryear{raz alon and
  contributors}{2025}]{berriai_litellm_issue_10106}
\begin{botherref}
\oauthor{\bsnm{raz-alon}},
\oauthor{\bsnm{contributors}}:
{[Bug]: Anthropic API throws Bad Request when user_id in metadata contains
  email or phone number}.
\url{https://github.com/BerriAI/litellm/issues/10106}.
GitHub issue 10106, opened 2025-04-17, accessed 2025-12-10
(2025)
\end{botherref}
\endbibitem

\bibitem[\protect\citeauthoryear{OpenAI}{2024}]{openai_needhelp_2024_349ad1}
\begin{botherref}
\oauthor{\bsnm{OpenAI}}:
Need Help: Facing OpenAI Usage Violation Due to User's Abuse - API.
\url{https://community.openai.com/t/need-help-facing-openai-usage-violation-due-to-users-abuse/1004947}
Accessed 2025-12-29
\end{botherref}
\endbibitem

\bibitem[\protect\citeauthoryear{Community}{2022}]{openaideve_apibanfr_2022_6e7ce6}
\begin{botherref}
\oauthor{\bsnm{Community}, \binits{O.D.}}:
API Ban from User Abuse - API - OpenAI Developer Community.
\url{https://community.openai.com/t/api-ban-from-user-abuse/25400}
Accessed 2025-12-29
\end{botherref}
\endbibitem

\bibitem[\protect\citeauthoryear{Community}{2024}]{openaicomm_anysugge_2024_873212}
\begin{botherref}
\oauthor{\bsnm{Community}, \binits{O.}}:
Any Suggestions for Preventing Openai API Abuse - API - OpenAI ...
\url{https://community.openai.com/t/any-suggestions-for-preventing-openai-api-abuse/1075315}
Accessed 2025-12-29
\end{botherref}
\endbibitem

\bibitem[\protect\citeauthoryear{OpenAI}{2022}]{openai_lessonsl_2022_863fc6}
\begin{botherref}
\oauthor{\bsnm{OpenAI}}:
Lessons Learned on Language Model Safety and Misuse.
\url{https://openai.com/index/language-model-safety-and-misuse/}
Accessed 2025-12-29
\end{botherref}
\endbibitem

\bibitem[\protect\citeauthoryear{Nandani et~al.}{2023}]{Nandani2023DACOS}
\begin{bchapter}
\bauthor{\bsnm{Nandani}, \binits{H.}},
\bauthor{\bsnm{Saad}, \binits{M.}},
\bauthor{\bsnm{Sharma}, \binits{T.}}:
\bctitle{{DACOS}: A manually annotated dataset of code smells}.
In: \bbtitle{Proceedings of the 38th IEEE/ACM International Conference on
  Automated Software Engineering (ASE)},
pp. \bfpage{1}--\blpage{12}
(\byear{2023}).
\doiurl{10.1109/MSR59073.2023.00067}
\end{bchapter}
\endbibitem

\bibitem[\protect\citeauthoryear{Cochran}{1977}]{cochran1977sampling}
\begin{bbook}
\bauthor{\bsnm{Cochran}, \binits{W.G.}}:
\bbtitle{Sampling Techniques},
\bedition{3}rd edn.
\bpublisher{John Wiley \& Sons},
\blocation{New York}
(\byear{1977}).
\bcomment{Chap. 5}
\end{bbook}
\endbibitem

\bibitem[\protect\citeauthoryear{Passi and Jackson}{2018}]{passi2018}
\begin{botherref}
\oauthor{\bsnm{Passi}, \binits{S.}},
\oauthor{\bsnm{Jackson}, \binits{S.J.}}:
Trust in data science: Collaboration, translation, and accountability in
  corporate data science projects.
Proc. ACM Hum.-Comput. Interact.
\textbf{2}(CSCW)
(2018)
\doiurl{10.1145/3274405}
\end{botherref}
\endbibitem

\bibitem[\protect\citeauthoryear{Livshits et~al.}{2015}]{liblit2005soundy}
\begin{barticle}
\bauthor{\bsnm{Livshits}, \binits{B.}},
\bauthor{\bsnm{Sridharan}, \binits{M.}},
\bauthor{\bsnm{Smaragdakis}, \binits{Y.}},
\bauthor{\bsnm{Lhot\'{a}k}, \binits{O.}},
\bauthor{\bsnm{Amaral}, \binits{J.N.}},
\bauthor{\bsnm{Chang}, \binits{B.-Y.E.}},
\bauthor{\bsnm{Guyer}, \binits{S.Z.}},
\bauthor{\bsnm{Khedker}, \binits{U.P.}},
\bauthor{\bsnm{M\o{}ller}, \binits{A.}},
\bauthor{\bsnm{Vardoulakis}, \binits{D.}}:
\batitle{In defense of soundiness: a manifesto}.
\bjtitle{Commun. ACM}
\bvolume{58}(\bissue{2}),
\bfpage{44}--\blpage{46}
(\byear{2015})
\doiurl{10.1145/2644805}
\end{barticle}
\endbibitem

\bibitem[\protect\citeauthoryear{Carvalho
  et~al.}{2019}]{Carvalho2019AndroidSmells}
\begin{barticle}
\bauthor{\bsnm{Carvalho}, \binits{S.G.}},
\bauthor{\bsnm{Aniche}, \binits{M.}},
\bauthor{\bsnm{Ver{\'i}ssimo}, \binits{J.}},
\bauthor{\bsnm{Garcia}, \binits{A.}},
\bauthor{\bsnm{Alves}, \binits{V.}},
\bauthor{\bsnm{Gheyi}, \binits{R.}}:
\batitle{An empirical catalog of code smells for the presentation layer of
  android apps}.
\bjtitle{Empirical Software Engineering}
\bvolume{24}(\bissue{6}),
\bfpage{3546}--\blpage{3586}
(\byear{2019})
\doiurl{10.1007/s10664-019-09768-9}
\end{barticle}
\endbibitem

\bibitem[\protect\citeauthoryear{Ning
  et~al.}{2024}]{ning2024definingdetectingdefectslarge}
\begin{botherref}
\oauthor{\bsnm{Ning}, \binits{K.}},
\oauthor{\bsnm{Chen}, \binits{J.}},
\oauthor{\bsnm{Zhang}, \binits{J.}},
\oauthor{\bsnm{Li}, \binits{W.}},
\oauthor{\bsnm{Wang}, \binits{Z.}},
\oauthor{\bsnm{Feng}, \binits{Y.}},
\oauthor{\bsnm{Zhang}, \binits{W.}},
\oauthor{\bsnm{Zheng}, \binits{Z.}}:
Defining and Detecting the Defects of the Large Language Model-based Autonomous
  Agents
(2024).
\url{https://arxiv.org/abs/2412.18371}
\end{botherref}
\endbibitem

\bibitem[\protect\citeauthoryear{Ke
  et~al.}{2025}]{ke2025surveyfrontiersllmreasoning}
\begin{botherref}
\oauthor{\bsnm{Ke}, \binits{Z.}},
\oauthor{\bsnm{Jiao}, \binits{F.}},
\oauthor{\bsnm{Ming}, \binits{Y.}},
\oauthor{\bsnm{Nguyen}, \binits{X.-P.}},
\oauthor{\bsnm{Xu}, \binits{A.}},
\oauthor{\bsnm{Long}, \binits{D.X.}},
\oauthor{\bsnm{Li}, \binits{M.}},
\oauthor{\bsnm{Qin}, \binits{C.}},
\oauthor{\bsnm{Wang}, \binits{P.}},
\oauthor{\bsnm{Savarese}, \binits{S.}},
\oauthor{\bsnm{Xiong}, \binits{C.}},
\oauthor{\bsnm{Joty}, \binits{S.}}:
A Survey of Frontiers in LLM Reasoning: Inference Scaling, Learning to Reason,
  and Agentic Systems
(2025).
\url{https://arxiv.org/abs/2504.09037}
\end{botherref}
\endbibitem

\bibitem[\protect\citeauthoryear{Winston and Just}{2025}]{11081716}
\begin{bchapter}
\bauthor{\bsnm{Winston}, \binits{C.}},
\bauthor{\bsnm{Just}, \binits{R.}}:
\bctitle{A taxonomy of failures in tool-augmented llms}.
In: \bbtitle{AST 2025},
pp. \bfpage{125}--\blpage{135}
(\byear{2025}).
\doiurl{10.1109/AST66626.2025.00019}
\end{bchapter}
\endbibitem

\bibitem[\protect\citeauthoryear{Cemri
  et~al.}{2025}]{cemri2025multiagentllmsystemsfail}
\begin{botherref}
\oauthor{\bsnm{Cemri}, \binits{M.}},
\oauthor{\bsnm{Pan}, \binits{M.Z.}},
\oauthor{\bsnm{Yang}, \binits{S.}},
\oauthor{\bsnm{Agrawal}, \binits{L.A.}},
\oauthor{\bsnm{Chopra}, \binits{B.}},
\oauthor{\bsnm{Tiwari}, \binits{R.}},
\oauthor{\bsnm{Keutzer}, \binits{K.}},
\oauthor{\bsnm{Parameswaran}, \binits{A.}},
\oauthor{\bsnm{Klein}, \binits{D.}},
\oauthor{\bsnm{Ramchandran}, \binits{K.}},
\oauthor{\bsnm{Zaharia}, \binits{M.}},
\oauthor{\bsnm{Gonzalez}, \binits{J.E.}},
\oauthor{\bsnm{Stoica}, \binits{I.}}:
Why Do Multi-Agent LLM Systems Fail?
(2025)
\end{botherref}
\endbibitem

\bibitem[\protect\citeauthoryear{Le~Jeune
  et~al.}{2025}]{le-jeune-etal-2025-realharm}
\begin{bchapter}
\bauthor{\bsnm{Le~Jeune}, \binits{P.}},
\bauthor{\bsnm{Liu}, \binits{J.}},
\bauthor{\bsnm{Rossi}, \binits{L.}},
\bauthor{\bsnm{Dora}, \binits{M.}}:
\bctitle{Realharm: A collection of real-world language model application
  failures}.
In: \bbtitle{LLMSEC 2025},
pp. \bfpage{87}--\blpage{100}
(\byear{2025})
\end{bchapter}
\endbibitem

\bibitem[\protect\citeauthoryear{Ronanki
  et~al.}{2024}]{ronanki2024promptsmellsomenundesirable}
\begin{botherref}
\oauthor{\bsnm{Ronanki}, \binits{K.}},
\oauthor{\bsnm{Cabrero-Daniel}, \binits{B.}},
\oauthor{\bsnm{Berger}, \binits{C.}}:
Prompt Smells: An Omen for Undesirable Generative AI Outputs
(2024).
\url{https://arxiv.org/abs/2401.12611}
\end{botherref}
\endbibitem

\bibitem[\protect\citeauthoryear{Agrawal
  et~al.}{2025}]{agrawal2025evaluatingperformancellminference}
\begin{botherref}
\oauthor{\bsnm{Agrawal}, \binits{A.}},
\oauthor{\bsnm{Kedia}, \binits{N.}},
\oauthor{\bsnm{Agarwal}, \binits{A.}},
\oauthor{\bsnm{Mohan}, \binits{J.}},
\oauthor{\bsnm{Kwatra}, \binits{N.}},
\oauthor{\bsnm{Kundu}, \binits{S.}},
\oauthor{\bsnm{Ramjee}, \binits{R.}},
\oauthor{\bsnm{Tumanov}, \binits{A.}}:
On Evaluating Performance of LLM Inference Serving Systems
(2025)
\end{botherref}
\endbibitem

\bibitem[\protect\citeauthoryear{Zhuo
  et~al.}{2025}]{zhuo2025identifyingmitigatingapimisuse}
\begin{botherref}
\oauthor{\bsnm{Zhuo}, \binits{T.Y.}},
\oauthor{\bsnm{He}, \binits{J.}},
\oauthor{\bsnm{Sun}, \binits{J.}},
\oauthor{\bsnm{Xing}, \binits{Z.}},
\oauthor{\bsnm{Lo}, \binits{D.}},
\oauthor{\bsnm{Grundy}, \binits{J.}},
\oauthor{\bsnm{Du}, \binits{X.}}:
Identifying and Mitigating API Misuse in Large Language Models
(2025)
\end{botherref}
\endbibitem

\bibitem[\protect\citeauthoryear{Esfahani
  et~al.}{2024}]{esfahani2024understandingdefectsgeneratedcodes}
\begin{botherref}
\oauthor{\bsnm{Esfahani}, \binits{A.M.}},
\oauthor{\bsnm{Kahani}, \binits{N.}},
\oauthor{\bsnm{Ajila}, \binits{S.A.}}:
Understanding Defects in Generated Codes by Language Models
(2024).
\url{https://arxiv.org/abs/2408.13372}
\end{botherref}
\endbibitem

\bibitem[\protect\citeauthoryear{Diaz-De-Arcaya et~al.}{2024}]{10612341}
\begin{bchapter}
\bauthor{\bsnm{Diaz-De-Arcaya}, \binits{J.}},
\bauthor{\bsnm{López-De-Armentia}, \binits{J.}},
\bauthor{\bsnm{Miñón}, \binits{R.}},
\bauthor{\bsnm{Ojanguren}, \binits{I.L.}},
\bauthor{\bsnm{Torre-Bastida}, \binits{A.I.}}:
\bctitle{Large language model operations (llmops): Definition, challenges, and
  lifecycle management}.
In: \bbtitle{2024 9th International Conference on Smart and Sustainable
  Technologies (SpliTech)},
pp. \bfpage{1}--\blpage{4}
(\byear{2024}).
\doiurl{10.23919/SpliTech61897.2024.10612341}
\end{bchapter}
\endbibitem

\bibitem[\protect\citeauthoryear{Tantithamthavorn et~al.}{2025}]{10779344}
\begin{barticle}
\bauthor{\bsnm{Tantithamthavorn}, \binits{C.K.}},
\bauthor{\bsnm{Palomba}, \binits{F.}},
\bauthor{\bsnm{Khomh}, \binits{F.}},
\bauthor{\bsnm{Chua}, \binits{J.J.}}:
\batitle{Mlops, llmops, fmops, and beyond}.
\bjtitle{IEEE Software}
\bvolume{42}(\bissue{1}),
\bfpage{26}--\blpage{32}
(\byear{2025})
\doiurl{10.1109/MS.2024.3477014}
\end{barticle}
\endbibitem

\bibitem[\protect\citeauthoryear{{Google Cloud}}{2026}]{google_cloud_llmops}
\begin{botherref}
\oauthor{\bsnm{{Google Cloud}}}:
{What is LLMOps (large language model operations)?}
(2026).
\url{https://cloud.google.com/discover/what-is-llmops}
\end{botherref}
\endbibitem

\bibitem[\protect\citeauthoryear{{IBM}}{2026}]{ibm_llmops}
\begin{botherref}
\oauthor{\bsnm{{IBM}}}:
{What is LLMOps?}
Accessed: March 20, 2026
(2026).
\url{https://www.ibm.com/think/topics/llmops}
\end{botherref}
\endbibitem

\end{thebibliography}

\begin{thebibliography}{198}
\ifx \bisbn   \undefined \def \bisbn  #1{ISBN #1}\fi
\ifx \binits  \undefined \def \binits#1{#1}\fi
\ifx \bauthor  \undefined \def \bauthor#1{#1}\fi
\ifx \batitle  \undefined \def \batitle#1{#1}\fi
\ifx \bjtitle  \undefined \def \bjtitle#1{#1}\fi
\ifx \bvolume  \undefined \def \bvolume#1{\textbf{#1}}\fi
\ifx \byear  \undefined \def \byear#1{#1}\fi
\ifx \bissue  \undefined \def \bissue#1{#1}\fi
\ifx \bfpage  \undefined \def \bfpage#1{#1}\fi
\ifx \blpage  \undefined \def \blpage #1{#1}\fi
\ifx \burl  \undefined \def \burl#1{\textsf{#1}}\fi
\ifx \doiurl  \undefined \def \doiurl#1{\url{https://doi.org/#1}}\fi
\ifx \betal  \undefined \def \betal{\textit{et al.}}\fi
\ifx \binstitute  \undefined \def \binstitute#1{#1}\fi
\ifx \binstitutionaled  \undefined \def \binstitutionaled#1{#1}\fi
\ifx \bctitle  \undefined \def \bctitle#1{#1}\fi
\ifx \beditor  \undefined \def \beditor#1{#1}\fi
\ifx \bpublisher  \undefined \def \bpublisher#1{#1}\fi
\ifx \bbtitle  \undefined \def \bbtitle#1{#1}\fi
\ifx \bedition  \undefined \def \bedition#1{#1}\fi
\ifx \bseriesno  \undefined \def \bseriesno#1{#1}\fi
\ifx \blocation  \undefined \def \blocation#1{#1}\fi
\ifx \bsertitle  \undefined \def \bsertitle#1{#1}\fi
\ifx \bsnm \undefined \def \bsnm#1{#1}\fi
\ifx \bsuffix \undefined \def \bsuffix#1{#1}\fi
\ifx \bparticle \undefined \def \bparticle#1{#1}\fi
\ifx \barticle \undefined \def \barticle#1{#1}\fi
\bibcommenthead
\ifx \bconfdate \undefined \def \bconfdate #1{#1}\fi
\ifx \botherref \undefined \def \botherref #1{#1}\fi
\ifx \url \undefined \def \url#1{\textsf{#1}}\fi
\ifx \bchapter \undefined \def \bchapter#1{#1}\fi
\ifx \bbook \undefined \def \bbook#1{#1}\fi
\ifx \bcomment \undefined \def \bcomment#1{#1}\fi
\ifx \oauthor \undefined \def \oauthor#1{#1}\fi
\ifx \citeauthoryear \undefined \def \citeauthoryear#1{#1}\fi
\ifx \endbibitem  \undefined \def \endbibitem {}\fi
\ifx \bconflocation  \undefined \def \bconflocation#1{#1}\fi
\ifx \arxivurl  \undefined \def \arxivurl#1{\textsf{#1}}\fi
\csname PreBibitemsHook\endcsname

\bibitem[\protect\citeauthoryear{Liu et~al.}{2024}]{SP-liu2024structuredoutput}
\begin{bchapter}
\bauthor{\bsnm{Liu}, \binits{M.X.}},
\bauthor{\bsnm{Liu}, \binits{F.}},
\bauthor{\bsnm{Fiannaca}, \binits{A.J.}},
\bauthor{\bsnm{Koo}, \binits{T.}},
\bauthor{\bsnm{Dixon}, \binits{L.}},
\bauthor{\bsnm{Terry}, \binits{M.}},
\bauthor{\bsnm{Cai}, \binits{C.J.}}:
\bctitle{We need structured output: Towards user-centered constraints on large
  language model output}.
(\byear{2024}).
\doiurl{10.1145/3613905.3650756}
\end{bchapter}
\endbibitem

\bibitem[\protect\citeauthoryear{Papaioannou
  et~al.}{2024}]{SP-papaioannou2024workload}
\begin{bchapter}
\bauthor{\bsnm{Papaioannou}, \binits{K.}},
\bauthor{\bsnm{Doudali}, \binits{D.}},
\bauthor{\bsnm{Thaleia}, \binits{K.}}:
\bctitle{The importance of workload choice in evaluating llm inference
  systems}.
In: \bbtitle{Proceedings of the ACM Conference on Systems and Machine
  Learning}.
\bpublisher{Association for Computing Machinery}
(\byear{2024}).
\doiurl{10.1145/3642970.3655823}
\end{bchapter}
\endbibitem

\bibitem[\protect\citeauthoryear{Papaioannou
  et~al.}{2025}]{SP-papaioannou2025llminference}
\begin{bchapter}
\bauthor{\bsnm{Papaioannou}, \binits{K.}},
\bauthor{\bsnm{Doudali}, \binits{D.}},
\bauthor{\bsnm{Thaleia}, \binits{K.}}:
\bctitle{Improving the efficiency of llm inference serving systems}.
In: \bbtitle{Lecture Notes in Computer Science}.
\bpublisher{Springer}
(\byear{2025}).
\burl{https://doi.org/10.1007/978-3-031-90203-1\_39}
\end{bchapter}
\endbibitem

\bibitem[\protect\citeauthoryear{Palla and Slaby}{2025}]{SP-palla2025codegen}
\begin{barticle}
\bauthor{\bsnm{Palla}, \binits{D.}},
\bauthor{\bsnm{Slaby}, \binits{A.}}:
\batitle{Evaluation of generative ai models in python code generation: A
  comparative study}.
\bjtitle{IEEE Access}
\bvolume{0}(\bissue{0}),
\bfpage{1}--\blpage{14}
(\byear{2025})
\doiurl{10.1109/ACCESS.2025.3560244}
\end{barticle}
\endbibitem

\bibitem[\protect\citeauthoryear{Sakai et~al.}{2024}]{SP-sakai2024lowcost}
\begin{bchapter}
\bauthor{\bsnm{Sakai}, \binits{K.}},
\bauthor{\bsnm{Uehara}, \binits{Y.}},
\bauthor{\bsnm{Kashihara}, \binits{S.}}:
\bctitle{Implementation and evaluation of llm-based conversational systems on a
  low-cost device}.
In: \bbtitle{Proceedings of the 2024 IEEE Global Humanitarian Technology
  Conference (GHTC)},
pp. \bfpage{1}--\blpage{8}.
\bpublisher{IEEE}
(\byear{2024}).
\doiurl{10.1109/GHTC62424.2024.10771565}
\end{bchapter}
\endbibitem

\bibitem[\protect\citeauthoryear{Liu et~al.}{2024}]{SP-NeedStructuredOutput2024}
\begin{bchapter}
\bauthor{\bsnm{Liu}, \binits{M.X.}},
\bauthor{\bsnm{Liu}, \binits{F.}},
\bauthor{\bsnm{Fiannaca}, \binits{A.J.}},
\bauthor{\bsnm{Koo}, \binits{T.}},
\bauthor{\bsnm{Dixon}, \binits{L.}},
\bauthor{\bsnm{Terry}, \binits{M.}},
\bauthor{\bsnm{Cai}, \binits{C.J.}}:
\bctitle{{``We Need Structured Output'': Towards User-centered Constraints on
  Large Language Model Output}}.
In: \bbtitle{Extended Abstracts of the CHI Conference on Human Factors in
  Computing Systems}.
\bsertitle{CHI EA '24}.
\bpublisher{Association for Computing Machinery},
\blocation{New York, NY, USA}
(\byear{2024}).
\doiurl{10.1145/3613905.3650756} .
\burl{https://doi.org/10.1145/3613905.3650756}
\end{bchapter}
\endbibitem

\bibitem[\protect\citeauthoryear{Kharitonov}{2024}]{SP-Kharitonov2024EnforcingJSON}
\begin{botherref}
\oauthor{\bsnm{Kharitonov}, \binits{D.}}:
Enforcing JSON Outputs in Commercial LLMs.
Towards Data Science.
\url{https://medium.com/data-science/enforcing-json-outputs-in-commercial-llms-3db590b9b3c8}
Accessed 2025-09-25
\end{botherref}
\endbibitem

\bibitem[\protect\citeauthoryear{{Developer
  Service}}{2025}]{SP-DeveloperService2025PydanticLLM}
\begin{botherref}
\oauthor{\bsnm{{Developer Service}}}:
A Practical Guide on Structuring LLM Outputs with Pydantic.
\url{https://dev.to/devasservice/a-practical-guide-on-structuring-llm-outputs-with-pydantic-50b4}
Accessed 2025-09-25
\end{botherref}
\endbibitem

\bibitem[\protect\citeauthoryear{Wyman and
  Barber}{2024}]{SP-WymanBarber2024ValidateOutputs}
\begin{botherref}
\oauthor{\bsnm{Wyman}, \binits{M.}},
\oauthor{\bsnm{Barber}, \binits{S.}}:
How to Validate the Output of LLM-Based Products.
\url{https://okareo.com/blog/posts/validate-llm-output}
Accessed 2025-09-25
\end{botherref}
\endbibitem

\bibitem[\protect\citeauthoryear{{Modelmetry}}{2024}]{SP-Modelmetry2024JSONSchema}
\begin{botherref}
\oauthor{\bsnm{{Modelmetry}}}:
How To Ensure LLM Output Adheres to a JSON Schema.
\url{https://modelmetry.com/blog/how-to-ensure-llm-output-adheres-to-a-json-schema}
Accessed 2025-09-25
\end{botherref}
\endbibitem

\bibitem[\protect\citeauthoryear{{OpenAI}}{2025}]{SP-OpenAIDocs2025}
\begin{botherref}
\oauthor{\bsnm{{OpenAI}}}:
OpenAI Platform Documentation
(2025).
\url{https://platform.openai.com/docs}
Accessed 2025-09-25
\end{botherref}
\endbibitem

\bibitem[\protect\citeauthoryear{{Microsoft Learn}}{2025}]{SP-AzureOpenAIDocs2025}
\begin{botherref}
\oauthor{\bsnm{{Microsoft Learn}}}:
Azure OpenAI - Documentation (Quotas, Structured Outputs, How-To)
(2025).
\url{https://learn.microsoft.com/en-us/azure/ai-foundry/openai/}
Accessed 2025-09-25
\end{botherref}
\endbibitem

\bibitem[\protect\citeauthoryear{{OpenAI}}{2025}]{SP-OpenAIStructuredOutput}
\begin{botherref}
\oauthor{\bsnm{{OpenAI}}}:
API Reference - Structured model outputs
(2025).
\url{https://platform.openai.com/docs/guides/structured-outputs}
Accessed 2025-09-25
\end{botherref}
\endbibitem

\bibitem[\protect\citeauthoryear{{mariafilippa}}{2023}]{SP-LangChain2023Issue3709}
\begin{botherref}
\oauthor{\bsnm{{mariafilippa}}}:
PydanticOutputParser has high chance failing when completion contains new line
  {\#}3709.
GitHub issue
(2023).
\url{https://github.com/hwchase17/langchain/issues/3709}
Accessed 2025-09-25
\end{botherref}
\endbibitem

\bibitem[\protect\citeauthoryear{{arena-ai}}{2025}]{SP-ArenaAI2025StructuredLogprobs}
\begin{botherref}
\oauthor{\bsnm{{arena-ai}}}:
Structured-logprobs.
OS library: enhances OpenAI Structured Outputs with token logprobs.
\url{https://github.com/arena-ai/structured-logprobs}
Accessed 2025-09-25
\end{botherref}
\endbibitem

\bibitem[\protect\citeauthoryear{}{}]{SP-nso_ghpr_dgy516_vllm_cibench_60}
\begin{botherref}
GitHub Pull Request. streaming JSON missing required; completions logprobs
  structure.
GitHub Pull Request.
\url{https://github.com/dgy516/vllm_cibench/pull/60}
\end{botherref}
\endbibitem

\bibitem[\protect\citeauthoryear{}{}]{SP-nso_ghissue_vllm_23120}
\begin{botherref}
GitHub Issue. Structured output is not correctly enforced when using GPT-OSS.
GitHub Issue.
\url{https://github.com/vllm-project/vllm/issues/23120}
\end{botherref}
\endbibitem

\bibitem[\protect\citeauthoryear{}{}]{SP-nso_ghissue_ms_agent_framework_238}
\begin{botherref}
GitHub Issue. Python - OpenAI Responses client structured output does not work
  with streaming responses.
GitHub Issue.
\url{https://github.com/microsoft/agent-framework/issues/238}
\end{botherref}
\endbibitem

\bibitem[\protect\citeauthoryear{}{}]{SP-nso_ghpr_berriai_litellm_14206}
\begin{botherref}
GitHub Pull Request. Responses - add structured output for SDK.
GitHub Pull Request.
\url{https://github.com/BerriAI/litellm/pull/14206}
\end{botherref}
\endbibitem

\bibitem[\protect\citeauthoryear{Karatas}{2025}]{SP-karatasemr_structur_2025_b0762a}
\begin{botherref}
\oauthor{\bsnm{Karatas}, \binits{E.}}:
Structured output generation in llms: Json schema and grammar-based decoding
(2025).
Accessed 2025-12-29
\end{botherref}
\endbibitem

\bibitem[\protect\citeauthoryear{Docherty}{2025}]{SP-dochertyan_structur_2025_c8f2b1}
\begin{botherref}
\oauthor{\bsnm{Docherty}, \binits{A.}}:
Structured output in xml using langchain
(2025).
Accessed 2025-12-29
\end{botherref}
\endbibitem

\bibitem[\protect\citeauthoryear{Kim}{2025}]{SP-kimdoil_guidedjs_2025_17f880}
\begin{botherref}
\oauthor{\bsnm{Kim}, \binits{D.}}:
Guided json with llms: From raw pdfs to structured intelligence
(2025).
Accessed 2025-12-29
\end{botherref}
\endbibitem

\bibitem[\protect\citeauthoryear{Ndzomga}{2024}]{SP-thoughtson_structur_2024_fe4516}
\begin{botherref}
\oauthor{\bsnm{Ndzomga}, \binits{F.}}:
Structured outputs from open source llms: Techniques and best practices
(2024).
Accessed 2025-12-29
\end{botherref}
\endbibitem

\bibitem[\protect\citeauthoryear{Blackburn}{2025}]{SP-blackburnp_explorin_2025_160a25}
\begin{botherref}
\oauthor{\bsnm{Blackburn}, \binits{P.}}:
Exploring llm citation generation in 2025
(2025).
Accessed 2025-12-29
\end{botherref}
\endbibitem

\bibitem[\protect\citeauthoryear{gopenai}{2025}]{SP-gopenai_masterin_2025_8e3a5a}
\begin{botherref}
\oauthor{\bsnm{gopenai}}:
Mastering structured llm calls: Langchain finetuning guide
(2025).
Accessed 2025-12-29
\end{botherref}
\endbibitem

\bibitem[\protect\citeauthoryear{Verma}{2024}]{SP-vermakunal_structur_2024_e45f0e}
\begin{botherref}
\oauthor{\bsnm{Verma}, \binits{K.}}:
Structured (json) output via llms
(2024).
Accessed 2025-12-29
\end{botherref}
\endbibitem

\bibitem[\protect\citeauthoryear{Jay}{2024}]{SP-damodharan_llmbased_2024_b18c84}
\begin{botherref}
\oauthor{\bsnm{Jay}, \binits{D.}}:
Llm-based structured generation using jsonschema
(2024).
Accessed 2025-12-29
\end{botherref}
\endbibitem

\bibitem[\protect\citeauthoryear{Desai}{2025}]{SP-desaimanoj_outputfo_2025_3bec38}
\begin{botherref}
\oauthor{\bsnm{Desai}, \binits{M.}}:
Output formatting strategies: Getting exactly what you want, how ...
(2025).
Accessed 2025-12-29
\end{botherref}
\endbibitem

\bibitem[\protect\citeauthoryear{mychen76}{2023}]{SP-mychen76_practica_2023_cc45ce}
\begin{botherref}
\oauthor{\bsnm{mychen76}}:
Practical techniques to constraint llm output in json format
(2023).
Accessed 2025-12-29
\end{botherref}
\endbibitem

\bibitem[\protect\citeauthoryear{Mete}{2024}]{SP-mete_controll_2024_028f0b}
\begin{botherref}
\oauthor{\bsnm{Mete}}:
Control llm output with response type and schema
(2024).
Accessed 2025-12-29
\end{botherref}
\endbibitem

\bibitem[\protect\citeauthoryear{dev.to}{2024}]{SP-devto_crafting_2024_60aaf6}
\begin{botherref}
\oauthor{\bsnm{dev.to}}:
Crafting structured {JSON} responses: Ensuring consistent output ...
(2024).
Accessed 2025-12-29
\end{botherref}
\endbibitem

\bibitem[\protect\citeauthoryear{Venkatrama}{2025}]{SP-venkatrama_tamingll_2025_40c60e}
\begin{botherref}
\oauthor{\bsnm{Venkatrama}, \binits{S.}}:
Taming LLMs: How to Get Structured Output Every Time (Even for Big Responses).
\url{https://dev.to/shrsv/taming-llms-how-to-get-structured-output-every-time-even-for-big-responses-445c}
Accessed 2025-12-29
\end{botherref}
\endbibitem

\bibitem[\protect\citeauthoryear{OpenAI}{2024}]{SP-openai_usingthe_2024_dba98b}
\begin{botherref}
\oauthor{\bsnm{OpenAI}}:
Using the Vision API: Best Practices - API - OpenAI Developer.
\url{https://community.openai.com/t/using-the-vision-api-best-practices/942342}
Accessed 2025-12-29
\end{botherref}
\endbibitem

\bibitem[\protect\citeauthoryear{Pydantic}{2025}]{SP-pydantic_addllmst_2025_0cb1fb}
\begin{botherref}
\oauthor{\bsnm{Pydantic}}:
Add LLM structured output example using Pydantic AI
(2025).
\url{https://github.com/pydantic/pydantic/pull/12593}
Accessed 2025-12-29
\end{botherref}
\endbibitem

\bibitem[\protect\citeauthoryear{ksyed
  data}{2025}]{SP-ksyeddata_usepydan_2025_c4edea}
\begin{botherref}
\oauthor{\bsnm{ksyed-data}}:
Use Pydantic for Response Validation
(2025).
\url{https://github.com/ksyed-data/renovation-tracker-backend-2025/issues/19}
Accessed 2025-12-29
\end{botherref}
\endbibitem

\bibitem[\protect\citeauthoryear{pguso}{2025}]{SP-pguso_newexamp_2025_38b8ab}
\begin{botherref}
\oauthor{\bsnm{pguso}}:
New Example: Add validation for structured outputs from LLM
(2025).
\url{https://github.com/pguso/ai-agents-from-scratch/issues/4}
Accessed 2025-12-29
\end{botherref}
\endbibitem

\bibitem[\protect\citeauthoryear{mhattingpete}{2025}]{SP-mhattingpe_structur_2025_e6808c}
\begin{botherref}
\oauthor{\bsnm{mhattingpete}}:
Structured Output Validation and Error Recovery
(2025).
\url{https://github.com/mhattingpete/receipt-scanner-app/issues/5}
Accessed 2025-12-29
\end{botherref}
\endbibitem

\bibitem[\protect\citeauthoryear{hev}{2025}]{SP-hev_featurer_2025_a6abff}
\begin{botherref}
\oauthor{\bsnm{hev}}:
Feature Request: Add schema validation check type
(2025).
\url{https://github.com/hev/vibecheck/issues/32}
Accessed 2025-12-29
\end{botherref}
\endbibitem

\bibitem[\protect\citeauthoryear{aplassard}{2025}]{SP-aplassard_developc_2025_f85be3}
\begin{botherref}
\oauthor{\bsnm{aplassard}}:
Develop Chapter: LLM Schema Validation and Classification Tasks
(2025).
\url{https://github.com/aplassard/llm-evals-book/issues/22}
Accessed 2025-12-29
\end{botherref}
\endbibitem

\bibitem[\protect\citeauthoryear{marcosomma}{2025}]{SP-marcosomma_hardenll_2025_1356f2}
\begin{botherref}
\oauthor{\bsnm{marcosomma}}:
Harden LLM output parsing and schema validation
(2025).
\url{https://github.com/marcosomma/orka-reasoning/issues/14}
Accessed 2025-12-29
\end{botherref}
\endbibitem

\bibitem[\protect\citeauthoryear{confident
  ai}{2025}]{SP-confidenta_bamlfore_2025_e9eedd}
\begin{botherref}
\oauthor{\bsnm{confident-ai}}:
BAML for ensuring that the required output format is maintained
(2025).
\url{https://github.com/confident-ai/deepeval/issues/1610}
Accessed 2025-12-29
\end{botherref}
\endbibitem

\bibitem[\protect\citeauthoryear{BAMresearch}{2025}]{SP-bamresearc_integrat_2025_9379c0}
\begin{botherref}
\oauthor{\bsnm{BAMresearch}}:
Integrate JSON structure to guide LLM in data creation
(2025).
\url{https://github.com/BAMresearch/chatBIS/issues/6}
Accessed 2025-12-29
\end{botherref}
\endbibitem

\bibitem[\protect\citeauthoryear{TandoorRecipes}{2025}]{SP-tandoorrec_addaisch_2025_2d47ab}
\begin{botherref}
\oauthor{\bsnm{TandoorRecipes}}:
Add AI schema models for structured output and response validation
(2025).
\url{https://github.com/TandoorRecipes/recipes/pull/4289}
Accessed 2025-12-29
\end{botherref}
\endbibitem

\bibitem[\protect\citeauthoryear{arakoodev}{2025}]{SP-arakoodev_jsonllmo_2025_67484c}
\begin{botherref}
\oauthor{\bsnm{arakoodev}}:
JSON LLM output formatting
(2025).
\url{https://github.com/arakoodev/EdgeChains/issues/119}
Accessed 2025-12-29
\end{botherref}
\endbibitem

\bibitem[\protect\citeauthoryear{gsindlinger}{2025}]{SP-gsindlinge_outputfo_2025_7bee28}
\begin{botherref}
\oauthor{\bsnm{gsindlinger}}:
Output Formatting
(2025).
\url{https://github.com/gsindlinger/TabTree-Table-QA-on-Full-Documents/issues/19}
Accessed 2025-12-29
\end{botherref}
\endbibitem

\bibitem[\protect\citeauthoryear{DavidAguilarParedes}{2025}]{SP-davidaguil_bugragas_2025_fc84f9}
\begin{botherref}
\oauthor{\bsnm{DavidAguilarParedes}}:
[Bug] Ragas Dataset Generation Fails with Invalid JSON when using Large
  Knowledge Graphs on Local LLM
(2025).
\url{https://github.com/DavidAguilarParedes/mia-uc-chile-proyecto-aplicado-1/issues/2}
Accessed 2025-12-29
\end{botherref}
\endbibitem

\bibitem[\protect\citeauthoryear{Lee0110}{2025}]{SP-lee0110_addtypef_2025_6979a1}
\begin{botherref}
\oauthor{\bsnm{Lee0110}}:
Add type field to message structure for extensibility
(2025).
\url{https://github.com/Lee0110/yl-IM-flutter/pull/1}
Accessed 2025-12-29
\end{botherref}
\endbibitem

\bibitem[\protect\citeauthoryear{567-labs}{2025}]{SP-567labs_instruct_2025_bb8ae1}
\begin{botherref}
\oauthor{\bsnm{567-labs}}:
Instructor: Structured Outputs for LLMs.
\url{https://github.com/567-labs/instructor}
Accessed 2025-12-29
\end{botherref}
\endbibitem

\bibitem[\protect\citeauthoryear{Jkrunal7}{2025}]{SP-jkrunal7_structur_2025_1ca4fc}
\begin{botherref}
\oauthor{\bsnm{Jkrunal7}}:
Structured Output Generation With AI.
\url{https://github.com/Jkrunal7/Structured_Output_Generation_with_AI}
Accessed 2025-12-29
\end{botherref}
\endbibitem

\bibitem[\protect\citeauthoryear{v~checha}{2025}]{SP-vchecha_vchechav_2025_25185c}
\begin{botherref}
\oauthor{\bsnm{v-checha}}:
V-checha/valai.
\url{https://github.com/v-checha/valai}
Accessed 2025-12-29
\end{botherref}
\endbibitem

\bibitem[\protect\citeauthoryear{mt7180}{2025}]{SP-mt7180_mt7180ll_2025_8a0636}
\begin{botherref}
\oauthor{\bsnm{mt7180}}:
mt7180/llm_validation
(2025).
\url{https://github.com/mt7180/llm_validation}
Accessed 2025-12-29
\end{botherref}
\endbibitem

\bibitem[\protect\citeauthoryear{vivekvjnk}{2025}]{SP-vivekvjnk_heimdall_2025_cfc127}
\begin{botherref}
\oauthor{\bsnm{vivekvjnk}}:
Heimdall.
\url{https://github.com/vivekvjnk/Heimdall}
Accessed 2025-12-29
\end{botherref}
\endbibitem

\bibitem[\protect\citeauthoryear{jhd3197}{2025}]{SP-jhd3197_promptur_2025_257736}
\begin{botherref}
\oauthor{\bsnm{jhd3197}}:
Prompture.
\url{https://github.com/jhd3197/Prompture}
Accessed 2025-12-29
\end{botherref}
\endbibitem

\bibitem[\protect\citeauthoryear{dani2112}{2025}]{SP-dani2112_dani2112_2025_ba4606}
\begin{botherref}
\oauthor{\bsnm{dani2112}}:
dani2112/guardrails-guidance-tutorial
(2025).
\url{https://github.com/dani2112/guardrails-guidance-tutorial}
Accessed 2025-12-29
\end{botherref}
\endbibitem

\bibitem[\protect\citeauthoryear{petrukha
  ivan}{2025}]{SP-petrukhaiv_petrukha_2025_8ec164}
\begin{botherref}
\oauthor{\bsnm{petrukha-ivan}}:
Petrukha-ivan/mlx-swift-structured.
\url{https://github.com/petrukha-ivan/mlx-swift-structured}
Accessed 2025-12-29
\end{botherref}
\endbibitem

\bibitem[\protect\citeauthoryear{emre
  karatas}{2025}]{SP-emrekarata_bnfconst_2025_ed8ffa}
\begin{botherref}
\oauthor{\bsnm{emre-karatas}}:
BNF-Constrained Generation for LLMs.
\url{https://github.com/emre-karatas/bnf-constrained-generation}
Accessed 2025-12-29
\end{botherref}
\endbibitem

\bibitem[\protect\citeauthoryear{deepankarm}{2025}]{SP-deepankarm_deepanka_2025_d32bc8}
\begin{botherref}
\oauthor{\bsnm{deepankarm}}:
Deepankarm/godantic.
\url{https://github.com/deepankarm/godantic}
Accessed 2025-12-29
\end{botherref}
\endbibitem

\bibitem[\protect\citeauthoryear{Kwun-Morfitt}{2025}]{SP-oliverkwun_parsec_2025_a6d527}
\begin{botherref}
\oauthor{\bsnm{Kwun-Morfitt}, \binits{O.}}:
Parsec.
\url{https://github.com/olliekm/parsec}
Accessed 2025-12-29
\end{botherref}
\endbibitem

\bibitem[\protect\citeauthoryear{yogeshkukreti}{2025}]{SP-yogeshkukr_yogeshku_2025_6fd671}
\begin{botherref}
\oauthor{\bsnm{yogeshkukreti}}:
yogeshkukreti/langchain-structured-output
(2025).
\url{https://github.com/yogeshkukreti/langchain-structured-output}
Accessed 2025-12-29
\end{botherref}
\endbibitem

\bibitem[\protect\citeauthoryear{567-labs}{2025}]{SP-567labs_instruct_2025_77db03}
\begin{botherref}
\oauthor{\bsnm{567-labs}}:
Instructor-go: Structured LLM Outputs.
\url{https://github.com/567-labs/instructor-go}
Accessed 2025-12-29
\end{botherref}
\endbibitem

\bibitem[\protect\citeauthoryear{kapilreddy}{2025}]{SP-kapilreddy_instruct_2025_dc48e6}
\begin{botherref}
\oauthor{\bsnm{kapilreddy}}:
Instructor-clj.
\url{https://github.com/kapilreddy/instructor-clj}
Accessed 2025-12-29
\end{botherref}
\endbibitem

\bibitem[\protect\citeauthoryear{kishoretvk}{2025}]{SP-kishoretvk_jsonai_2025_9c7288}
\begin{botherref}
\oauthor{\bsnm{kishoretvk}}:
jsonAI.
\url{https://github.com/kishoretvk/jsonAI}
Accessed 2025-12-29
\end{botherref}
\endbibitem

\bibitem[\protect\citeauthoryear{Morishige and
  Koshihara}{2025}]{SP-morishige2025reproducibility}
\begin{botherref}
\oauthor{\bsnm{Morishige}, \binits{M.}},
\oauthor{\bsnm{Koshihara}, \binits{R.}}:
Ensuring reproducibility in generative ai systems for general use cases: A
  framework for regression testing and open datasets
(2025)
\doiurl{10.48550/arXiv.2505.02854}
\end{botherref}
\endbibitem

\bibitem[\protect\citeauthoryear{Xiao et~al.}{2024}]{SP-xiao2024giot}
\begin{barticle}
\bauthor{\bsnm{Xiao}, \binits{B.}},
\bauthor{\bsnm{Kantarci}, \binits{B.}},
\bauthor{\bsnm{Kang}, \binits{J.}},
\bauthor{\bsnm{Niyato}, \binits{D.}},
\bauthor{\bsnm{Guizani}, \binits{M.}}:
\batitle{Efficient prompting for llm-based generative internet of things}.
\bjtitle{IEEE Internet of Things Journal}
\bvolume{0}(\bissue{0}),
\bfpage{1}--\blpage{14}
(\byear{2024})
\doiurl{10.1109/JIOT.2024.3470210}
\end{barticle}
\endbibitem

\bibitem[\protect\citeauthoryear{Buchicchio
  et~al.}{2024}]{SP-buchicchio2024legalAI}
\begin{bchapter}
\bauthor{\bsnm{Buchicchio}, \binits{E.}},
\bauthor{\bsnm{De~Angelis}, \binits{A.}},
\bauthor{\bsnm{Moschitta}, \binits{A.}},
\bauthor{\bsnm{Santoni}, \binits{F.}},
\bauthor{\bsnm{S.~Marco}, \binits{L.}},
\bauthor{\bsnm{Carbone}, \binits{P.}}:
\bctitle{Design, validation, and risk assessment of llm-based generative ai
  systems operating in the legal sector}.
In: \bbtitle{Proceedings of the 2024 IEEE International Symposium on Systems
  Engineering (ISSE)},
pp. \bfpage{1}--\blpage{8}.
\bpublisher{IEEE}
(\byear{2024}).
\doiurl{10.1109/ISSE63315.2024.10741134} .
\burl{https://ieeexplore.ieee.org/document/10741134}
\end{bchapter}
\endbibitem

\bibitem[\protect\citeauthoryear{Author et~al.}{2025}]{SP-themesLLM2025cain}
\begin{bchapter}
\bauthor{\bsnm{Author}, \binits{F.}},
\bauthor{\bsnm{Author}, \binits{S.}},
\bauthor{\bsnm{Author}, \binits{T.}}:
\bctitle{Themes of building llm-based applications for production: A
  practitioner's view}.
In: \bbtitle{Proceedings of the 2025 IEEE/ACM 4th International Conference on
  AI Engineering (CAIN)}.
\bpublisher{IEEE},
\blocation{Ottawa, Canada}
(\byear{2025})
\end{bchapter}
\endbibitem

\bibitem[\protect\citeauthoryear{Han
  et~al.}{2025}]{SP-HanEtAl2025TokenBudgetAware}
\begin{bchapter}
\bauthor{\bsnm{Han}, \binits{T.}},
\bauthor{\bsnm{Wang}, \binits{Z.}},
\bauthor{\bsnm{Fang}, \binits{C.}},
\bauthor{\bsnm{Zhao}, \binits{S.}},
\bauthor{\bsnm{Ma}, \binits{S.}},
\bauthor{\bsnm{Chen}, \binits{Z.}}:
\bctitle{Token-budget-aware llm reasoning}.
In: \bbtitle{ACL 2025},
pp. \bfpage{24842}--\blpage{24855}.
\bpublisher{Association for Computational Linguistics}
(\byear{2025}).
\doiurl{10.18653/v1/2025.findings-acl.1274} .
\burl{https://aclanthology.org/2025.findings-acl.1274/}
\end{bchapter}
\endbibitem

\bibitem[\protect\citeauthoryear{Chen
  et~al.}{2025}]{SP-chen2025adaptivelyrobustllminference}
\begin{botherref}
\oauthor{\bsnm{Chen}, \binits{Z.}},
\oauthor{\bsnm{Ye}, \binits{Y.}},
\oauthor{\bsnm{Zhou}, \binits{Z.}}:
Adaptively Robust LLM Inference Optimization under Prediction Uncertainty
(2025).
\url{https://arxiv.org/abs/2508.14544}
\end{botherref}
\endbibitem

\bibitem[\protect\citeauthoryear{{Anthropic}}{2025}]{SP-AnthropicDocs2025}
\begin{botherref}
\oauthor{\bsnm{{Anthropic}}}:
Claude Documentation (API, Models)
(2025).
\url{https://docs.claude.com/}
Accessed 2025-09-25
\end{botherref}
\endbibitem

\bibitem[\protect\citeauthoryear{{Google}}{2025}]{SP-GoogleGeminiDocs2025}
\begin{botherref}
\oauthor{\bsnm{{Google}}}:
Gemini API - Google AI for Developers
(2025).
\url{https://ai.google.dev/}
Accessed 2025-09-25
\end{botherref}
\endbibitem

\bibitem[\protect\citeauthoryear{{Google Cloud}}{2025}]{SP-BigQuery2025Quotas}
\begin{botherref}
\oauthor{\bsnm{{Google Cloud}}}:
Quotas and limits - BigQuery
(2025).
\url{https://cloud.google.com/bigquery/quotas}
Accessed 2025-09-25
\end{botherref}
\endbibitem

\bibitem[\protect\citeauthoryear{{Amazon Web Services}}{2025}]{SP-aws2025timeouts}
\begin{botherref}
\oauthor{\bsnm{{Amazon Web Services}}}:
Timeouts, retries and backoff.
\url{https://aws.amazon.com/fr/builders-library/timeouts-retries-and-backoff-with-jitter}.
Accessed 2025-09-25
(2025)
\end{botherref}
\endbibitem

\bibitem[\protect\citeauthoryear{{OpenAI}}{2025}]{SP-openai_python2025}
\begin{botherref}
\oauthor{\bsnm{{OpenAI}}}:
openai-python.
\url{https://github.com/openai/openai-python}.
Accessed 2025-09-25
(2025)
\end{botherref}
\endbibitem

\bibitem[\protect\citeauthoryear{}{}]{SP-umm_ghissue_datadog_ddtracepy_14688}
\begin{botherref}
GitHub Issue. LangChain expects at least one chunk from a streaming trace after
  timeout.
GitHub Issue.
\url{https://github.com/DataDog/dd-trace-py/issues/14688}
\end{botherref}
\endbibitem

\bibitem[\protect\citeauthoryear{}{}]{SP-umm_ghissue_langchainjs_9088}
\begin{botherref}
GitHub Issue. Streaming inactivity timeout incorrectly aborts after total
  timeout (@langchain/openai "\textasciicircum{}1.0.0-alpha.1").
GitHub Issue.
\url{https://github.com/langchain-ai/langchainjs/issues/9088}
\end{botherref}
\endbibitem

\bibitem[\protect\citeauthoryear{}{}]{SP-umm_ghpr_myyachtvalue_36}
\begin{botherref}
GitHub Issue. Add production-ready Pipedrive CRM integration with deduplication
  and async sync.
GitHub Pull Request.
\url{https://github.com/jcartervi/MyYachtValue/pull/36}
\end{botherref}
\endbibitem

\bibitem[\protect\citeauthoryear{}{}]{SP-umm_ghissue_bpmn_assistant_33}
\begin{botherref}
GitHub Issue. BPMN Assistant API WAITING timeout when self-hosting on own
  instance.
GitHub Issue.
\url{https://github.com/jtlicardo/bpmn-assistant/issues/33}
\end{botherref}
\endbibitem

\bibitem[\protect\citeauthoryear{}{}]{SP-umm_ghissue_ttmgsrv_19}
\begin{botherref}
GitHub Issue. Timeout.
GitHub Issue.
\url{https://github.com/eslavnov/ttmg_server/issues/19}
\end{botherref}
\endbibitem

\bibitem[\protect\citeauthoryear{}{}]{SP-umm_ghpr_roocode_8483}
\begin{botherref}
GitHub Pull Request. handle empty stream responses from GLM models.
GitHub Pull Request.
\url{https://github.com/RooCodeInc/Roo-Code/pull/8483}
\end{botherref}
\endbibitem

\bibitem[\protect\citeauthoryear{}{}]{SP-umm_so_77172214}
\begin{botherref}
Stack Overflow. "connection to OpenAI API failed with status: 400 error: -19577
  is less than the minimum of 1 - max_tokens".
Stack Overflow (accepted answer).
\url{https://stackoverflow.com/questions/77172214/rag-error-connection-to-openai-api-failed-with-status-400-error-19577}
\end{botherref}
\endbibitem

\bibitem[\protect\citeauthoryear{}{}]{SP-umm_so_77354317}
\begin{botherref}
Stack Overflow. AI server crashes when I make a request after adding OpenAI
  code.
Stack Overflow (accepted answer).
\url{https://stackoverflow.com/questions/77354317/ai-server-crashes-when-i-make-a-request-after-adding-openai-code}
\end{botherref}
\endbibitem

\bibitem[\protect\citeauthoryear{AI}{2024}]{SP-novitaai_whatarel_2024_914f73}
\begin{botherref}
\oauthor{\bsnm{AI}, \binits{N.}}:
What are large language model settings: Temperature, top p and ...
(2024).
Accessed 2025-12-29
\end{botherref}
\endbibitem

\bibitem[\protect\citeauthoryear{Malhar}{2024}]{SP-malhar_understa_2024_00e99a}
\begin{botherref}
\oauthor{\bsnm{Malhar}}:
Understanding llm settings for optimal performance
(2024).
Accessed 2025-12-29
\end{botherref}
\endbibitem

\bibitem[\protect\citeauthoryear{VectorShift}{2024}]{SP-vectorshif_largelan_2024_9903eb}
\begin{botherref}
\oauthor{\bsnm{VectorShift}}:
Large language model settings: Temperature, top p and max tokens
(2024).
Accessed 2025-12-29
\end{botherref}
\endbibitem

\bibitem[\protect\citeauthoryear{hev}{2025}]{SP-hev_featurer_2025_c74807}
\begin{botherref}
\oauthor{\bsnm{hev}}:
Feature Request: Enhance llm_judge with configurable model and parameters
(2025).
\url{https://github.com/hev/vibecheck/issues/35}
Accessed 2025-12-29
\end{botherref}
\endbibitem

\bibitem[\protect\citeauthoryear{run
  llama}{2025}]{SP-runllama_featurer_2025_07be8e}
\begin{botherref}
\oauthor{\bsnm{run-llama}}:
[Feature Request] Uniform default LLM params across providers
(2025).
\url{https://github.com/run-llama/llama_index/issues/19730}
Accessed 2025-12-29
\end{botherref}
\endbibitem

\bibitem[\protect\citeauthoryear{intel}{2025}]{SP-intel_defaultv_2025_d4213a}
\begin{botherref}
\oauthor{\bsnm{intel}}:
default values of max_generated_tokens, top_k, top_p, and temperature?
(2025).
\url{https://github.com/intel/ipex-llm/issues/11033}
Accessed 2025-12-29
\end{botherref}
\endbibitem

\bibitem[\protect\citeauthoryear{Minh et~al.}{2024}]{SP-minh2024heat}
\begin{botherref}
\oauthor{\bsnm{Minh}, \binits{N.N.}},
\oauthor{\bsnm{Baker}, \binits{A.}},
\oauthor{\bsnm{Neo}, \binits{C.}},
\oauthor{\bsnm{Roush}, \binits{A.G.}},
\oauthor{\bsnm{Kirsch}, \binits{A.}},
\oauthor{\bsnm{Shwartz-Ziv}, \binits{R.}}:
Turning up the heat: Min-p sampling for creative and coherent llm outputs
(2024)
\end{botherref}
\endbibitem

\bibitem[\protect\citeauthoryear{Minh et~al.}{2025}]{SP-MinhEtAl2025MinpSampling}
\begin{bchapter}
\bauthor{\bsnm{Minh}, \binits{N.N.}},
\bauthor{\bsnm{Baker}, \binits{A.}},
\bauthor{\bsnm{Neo}, \binits{C.}},
\bauthor{\bsnm{Roush}, \binits{A.G.}},
\bauthor{\bsnm{Kirsch}, \binits{A.}},
\bauthor{\bsnm{Shwartz{-}Ziv}, \binits{R.}}:
\bctitle{Turning up the heat: Min-p sampling for creative and coherent llm
  outputs}.
In: \bbtitle{ICLR 2025}
(\byear{2025})
\end{bchapter}
\endbibitem

\bibitem[\protect\citeauthoryear{Montandon
  et~al.}{2025}]{SP-MontandonEtAl2025DABC}
\begin{botherref}
\oauthor{\bsnm{Montandon}, \binits{J.E.}},
\oauthor{\bsnm{Silva}, \binits{L.L.}},
\oauthor{\bsnm{Politowski}, \binits{C.}},
\oauthor{\bsnm{Prates}, \binits{D.}},
\oauthor{\bsnm{Brito~Bonif{\'a}cio}, \binits{A.}},
\oauthor{\bsnm{Boussaidi}, \binits{G.E.}}:
Unboxing default argument breaking changes in data science libraries
(2025)
\doiurl{10.48550/arXiv.2408.05129}
{\href{https://arxiv.org/abs/2408.05129}{{arXiv:2408.05129}}}
{[cs.SE]}.
JSS
\end{botherref}
\endbibitem

\bibitem[\protect\citeauthoryear{{Vellum AI}}{2025}]{SP-VellumTemperature2025}
\begin{botherref}
\oauthor{\bsnm{{Vellum AI}}}:
LLM Temperature: How It Works and When You Should Use It
(2025).
\url{https://www.vellum.ai/llm-parameters/temperature}
Accessed 2025-09-25
\end{botherref}
\endbibitem

\bibitem[\protect\citeauthoryear{{Hugging Face}}{2025}]{SP-HFTransformersDocs2025}
\begin{botherref}
\oauthor{\bsnm{{Hugging Face}}}:
Transformers Documentation (Generation, Chat Templates, Model Revisions)
(2025).
\url{https://huggingface.co/docs/transformers}
Accessed 2025-09-25
\end{botherref}
\endbibitem

\bibitem[\protect\citeauthoryear{{Ollama}}{2025}]{SP-OllamaModelfile2025}
\begin{botherref}
\oauthor{\bsnm{{Ollama}}}:
Modelfile: Valid parameters and values
(2025).
\url{https://github.com/ollama/ollama/blob/main/docs/modelfile.md#valid-parameters-and-values}
Accessed 2025-09-25
\end{botherref}
\endbibitem

\bibitem[\protect\citeauthoryear{}{}]{SP-tnes_ghissue_vllm_26806}
\begin{botherref}
GitHub Issue/PR. MCP-USE with VLLM gpt-oss:20b via ChatOpenAI.
GitHub Issue/PR.
\url{https://github.com/vllm-project/vllm/issues/26806}
\end{botherref}
\endbibitem

\bibitem[\protect\citeauthoryear{}{}]{SP-tnes_ghissue_langfuse_9566}
\begin{botherref}
GitHub Issue/PR. When streaming responses with the OpenAI Responses API,
  temperature is not captured correctly.
GitHub Issue/PR.
\url{https://github.com/langfuse/langfuse/issues/9566}
\end{botherref}
\endbibitem

\bibitem[\protect\citeauthoryear{Wang}{2025}]{SP-wangkelsey_acompreh_2025_7d4ce5}
\begin{botherref}
\oauthor{\bsnm{Wang}, \binits{K.}}:
A comprehensive guide to llm temperature
(2025).
Accessed 2025-12-29
\end{botherref}
\endbibitem

\bibitem[\protect\citeauthoryear{Pathak}{2024}]{SP-pathakteju_setthete_2024_7a1bf6}
\begin{botherref}
\oauthor{\bsnm{Pathak}, \binits{T.}}:
Set the temperature right: Optimizing temperature settings in prompt
  engineering
(2024).
Accessed 2025-12-29
\end{botherref}
\endbibitem

\bibitem[\protect\citeauthoryear{kaviyadharishini21}{2024}]{SP-kaviyadhar_impactof_2024_5f226d}
\begin{botherref}
\oauthor{\bsnm{kaviyadharishini21}}:
Impact of temperature on large language models (llms)
(2024).
Accessed 2025-12-29
\end{botherref}
\endbibitem

\bibitem[\protect\citeauthoryear{Balarabe}{2025}]{SP-tahirbalar_understa_2025_941da5}
\begin{botherref}
\oauthor{\bsnm{Balarabe}, \binits{T.}}:
Understanding llm temperature: Creativity vs. consistency
(2025).
Accessed 2025-12-29
\end{botherref}
\endbibitem

\bibitem[\protect\citeauthoryear{u/Extreme
  Wall9508}{2025}]{SP-uextremewa_whatisth_2025_d27670}
\begin{botherref}
\oauthor{\bsnm{u/Extreme-Wall9508}}:
What Is the Right Temperature Parameter to Work With? : r/LLMDevs.
\url{https://www.reddit.com/r/LLMDevs/comments/1i3113d/what_is_the_right_temperature_parameter_to_work/}
Accessed 2025-12-29
\end{botherref}
\endbibitem

\bibitem[\protect\citeauthoryear{Herrmannova}{2024}]{SP-herrmannov_temperat_2024_1c137a}
\begin{botherref}
\oauthor{\bsnm{Herrmannova}, \binits{D.}}:
Temperature sampling
(2024).
Accessed 2025-12-29
\end{botherref}
\endbibitem

\bibitem[\protect\citeauthoryear{Šubonis}{2025}]{SP-ubonismart_zerotemp_2025_15a9b4}
\begin{botherref}
\oauthor{\bsnm{Šubonis}, \binits{M.}}:
Zero temperature randomness in llms
(2025).
Accessed 2025-12-29
\end{botherref}
\endbibitem

\bibitem[\protect\citeauthoryear{Patel}{2025}]{SP-patelashis_llmtempe_2025_da6628}
\begin{botherref}
\oauthor{\bsnm{Patel}, \binits{A.}}:
"LLM Temperature” Is NOT the Little Slider You Think It Is. Ever ...
\url{https://www.linkedin.com/posts/ashishpatel2604_llm-temperature-is-not-the-little-slider-activity-7391720132559118336-2D6C}
Accessed 2025-12-29
\end{botherref}
\endbibitem

\bibitem[\protect\citeauthoryear{Face}{2025}]{SP-huggingfac_thedocum_2025_b77968}
\begin{botherref}
\oauthor{\bsnm{Face}, \binits{H.}}:
The document of generation seems to wrongly describe the default value of
  top_p, top_k and temperature
(2025).
\url{https://github.com/huggingface/transformers/issues/35045}
Accessed 2025-12-29
\end{botherref}
\endbibitem

\bibitem[\protect\citeauthoryear{OpenHands}{2025}]{SP-openhands_bugazure_2025_cfd769}
\begin{botherref}
\oauthor{\bsnm{OpenHands}}:
Bug: Azure GPT-5 fails due to hardcoded temperature in responses_options.py
(2025).
\url{https://github.com/OpenHands/software-agent-sdk/issues/986}
Accessed 2025-12-29
\end{botherref}
\endbibitem

\bibitem[\protect\citeauthoryear{micz}{2025}]{SP-micz_exposemo_2025_d3e768}
\begin{botherref}
\oauthor{\bsnm{micz}}:
Expose model temperature control in ThunderAI connection settings (global
  default + optional per-prompt override)
(2025).
\url{https://github.com/micz/ThunderAI/issues/561}
Accessed 2025-12-29
\end{botherref}
\endbibitem

\bibitem[\protect\citeauthoryear{penguoir}{2025}]{SP-penguoir_allowuse_2025_ee4b45}
\begin{botherref}
\oauthor{\bsnm{penguoir}}:
Allow user-defined temperature
(2025).
\url{https://github.com/penguoir/active_cortex/issues/3}
Accessed 2025-12-29
\end{botherref}
\endbibitem

\bibitem[\protect\citeauthoryear{Pepelespooder}{2025}]{SP-pepelespoo_fixdarkm_2025_0ad199}
\begin{botherref}
\oauthor{\bsnm{Pepelespooder}}:
Fix Darkmoon temperature handling to respect user values and prevent dirty
  state
(2025).
\url{https://github.com/Pepelespooder/BambuStudio/pull/13}
Accessed 2025-12-29
\end{botherref}
\endbibitem

\bibitem[\protect\citeauthoryear{haktancetin}{2025}]{SP-haktanceti_llmrando_2025_ec24a3}
\begin{botherref}
\oauthor{\bsnm{haktancetin}}:
LLM Randomness
(2025).
\url{https://github.com/haktancetin/496_CookBuddyProject/issues/24}
Accessed 2025-12-29
\end{botherref}
\endbibitem

\bibitem[\protect\citeauthoryear{{OpenAI Developer
  Community}}{2024}]{SP-openai_temp0_forum_2024}
\begin{botherref}
\oauthor{\bsnm{{OpenAI Developer Community}}}:
Clarifications on Setting Temperature = 0.
Discussion thread accessed 2025-12-09.
\url{https://community.openai.com/t/clarifications-on-setting-temperature-0/886447}
\end{botherref}
\endbibitem

\bibitem[\protect\citeauthoryear{Institute}{2024}]{SP-promptengineering_temperature_top_p_2024}
\begin{botherref}
\oauthor{\bsnm{Institute}, \binits{P.E.}}:
Complete Guide to Prompt Engineering with Temperature and Top-p.
Accessed: 2025-12-31
(2024).
\url{https://promptengineering.org/prompt-engineering-with-temperature-and-top-p/}
\end{botherref}
\endbibitem

\bibitem[\protect\citeauthoryear{Reyes et~al.}{2024}]{SP-reyes2024bump}
\begin{bchapter}
\bauthor{\bsnm{Reyes}, \binits{F.}},
\bauthor{\bsnm{Gamage}, \binits{Y.}},
\bauthor{\bsnm{Skoglund}, \binits{G.}},
\bauthor{\bsnm{Baudry}, \binits{B.}},
\bauthor{\bsnm{Monperrus}, \binits{M.}}:
\bctitle{Bump: A benchmark of reproducible breaking dependency updates}.
(\byear{2024}).
\doiurl{10.48550/arXiv.2401.09906} .
\burl{https://doi.org/10.48550/arXiv.2401.09906}
\end{bchapter}
\endbibitem

\bibitem[\protect\citeauthoryear{Venturini
  et~al.}{2023}]{SP-venturini2023depended}
\begin{botherref}
\oauthor{\bsnm{Venturini}, \binits{D.}},
\oauthor{\bsnm{Cogo}, \binits{F.R.}},
\oauthor{\bsnm{Polato}, \binits{I.}},
\oauthor{\bsnm{Gerosa}, \binits{M.A.}},
\oauthor{\bsnm{Wiese}, \binits{I.S.}}:
I depended on you and you broke me: An empirical study of manifesting breaking
  changes in client packages
(2023)
\doiurl{10.48550/arXiv.2301.04563}
\end{botherref}
\endbibitem

\bibitem[\protect\citeauthoryear{Montandon
  et~al.}{2024}]{SP-montandon2024defaultargs}
\begin{botherref}
\oauthor{\bsnm{Montandon}, \binits{J.E.}},
\oauthor{\bsnm{Silva}, \binits{L.L.}},
\oauthor{\bsnm{Politowski}, \binits{C.}},
\oauthor{\bsnm{Prates}, \binits{D.}},
\oauthor{\bsnm{Bonifacio}, \binits{A.B.}},
\oauthor{\bsnm{Boussaidi}, \binits{G.E.}}:
Unboxing default argument breaking changes in data science libraries
(2024)
\doiurl{10.48550/arXiv.2408.05129}
\end{botherref}
\endbibitem

\bibitem[\protect\citeauthoryear{da~Silva~Sim{\~o}es and
  Venson}{2024}]{SP-simoes2024evaluating}
\begin{botherref}
\oauthor{\bsnm{Silva~Sim{\~o}es}, \binits{I.R.}},
\oauthor{\bsnm{Venson}, \binits{E.}}:
Evaluating source code quality with large language models: a comparative study,
103--113
(2024)
\doiurl{10.1145/3701625.3701650}
\end{botherref}
\endbibitem

\bibitem[\protect\citeauthoryear{Wilson
  et~al.}{2014}]{SP-WilsonEtAl2014BestPractices}
\begin{barticle}
\bauthor{\bsnm{Wilson}, \binits{G.}},
\bauthor{\bsnm{Aruliah}, \binits{D.A.}},
\bauthor{\bsnm{Brown}, \binits{C.T.}},
\bauthor{\bsnm{Chue~Hong}, \binits{N.P.}},
\bauthor{\bsnm{Davis}, \binits{M.}},
\bauthor{\bsnm{Guy}, \binits{R.T.}},
\bauthor{\bsnm{Haddock}, \binits{S.H.D.}},
\bauthor{\bsnm{Huff}, \binits{K.D.}},
\bauthor{\bsnm{Mitchell}, \binits{I.M.}},
\bauthor{\bsnm{Plumbley}, \binits{M.D.}},
\bauthor{\bsnm{Waugh}, \binits{B.}},
\bauthor{\bsnm{White}, \binits{E.P.}},
\bauthor{\bsnm{Wilson}, \binits{P.}}:
\batitle{Best practices for scientific computing}.
\bjtitle{PLOS Biology}
\bvolume{12}(\bissue{1}),
\bfpage{1001745}
(\byear{2014})
\end{barticle}
\endbibitem

\bibitem[\protect\citeauthoryear{Venturini
  et~al.}{2023}]{SP-VenturiniEtAl2023IDepended}
\begin{botherref}
\oauthor{\bsnm{Venturini}, \binits{D.}},
\oauthor{\bsnm{Cogo}, \binits{F.R.}},
\oauthor{\bsnm{Polato}, \binits{I.}},
\oauthor{\bsnm{Gerosa}, \binits{M.A.}},
\oauthor{\bsnm{Wiese}, \binits{I.S.}}:
I depended on you and you broke me: An empirical study of manifesting breaking
  changes in client packages
(2023)
\doiurl{10.48550/arXiv.2301.04563}
{\href{https://arxiv.org/abs/2301.04563}{{arXiv:2301.04563}}}.
TOSEM, 2023
\end{botherref}
\endbibitem

\bibitem[\protect\citeauthoryear{Reyes et~al.}{2024}]{SP-ReyesEtAl2024BUMP}
\begin{bchapter}
\bauthor{\bsnm{Reyes}, \binits{F.}},
\bauthor{\bsnm{Gamage}, \binits{Y.}},
\bauthor{\bsnm{Skoglund}, \binits{G.}},
\bauthor{\bsnm{Baudry}, \binits{B.}},
\bauthor{\bsnm{Monperrus}, \binits{M.}}:
\bctitle{Bump: A benchmark of reproducible breaking dependency updates}.
In: \bbtitle{SANER 2024}
(\byear{2024}).
\doiurl{10.48550/arXiv.2401.09906} .
\burl{https://arxiv.org/abs/2401.09906}
Accessed 2025-09-25
\end{bchapter}
\endbibitem

\bibitem[\protect\citeauthoryear{Morishige and Koshihara}{2025}]{SP-Morishige}
\begin{bchapter}
\bauthor{\bsnm{Morishige}, \binits{M.}},
\bauthor{\bsnm{Koshihara}, \binits{R.}}:
\bctitle{Ensuring reproducibility in generative ai systems for general use
  cases: A framework for regression testing and open datasets}.
(\byear{2025}).
\doiurl{10.48550/arXiv.2505.02854}
\end{bchapter}
\endbibitem

\bibitem[\protect\citeauthoryear{{Microsoft
  Learn}}{2025}]{SP-Microsoft2025FoundationLifecycle}
\begin{botherref}
\oauthor{\bsnm{{Microsoft Learn}}}:
Design to Support Foundation Model Life Cycles
(2025).
\url{https://learn.microsoft.com/en-us/azure/architecture/ai-ml/guide/manage-foundation-models-lifecycle}
Accessed 2025-09-25
\end{botherref}
\endbibitem

\bibitem[\protect\citeauthoryear{{OpenRouter}}{2025}]{SP-OpenRouterHome2025}
\begin{botherref}
\oauthor{\bsnm{{OpenRouter}}}:
OpenRouter.ai - One API for Any Model.
\url{https://openrouter.ai/}
Accessed 2025-09-25
\end{botherref}
\endbibitem

\bibitem[\protect\citeauthoryear{Jeong et~al.}{2025}]{SP-jeong2025systemmsg}
\begin{botherref}
\oauthor{\bsnm{Jeong}, \binits{M.}},
\oauthor{\bsnm{Cho}, \binits{J.}},
\oauthor{\bsnm{Khang}, \binits{M.}},
\oauthor{\bsnm{Jung}, \binits{D.}},
\oauthor{\bsnm{Hong}, \binits{T.}}:
System message generation for user preferences using open-source models
(2025)
\end{botherref}
\endbibitem

\bibitem[\protect\citeauthoryear{Neumann et~al.}{2025}]{SP-neumann2025position}
\begin{bchapter}
\bauthor{\bsnm{Neumann}, \binits{A.}},
\bauthor{\bsnm{Kirsten}, \binits{E.}},
\bauthor{\bsnm{Zafar}, \binits{M.B.}},
\bauthor{\bsnm{Singh}, \binits{J.}}:
\bctitle{Position is power: System prompts as a mechanism of bias in large
  language models}.
(\byear{2025}).
\doiurl{10.1145/3715275.3732038} .
\burl{https://doi.org/10.1145/3715275.3732038}
\end{bchapter}
\endbibitem

\bibitem[\protect\citeauthoryear{Jeong
  et~al.}{2025}]{SP-jeong2025messagegenerationuserpreferences}
\begin{botherref}
\oauthor{\bsnm{Jeong}, \binits{M.}},
\oauthor{\bsnm{Cho}, \binits{J.}},
\oauthor{\bsnm{Khang}, \binits{M.}},
\oauthor{\bsnm{Jung}, \binits{D.}},
\oauthor{\bsnm{Hong}, \binits{T.}}:
System Message Generation for User Preferences using Open-Source Models
(2025).
\url{https://arxiv.org/abs/2502.11330}
\end{botherref}
\endbibitem

\bibitem[\protect\citeauthoryear{Neumann et~al.}{2025}]{SP-Neumann_2025}
\begin{bchapter}
\bauthor{\bsnm{Neumann}, \binits{A.}},
\bauthor{\bsnm{Kirsten}, \binits{E.}},
\bauthor{\bsnm{Zafar}, \binits{M.B.}},
\bauthor{\bsnm{Singh}, \binits{J.}}:
\bctitle{Position is power: System prompts as a mechanism of bias in large
  language models (llms)}.
In: \bbtitle{Proceedings of the 2025 ACM Conference on Fairness,
  Accountability, and Transparency}.
\bsertitle{FAccT '25},
pp. \bfpage{573}--\blpage{598}.
\bpublisher{ACM}
(\byear{2025}).
\doiurl{10.1145/3715275.3732038} .
\burl{http://dx.doi.org/10.1145/3715275.3732038}
\end{bchapter}
\endbibitem

\bibitem[\protect\citeauthoryear{Cleary}{2025}]{SP-PromptHub2025SystemMessages}
\begin{botherref}
\oauthor{\bsnm{Cleary}, \binits{D.}}:
System Messages: Best Practices, Real-world Experiments \& Prompt Injection
  Protectors.
PromptHub Blog.
\url{https://www.prompthub.us/blog/everything-system-messages-how-to-use-them-real-world-experiments-prompt-injection-protectors}
Accessed 2025-09-25
\end{botherref}
\endbibitem

\bibitem[\protect\citeauthoryear{{cyz3a5c0v1}}{2023}]{SP-StackOverflow2023SystemRoleUseCase}
\begin{botherref}
\oauthor{\bsnm{{cyz3a5c0v1}}}:
What is the use case of System role
(2023).
\url{https://stackoverflow.com/questions/76272624/what-is-the-use-case-of-system-role}
Accessed 2025-09-25
\end{botherref}
\endbibitem

\bibitem[\protect\citeauthoryear{OpenAI}{2023}]{SP-openai_whatexac_2023_59a966}
\begin{botherref}
\oauthor{\bsnm{OpenAI}}:
What Exactly Does a System Msg Do?
\url{https://community.openai.com/t/what-exactly-does-a-system-msg-do/459409}
Accessed 2025-12-29
\end{botherref}
\endbibitem

\bibitem[\protect\citeauthoryear{Community}{2024}]{SP-openaicomm_question_2024_bdb2f3}
\begin{botherref}
\oauthor{\bsnm{Community}, \binits{O.}}:
Question Around the Structure of System, User, Assistant - API.
\url{https://community.openai.com/t/question-around-the-structure-of-system-user-assistant/581078}
Accessed 2025-12-29
\end{botherref}
\endbibitem

\bibitem[\protect\citeauthoryear{OpenAI}{2023a}]{SP-openai_understa_2023_f4b4d0}
\begin{botherref}
\oauthor{\bsnm{OpenAI}}:
Understanding Role Management in OpenAI's API: Two Methods Compared.
\url{https://community.openai.com/t/understanding-role-management-in-openais-api-two-methods-compared/253289}
Accessed 2025-12-29
\end{botherref}
\endbibitem

\bibitem[\protect\citeauthoryear{OpenAI}{2023b}]{SP-openai_thesyste_2023_d84a9a}
\begin{botherref}
\oauthor{\bsnm{OpenAI}}:
The “system” Role - How It Influences the Chat Behavior - API.
\url{https://community.openai.com/t/the-system-role-how-it-influences-the-chat-behavior/87353}
Accessed 2025-12-29
\end{botherref}
\endbibitem

\bibitem[\protect\citeauthoryear{OpenAI}{2023c}]{SP-openai_providin_2023_e5a2fb}
\begin{botherref}
\oauthor{\bsnm{OpenAI}}:
Providing Context to the Chat API Before a Conversation - Prompting ...
\url{https://community.openai.com/t/providing-context-to-the-chat-api-before-a-conversation/195853}
Accessed 2025-12-29
\end{botherref}
\endbibitem

\bibitem[\protect\citeauthoryear{dan_43009}{2024}]{SP-dan43009_thediffe_2024_06451f}
\begin{botherref}
\oauthor{\bsnm{dan_43009}}:
The difference between system messages and user messages in ...
(2024).
Accessed 2025-12-29
\end{botherref}
\endbibitem

\bibitem[\protect\citeauthoryear{Hakim}{2025}]{SP-hakimmudas_masterin_2025_22b6f4}
\begin{botherref}
\oauthor{\bsnm{Hakim}, \binits{M.}}:
Mastering prompt engineering: A guide to system, user, and ...
(2025).
Accessed 2025-12-29
\end{botherref}
\endbibitem

\bibitem[\protect\citeauthoryear{u/thexdroid}{2024}]{SP-uthexdroid_whatisth_2024_a14eae}
\begin{botherref}
\oauthor{\bsnm{u/thexdroid}}:
What Is the Correct Way to Use System Message Roles? : R/ollama.
\url{https://www.reddit.com/r/ollama/comments/1dgw4w9/what_is_the_correct_way_to_use_system_message/}
Accessed 2025-12-29
\end{botherref}
\endbibitem

\bibitem[\protect\citeauthoryear{lgabs}{2025}]{SP-lgabs_usesyste_2025_b78ffb}
\begin{botherref}
\oauthor{\bsnm{lgabs}}:
use System, Human and AI patterns for prompts roles for OpenAI chat models
(2025).
\url{https://github.com/lgabs/dialog/issues/16}
Accessed 2025-12-29
\end{botherref}
\endbibitem

\bibitem[\protect\citeauthoryear{Wen et~al.}{2025}]{SP-wen2025budgetthinker}
\begin{botherref}
\oauthor{\bsnm{Wen}, \binits{H.}},
\oauthor{\bsnm{Wu}, \binits{X.}},
\oauthor{\bsnm{Sun}, \binits{Y.}},
\oauthor{\bsnm{Zhang}, \binits{F.}},
\oauthor{\bsnm{Chen}, \binits{L.}},
\oauthor{\bsnm{Wang}, \binits{J.}},
\oauthor{\bsnm{Liu}, \binits{Y.}},
\oauthor{\bsnm{Liu}, \binits{Y.}},
\oauthor{\bsnm{Zhang}, \binits{Y.}},
\oauthor{\bsnm{Li}, \binits{Y.}}:
Budgetthinker: Empowering budget-aware llm reasoning with control tokens
(2025)
\doiurl{10.48550/arXiv.2508.17196}
\end{botherref}
\endbibitem

\bibitem[\protect\citeauthoryear{Wang et~al.}{2024}]{SP-wang2024reasoningbudget}
\begin{bchapter}
\bauthor{\bsnm{Wang}, \binits{J.}},
\bauthor{\bsnm{Jain}, \binits{S.}},
\bauthor{\bsnm{Zhang}, \binits{D.}},
\bauthor{\bsnm{Ray}, \binits{B.}},
\bauthor{\bsnm{Kumar}, \binits{V.}},
\bauthor{\bsnm{Athiwaratkun}, \binits{B.}}:
\bctitle{Reasoning in token economies: Budget-aware evaluation of llm reasoning
  strategies}.
(\byear{2024}).
\doiurl{10.18653/v1/2024.emnlp-main.1112} .
\burl{https://aclanthology.org/2024.emnlp-main.1112/}
\end{bchapter}
\endbibitem

\bibitem[\protect\citeauthoryear{Chen et~al.}{2025}]{SP-chen2025robustinference}
\begin{botherref}
\oauthor{\bsnm{Chen}, \binits{Z.}},
\oauthor{\bsnm{Ye}, \binits{Y.}},
\oauthor{\bsnm{Zhou}, \binits{Z.}}:
Adaptively robust llm inference optimization under prediction uncertainty
(2025)
\end{botherref}
\endbibitem

\bibitem[\protect\citeauthoryear{Han et~al.}{2025}]{SP-han2025tokenbudget}
\begin{bchapter}
\bauthor{\bsnm{Han}, \binits{T.}},
\bauthor{\bsnm{Wang}, \binits{Z.}},
\bauthor{\bsnm{Fang}, \binits{C.}},
\bauthor{\bsnm{Zhao}, \binits{S.}},
\bauthor{\bsnm{Ma}, \binits{S.}},
\bauthor{\bsnm{Chen}, \binits{Z.}}:
\bctitle{Token-budget-aware llm reasoning}.
(\byear{2025}).
\doiurl{10.18653/v1/2025.findings-acl} .
\burl{https://doi.org/10.18653/v1/2025.findings-acl}
\end{bchapter}
\endbibitem

\bibitem[\protect\citeauthoryear{Lee et~al.}{2025}]{SP-lee2025ergo}
\begin{botherref}
\oauthor{\bsnm{Lee}, \binits{J.}},
\oauthor{\bsnm{Shin}, \binits{W.}},
\oauthor{\bsnm{Yang}, \binits{S.}},
\oauthor{\bsnm{Song}, \binits{K.-U.}},
\oauthor{\bsnm{Lim}, \binits{D.}},
\oauthor{\bsnm{Kim}, \binits{J.}},
\oauthor{\bsnm{Kim}, \binits{T.-H.}},
\oauthor{\bsnm{Kim}, \binits{B.-K.}}:
Ergo: Efficient high-resolution visual understanding for vision-language models
(2025)
\doiurl{10.48550/arXiv.2509.21991}
\end{botherref}
\endbibitem

\bibitem[\protect\citeauthoryear{Vasu et~al.}{2024}]{SP-vasu2024fastvlm}
\begin{botherref}
\oauthor{\bsnm{Vasu}, \binits{P.K.A.}},
\oauthor{\bsnm{Faghri}, \binits{F.}},
\oauthor{\bsnm{Li}, \binits{C.-L.}},
\oauthor{\bsnm{Koc}, \binits{C.}},
\oauthor{\bsnm{True}, \binits{N.}},
\oauthor{\bsnm{Antony}, \binits{A.}},
\oauthor{\bsnm{Santhanam}, \binits{G.}},
\oauthor{\bsnm{Gabriel}, \binits{J.}},
\oauthor{\bsnm{Grasch}, \binits{P.}},
\oauthor{\bsnm{Tuzel}, \binits{O.}},
\oauthor{\bsnm{Pouransari}, \binits{H.}}:
Fastvlm: Efficient vision encoding for vision language models
(2024)
\doiurl{10.48550/arXiv.2412.13303}
\end{botherref}
\endbibitem

\bibitem[\protect\citeauthoryear{Qian et~al.}{2025}]{SP-qian2025zoomer}
\begin{botherref}
\oauthor{\bsnm{Qian}, \binits{J.}},
\oauthor{\bsnm{Wang}, \binits{C.}},
\oauthor{\bsnm{Yang}, \binits{Y.}},
\oauthor{\bsnm{Zhang}, \binits{C.}},
\oauthor{\bsnm{Jiang}, \binits{H.}},
\oauthor{\bsnm{Luo}, \binits{X.}},
\oauthor{\bsnm{Kang}, \binits{Y.}},
\oauthor{\bsnm{Lin}, \binits{Q.}},
\oauthor{\bsnm{Zhang}, \binits{A.}},
\oauthor{\bsnm{Jiang}, \binits{S.}},
\oauthor{\bsnm{Cao}, \binits{T.}},
\oauthor{\bsnm{Mao}, \binits{T.}},
\oauthor{\bsnm{Banerjee}, \binits{S.}},
\oauthor{\bsnm{Liu}, \binits{G.}},
\oauthor{\bsnm{Rajmohan}, \binits{S.}},
\oauthor{\bsnm{Zhang}, \binits{D.}},
\oauthor{\bsnm{Yang}, \binits{Y.}},
\oauthor{\bsnm{Zhang}, \binits{Q.}},
\oauthor{\bsnm{Qiu}, \binits{L.}}:
Zoomer: Adaptive image focus optimization for black-box mllm
(2025)
\doiurl{10.48550/arXiv.2505.00742}
\end{botherref}
\endbibitem

\bibitem[\protect\citeauthoryear{{Microsoft
  Corporation}}{2025}]{SP-azure_openai_temperature_topp}
\begin{botherref}
\oauthor{\bsnm{{Microsoft Corporation}}}:
Azure OpenAI in Microsoft Foundry Models REST API.
Accessed 2025-12-09.
\url{https://learn.microsoft.com/en-us/azure/ai-foundry/openai/latest?view=foundry-classic}
\end{botherref}
\endbibitem

\bibitem[\protect\citeauthoryear{{Anthropic}}{2025}]{SP-anthropic_claude_temperature_topp}
\begin{botherref}
\oauthor{\bsnm{{Anthropic}}}:
Create a Text Completion Claude API Reference.
Accessed 2025-12-09.
\url{https://docs.anthropic.com/claude/reference/complete_post}
\end{botherref}
\endbibitem

\bibitem[\protect\citeauthoryear{Willison}{2025}]{SP-willison_llm_anthropic_2025}
\begin{botherref}
\oauthor{\bsnm{Willison}, \binits{S.}}:
Llm-anthropic.
Python package version 0.23 accessed 2025-12-09.
\url{https://pypi.org/project/llm-anthropic/}
\end{botherref}
\endbibitem

\bibitem[\protect\citeauthoryear{{n8n.io}}{2025}]{SP-n8n_anthropic_temp_topp_issue_2025}
\begin{botherref}
\oauthor{\bsnm{{n8n.io}}}:
AI Agent with Anthropic Models Fails Temperature and Top_p Cannot Both Be
  Specified for this Model.
GitHub issue accessed 2025-12-09.
\url{https://github.com/n8n-io/n8n/issues/18304}
\end{botherref}
\endbibitem

\bibitem[\protect\citeauthoryear{Medium}{2024}]{SP-medium_settingt_2024_5a6dd8}
\begin{botherref}
\oauthor{\bsnm{Medium}}:
Setting top-k, top-p and temperature in llms
(2024).
Accessed 2025-12-29
\end{botherref}
\endbibitem

\bibitem[\protect\citeauthoryear{Albert}{2024}]{SP-albert_largelan_2024_7da749}
\begin{botherref}
\oauthor{\bsnm{Albert}}:
Large language model settings: Temperature, top p and max ...
(2024).
Accessed 2025-12-29
\end{botherref}
\endbibitem

\bibitem[\protect\citeauthoryear{RepoWise}{2025}]{SP-repowise_ensurede_2025_3759a4}
\begin{botherref}
\oauthor{\bsnm{RepoWise}}:
Ensure Deterministic \& Fully Consistent LLM Output
(2025).
\url{https://github.com/RepoWise/backend/issues/11}
Accessed 2025-12-29
\end{botherref}
\endbibitem

\bibitem[\protect\citeauthoryear{mpfaffenberger}{2025}]{SP-mpfaffenbe_featurer_2025_639588}
\begin{botherref}
\oauthor{\bsnm{mpfaffenberger}}:
Feature Request: Add Recommended Sampling Parameters for GLM Model
(2025).
\url{https://github.com/mpfaffenberger/code_puppy/issues/70}
Accessed 2025-12-29
\end{botherref}
\endbibitem

\bibitem[\protect\citeauthoryear{{OpenAI}}{2025}]{SP-openai_safety_best_practices}
\begin{botherref}
\oauthor{\bsnm{{OpenAI}}}:
Safety best practices.
\url{https://platform.openai.com/docs/guides/safety-best-practices}.
Accessed 2025-12-10
(2025)
\end{botherref}
\endbibitem

\bibitem[\protect\citeauthoryear{{Google
  DeepMind}}{2025}]{SP-google_gemini_safety_guidance}
\begin{botherref}
\oauthor{\bsnm{{Google DeepMind}}}:
Safety guidance.
\url{https://ai.google.dev/gemini-api/docs/safety-guidance}.
Accessed 2025-12-10
(2025)
\end{botherref}
\endbibitem

\bibitem[\protect\citeauthoryear{{Anthropic}}{2025}]{SP-anthropic_claude_messages_create_python}
\begin{botherref}
\oauthor{\bsnm{{Anthropic}}}:
Create a Message.
\url{https://platform.claude.com/docs/en/api/python/messages/create}.
Accessed 2025-12-10
(2025)
\end{botherref}
\endbibitem

\bibitem[\protect\citeauthoryear{raz alon and
  contributors}{2025}]{SP-berriai_litellm_issue_10106}
\begin{botherref}
\oauthor{\bsnm{raz-alon}},
\oauthor{\bsnm{contributors}}:
{[Bug]: Anthropic API throws Bad Request when user_id in metadata contains
  email or phone number}.
\url{https://github.com/BerriAI/litellm/issues/10106}.
GitHub issue 10106, opened 2025-04-17, accessed 2025-12-10
(2025)
\end{botherref}
\endbibitem

\bibitem[\protect\citeauthoryear{OpenAI}{2024}]{SP-openai_needhelp_2024_349ad1}
\begin{botherref}
\oauthor{\bsnm{OpenAI}}:
Need Help: Facing OpenAI Usage Violation Due to User's Abuse - API.
\url{https://community.openai.com/t/need-help-facing-openai-usage-violation-due-to-users-abuse/1004947}
Accessed 2025-12-29
\end{botherref}
\endbibitem

\bibitem[\protect\citeauthoryear{Community}{2022}]{SP-openaideve_apibanfr_2022_6e7ce6}
\begin{botherref}
\oauthor{\bsnm{Community}, \binits{O.D.}}:
API Ban from User Abuse - API - OpenAI Developer Community.
\url{https://community.openai.com/t/api-ban-from-user-abuse/25400}
Accessed 2025-12-29
\end{botherref}
\endbibitem

\bibitem[\protect\citeauthoryear{Community}{2024}]{SP-openaicomm_anysugge_2024_873212}
\begin{botherref}
\oauthor{\bsnm{Community}, \binits{O.}}:
Any Suggestions for Preventing Openai API Abuse - API - OpenAI ...
\url{https://community.openai.com/t/any-suggestions-for-preventing-openai-api-abuse/1075315}
Accessed 2025-12-29
\end{botherref}
\endbibitem

\bibitem[\protect\citeauthoryear{OpenAI}{2022}]{SP-openai_lessonsl_2022_863fc6}
\begin{botherref}
\oauthor{\bsnm{OpenAI}}:
Lessons Learned on Language Model Safety and Misuse.
\url{https://openai.com/index/language-model-safety-and-misuse/}
Accessed 2025-12-29
\end{botherref}
\endbibitem

\bibitem[\protect\citeauthoryear{OpenAI}{2024}]{SP-openai_apipolic_2024_325242}
\begin{botherref}
\oauthor{\bsnm{OpenAI}}:
API Policy Violation Warning - OpenAI Developer Community.
\url{https://community.openai.com/t/api-policy-violation-warning-advice-on-how-to-best-resolve/788354}
Accessed 2025-12-29
\end{botherref}
\endbibitem

\bibitem[\protect\citeauthoryear{Anthropic}{2025}]{SP-anthropic_detectin_2025_f8b326}
\begin{botherref}
\oauthor{\bsnm{Anthropic}}:
Detecting and Countering Malicious Uses of Claude.
\url{https://www.anthropic.com/news/detecting-and-countering-malicious-uses-of-claude-march-2025}
Accessed 2025-12-29
\end{botherref}
\endbibitem

\bibitem[\protect\citeauthoryear{Tech}{2023}]{SP-nerdfortec_safeguar_2023_1af5bd}
\begin{botherref}
\oauthor{\bsnm{Tech}, \binits{N.F.}}:
Safeguarding your ai: Best practices for securing your openai api
(2023).
Accessed 2025-12-29
\end{botherref}
\endbibitem

\bibitem[\protect\citeauthoryear{de~Oliveira
  et~al.}{2025}]{SP-deOliveira2025trust}
\begin{botherref}
\oauthor{\bsnm{Oliveira}, \binits{A.C.}},
\oauthor{\bsnm{Azevedo}, \binits{J.P.C.}},
\oauthor{\bsnm{Ruback}, \binits{L.}},
\oauthor{\bsnm{Moreira}, \binits{R.}},
\oauthor{\bsnm{Teixeira}, \binits{S.S.}},
\oauthor{\bsnm{Teles}, \binits{A.S.}}:
Effect of explainable artificial intelligence on trust of mental health
  professionals in an ai-based system for suicide prevention.
IEEE Access,
1--14
(2025)
\doiurl{10.1109/ACCESS.2025.3556245}
\end{botherref}
\endbibitem

\bibitem[\protect\citeauthoryear{Ijas
  et~al.}{2024}]{SP-IjasJoRaj2024XAI_LLM_TransparencyTrust}
\begin{bchapter}
\bauthor{\bsnm{Ijas}, \binits{A.H.}},
\bauthor{\bsnm{Jo}, \binits{A.A.}},
\bauthor{\bsnm{Raj}, \binits{E.D.}}:
\bctitle{Exploring explainable ai in large language models: Enhancing
  transparency and trust}.
In: \bbtitle{2024 11th International Conference on Advances in Computing and
  Communications (ICACC)},
\bconflocation{Kochi, India},
pp. \bfpage{1}--\blpage{7}
(\byear{2024}).
\doiurl{10.1109/ICACC63692.2024.10845370} .
\bcomment{IEEE. Held 6--8 November 2024}
\end{bchapter}
\endbibitem

\bibitem[\protect\citeauthoryear{Na et~al.}{2024}]{SP-na2024llmcpus}
\begin{bchapter}
\bauthor{\bsnm{Na}, \binits{S.}},
\bauthor{\bsnm{Jeong}, \binits{G.}},
\bauthor{\bsnm{Ahn}, \binits{B.H.}},
\bauthor{\bsnm{Young}, \binits{J.}},
\bauthor{\bsnm{Krishna}, \binits{T.}},
\bauthor{\bsnm{Kim}, \binits{H.}}:
\bctitle{Understanding performance implications of llm inference on cpus}.
In: \bbtitle{Proceedings of the 2024 IEEE International Symposium on Workload
  Characterization (IISWC)},
pp. \bfpage{1}--\blpage{12}.
\bpublisher{IEEE}
(\byear{2024}).
\doiurl{10.1109/IISWC63097.2024.00024} .
\burl{https://ieeexplore.ieee.org/document/10763564}
\end{bchapter}
\endbibitem

\bibitem[\protect\citeauthoryear{Karthikeyan}{2025}]{SP-karthikeya_openaiap_2025_4fab36}
\begin{botherref}
\oauthor{\bsnm{Karthikeyan}}:
Openai api security: Managing ai risk in chatbots
(2025).
Accessed 2025-12-29
\end{botherref}
\endbibitem

\bibitem[\protect\citeauthoryear{ollama}{2025}]{SP-ollama_ollamaol_2025_6af93e}
\begin{botherref}
\oauthor{\bsnm{ollama}}:
Ollama/ollama.
\url{https://github.com/ollama/ollama}
Accessed 2025-12-29
\end{botherref}
\endbibitem

\bibitem[\protect\citeauthoryear{owainlewis}{2025}]{SP-owainlewis_owainlew_2025_c9fdf0}
\begin{botherref}
\oauthor{\bsnm{owainlewis}}:
owainlewis/awesome-artificial-intelligence
(2025).
\url{https://github.com/owainlewis/awesome-artificial-intelligence}
Accessed 2025-12-29
\end{botherref}
\endbibitem

\bibitem[\protect\citeauthoryear{humanlayer}{2025}]{SP-humanlayer_12factor_2025_66d41b}
\begin{botherref}
\oauthor{\bsnm{humanlayer}}:
12-Factor Agents - Principles for building reliable LLM applications
(2025).
\url{https://github.com/humanlayer/12-factor-agents}
Accessed 2025-12-29
\end{botherref}
\endbibitem

\bibitem[\protect\citeauthoryear{DeepSeek-AI}{2025}]{SP-deepseekai_deepseek_2025_767b4a}
\begin{botherref}
\oauthor{\bsnm{DeepSeek-AI}}:
DeepSeek-R1.
\url{https://github.com/deepseek-ai/DeepSeek-R1}
Accessed 2025-12-29
\end{botherref}
\endbibitem

\bibitem[\protect\citeauthoryear{OpenAI}{2025}]{SP-openai_gpt41cha_2025_3482c8}
\begin{botherref}
\oauthor{\bsnm{OpenAI}}:
GPT 4.1 Character Encoding Issues? - Bugs - OpenAI Developer.
\url{https://community.openai.com/t/gpt-4-1-character-encoding-issues/1236017}
Accessed 2025-12-29
\end{botherref}
\endbibitem

\bibitem[\protect\citeauthoryear{OpenAI}{2024}]{SP-openai_howtopas_2024_f80a3f}
\begin{botherref}
\oauthor{\bsnm{OpenAI}}:
How to Pass Conversation History Back to the API - API - OpenAI.
\url{https://community.openai.com/t/how-to-pass-conversation-history-back-to-the-api/697083}
Accessed 2025-12-29
\end{botherref}
\endbibitem

\bibitem[\protect\citeauthoryear{quic}{2025}]{SP-quic_llmoutpu_2025_6021e8}
\begin{botherref}
\oauthor{\bsnm{quic}}:
LLM Output Text formatting Issue
(2025).
\url{https://github.com/quic/cloud-ai-sdk/issues/6}
Accessed 2025-12-29
\end{botherref}
\endbibitem

\bibitem[\protect\citeauthoryear{alltuner}{2025}]{SP-alltuner_adderror_2025_1161a0}
\begin{botherref}
\oauthor{\bsnm{alltuner}}:
Add error handling for LLM response parsing
(2025).
\url{https://github.com/alltuner/blogtuner/issues/43}
Accessed 2025-12-29
\end{botherref}
\endbibitem

\bibitem[\protect\citeauthoryear{Skyvern-AI}{2025}]{SP-skyvernai_alwaysca_2025_835ac2}
\begin{botherref}
\oauthor{\bsnm{Skyvern-AI}}:
always capture llm artifacts
(2025).
\url{https://github.com/Skyvern-AI/skyvern/pull/4284}
Accessed 2025-12-29
\end{botherref}
\endbibitem

\bibitem[\protect\citeauthoryear{jerseycheese}{2025}]{SP-jerseychee_advanced_2025_c58498}
\begin{botherref}
\oauthor{\bsnm{jerseycheese}}:
Advanced provider settings and generation parameters
(2025).
\url{https://github.com/jerseycheese/Narraitor/issues/899}
Accessed 2025-12-29
\end{botherref}
\endbibitem

\bibitem[\protect\citeauthoryear{docling
  project}{2025}]{SP-doclingpro_vlmpipel_2025_d71e58}
\begin{botherref}
\oauthor{\bsnm{docling-project}}:
VLM pipeline hangs indefinitely during document processing with Transformers
  backend on NVIDIA GPU
(2025).
\url{https://github.com/docling-project/docling/issues/2472}
Accessed 2025-12-29
\end{botherref}
\endbibitem

\bibitem[\protect\citeauthoryear{Blaizzy}{2025}]{SP-blaizzy_commanda_2025_bc5bc9}
\begin{botherref}
\oauthor{\bsnm{Blaizzy}}:
Command A Vision responses contain only "<PAD>" tokens
(2025).
\url{https://github.com/Blaizzy/mlx-vlm/issues/487}
Accessed 2025-12-29
\end{botherref}
\endbibitem

\bibitem[\protect\citeauthoryear{ai~dynamo}{2025}]{SP-aidynamo_featgene_2025_304aac}
\begin{botherref}
\oauthor{\bsnm{ai-dynamo}}:
feat: generalize VLM embedding extraction
(2025).
\url{https://github.com/ai-dynamo/dynamo/pull/1388}
Accessed 2025-12-29
\end{botherref}
\endbibitem

\bibitem[\protect\citeauthoryear{sgl
  project}{2025}]{SP-sglproject_docsaddt_2025_13d3d3}
\begin{botherref}
\oauthor{\bsnm{sgl-project}}:
docs: Add token-in-token-out examples for LLM and VLM engines
(2025).
\url{https://github.com/sgl-project/sglang/pull/7262}
Accessed 2025-12-29
\end{botherref}
\endbibitem

\bibitem[\protect\citeauthoryear{M5Stack}{2025}]{SP-m5stack_sampling_2025_aed1a0}
\begin{botherref}
\oauthor{\bsnm{M5Stack}}:
Sampling parameters (temperature, top_k, top_p, etc.) not applied in LLM module
  (standalone mode)
(2025).
\url{https://github.com/m5stack/StackFlow/issues/22}
Accessed 2025-12-29
\end{botherref}
\endbibitem

\bibitem[\protect\citeauthoryear{mistralai}{2025}]{SP-mistralai_featurer_2025_719ae5}
\begin{botherref}
\oauthor{\bsnm{mistralai}}:
Feature request: Notify user when LLM response lacks content body
(2025).
\url{https://github.com/mistralai/mistral-vibe/issues/130}
Accessed 2025-12-29
\end{botherref}
\endbibitem

\bibitem[\protect\citeauthoryear{HouseOfBetterAuth}{2025}]{SP-houseofbet_implemen_2025_795c25}
\begin{botherref}
\oauthor{\bsnm{HouseOfBetterAuth}}:
Implement conversation history/context for chat LLM calls
(2025).
\url{https://github.com/HouseOfBetterAuth/nuxt-better-auth-saas/issues/1}
Accessed 2025-12-29
\end{botherref}
\endbibitem

\bibitem[\protect\citeauthoryear{ollama}{2025}]{SP-ollama_ollamafr_2025_ad8643}
\begin{botherref}
\oauthor{\bsnm{ollama}}:
Ollama freezes when specifying chat roles for some models.
(2025).
\url{https://github.com/ollama/ollama/issues/7003}
Accessed 2025-12-29
\end{botherref}
\endbibitem

\bibitem[\protect\citeauthoryear{spring
  projects}{2025}]{SP-springproj_chatmemo_2025_6ce4a7}
\begin{botherref}
\oauthor{\bsnm{spring-projects}}:
Chat memory advisor causes message ordering problem with default system message
  prompt
(2025).
\url{https://github.com/spring-projects/spring-ai/issues/4170}
Accessed 2025-12-29
\end{botherref}
\endbibitem

\bibitem[\protect\citeauthoryear{alexyang0826}{2025}]{SP-alexyang08_inferenc_2025_518dd3}
\begin{botherref}
\oauthor{\bsnm{alexyang0826}}:
Inference-Time Techniques for LLM Reasoning
(2025).
\url{https://github.com/alexyang0826/ADSL_Sumer_School_2025/issues/6}
Accessed 2025-12-29
\end{botherref}
\endbibitem

\bibitem[\protect\citeauthoryear{Praveen76}{2025}]{SP-praveen76_llmsapiu_2025_4f93e7}
\begin{botherref}
\oauthor{\bsnm{Praveen76}}:
LLMs-API-Usage-Best-Practices
(2025).
\url{https://github.com/Praveen76/LLMs-API-Usage-Best-Practices}
Accessed 2025-12-29
\end{botherref}
\endbibitem

\bibitem[\protect\citeauthoryear{feibaoBob}{2025}]{SP-feibaobob_llmbased_2025_b51625}
\begin{botherref}
\oauthor{\bsnm{feibaoBob}}:
LLM-based_CAMD.
\url{https://github.com/feibaoBob/LLM-based_CAMD}
Accessed 2025-12-29
\end{botherref}
\endbibitem

\bibitem[\protect\citeauthoryear{andrewyng}{2025}]{SP-andrewyng_aisuite_2025_1b4871}
\begin{botherref}
\oauthor{\bsnm{andrewyng}}:
Aisuite.
\url{https://github.com/andrewyng/aisuite}
Accessed 2025-12-29
\end{botherref}
\endbibitem

\bibitem[\protect\citeauthoryear{0x6f677548}{2025}]{SP-0x6f677548_unicodei_2025_1e6bb6}
\begin{botherref}
\oauthor{\bsnm{0x6f677548}}:
Unicode Injection Proof of Concept
(2025).
\url{https://github.com/0x6f677548/unicode-injection}
Accessed 2025-12-29
\end{botherref}
\endbibitem

\bibitem[\protect\citeauthoryear{DrPwner}{2025}]{SP-drpwner_promptsn_2025_d00848}
\begin{botherref}
\oauthor{\bsnm{DrPwner}}:
PromptSniffer.
\url{https://github.com/DrPwner/PromptSniffer}
Accessed 2025-12-29
\end{botherref}
\endbibitem

\bibitem[\protect\citeauthoryear{pablo
  chacon}{2025}]{SP-pablochaco_adversar_2025_ca6c98}
\begin{botherref}
\oauthor{\bsnm{pablo-chacon}}:
Adversarial LLM Threats
(2025).
\url{https://github.com/pablo-chacon/Adversarial-LLM-Threats}
Accessed 2025-12-29
\end{botherref}
\endbibitem

\bibitem[\protect\citeauthoryear{ridpath}{2025}]{SP-ridpath_llmvulne_2025_371dfc}
\begin{botherref}
\oauthor{\bsnm{ridpath}}:
LLM Vulnerability Scanner.
\url{https://github.com/ridpath/llm-vuln-scanner}
Accessed 2025-12-29
\end{botherref}
\endbibitem

\bibitem[\protect\citeauthoryear{grafbase}{2025}]{SP-grafbase_grafbase_2025_9f0e32}
\begin{botherref}
\oauthor{\bsnm{grafbase}}:
Grafbase/nexus.
\url{https://github.com/grafbase/nexus}
Accessed 2025-12-29
\end{botherref}
\endbibitem

\bibitem[\protect\citeauthoryear{Libr-AI}{2025}]{SP-librai_libraiop_2025_13948c}
\begin{botherref}
\oauthor{\bsnm{Libr-AI}}:
Libr-AI/OpenRedTeaming
(2025).
\url{https://github.com/Libr-AI/OpenRedTeaming}
Accessed 2025-12-29
\end{botherref}
\endbibitem

\bibitem[\protect\citeauthoryear{HqWu-HITCS}{2025}]{SP-hqwuhitcs_awesomel_2025_abbaf2}
\begin{botherref}
\oauthor{\bsnm{HqWu-HITCS}}:
Awesome-LLM-Survey
(2025).
\url{https://github.com/HqWu-HITCS/Awesome-LLM-Survey}
Accessed 2025-12-29
\end{botherref}
\endbibitem

\bibitem[\protect\citeauthoryear{Gong
  et~al.}{2025}]{SP-gongyichen_figstepj_2025_5e6a3d}
\begin{botherref}
\oauthor{\bsnm{Gong}, \binits{Y.}},
\oauthor{\bsnm{Ran}, \binits{D.}},
\oauthor{\bsnm{Liu}, \binits{J.}},
\oauthor{\bsnm{Wang}, \binits{C.}},
\oauthor{\bsnm{Cong}, \binits{T.}},
\oauthor{\bsnm{Wang}, \binits{A.}},
\oauthor{\bsnm{Duan}, \binits{S.}},
\oauthor{\bsnm{Wang}, \binits{X.}}:
FigStep: Jailbreaking Large Vision-language Models Via Typographic Visual
  Prompts.
\url{https://github.com/CryptoAILab/FigStep}
Accessed 2025-12-29
\end{botherref}
\endbibitem

\bibitem[\protect\citeauthoryear{0xAIDR}{2025}]{SP-0xaidr_aidrbast_2025_bcfdcc}
\begin{botherref}
\oauthor{\bsnm{0xAIDR}}:
AIDR Bastion.
\url{https://github.com/0xAIDR/AIDR-Bastion}
Accessed 2025-12-29
\end{botherref}
\endbibitem

\end{thebibliography}
\end{document}